\documentclass{ws-ijmpa}
\usepackage{here}
\def\al{\alpha}
\def\be{\beta}
\def\ga{\gamma}

\def\th{\theta}

\def\ps{\psi}

\def\cL{{\cal L}}

\def\fr#1#2{{{#1} \over {#2}}}
\def\frac#1#2{\textstyle{{{#1} \over {#2}}}}

\def\prt{\partial}

\def\half{{\textstyle{1\over 2}}}
\def\lsim{\mathrel{\rlap{\lower4pt\hbox{\hskip1pt$\sim$}}
    \raise1pt\hbox{$<$}}}
\def\gsim{\mathrel{\rlap{\lower4pt\hbox{\hskip1pt$\sim$}}
    \raise1pt\hbox{$>$}}}

\def\lrDmu{\stackrel{\leftrightarrow}{D_\mu}}

\def\lrDbe{\stackrel{\leftrightarrow}{D_\be}}
\def\lrhatDmu{\stackrel{\leftrightarrow}{\widehat D_\mu}}

\newcommand{\beq}{\begin{equation}}
\newcommand{\eeq}{\end{equation}}
\newcommand{\bea}{\begin{eqnarray}}
\newcommand{\eea}{\end{eqnarray}}

\begin{document}

\markboth{N.E. Mavromatos}
{String Quantum Gravity, LIV and Gamma-Ray Astronomy}

%
\catchline{}{}{}{}{}
%

\title{STRING QUANTUM GRAVITY, LORENTZ-INVARIANCE VIOLATION  AND GAMMA-RAY ASTRONOMY\footnote{Invited Review}
}

\author{NICK E. MAVROMATOS
}

\address{CERN, Theory Division, CH-1211 Geneva 23, Switzerland; \\ On leave from: Department of Physics, Theoretical Physics, King's College London, Strand\\
London, WC2R 2LS, UK
}

%

\maketitle

\begin{history}
\received{17th August 2010}
\end{history}

\begin{abstract}
 In the first part of the review, I discuss ways of obtaining Lorentz-Invariance-Violating (LIV) space-time foam in the modern context of string theory, involving brane world scenarios. The foamy structures are provided by lower-dimensional background brane defects in a D3-brane Universe, whose density is a free parameter to be constrained phenomenologically. Such constraining can be provided by  high energy gamma-ray photon tests, including ultra-high energy/infrared photon-photon scattering.
In the second part, I analyze the currently available data from MAGIC and FERMI Telescopes on delayed cosmic photon arrivals in this context. It is understood of course that conventional Astrophysics source effects, which currently are far from being understood, might be the dominant reason for the observed delayed arrivals.
I also discuss how the stringent constraints from studies of synchrotron-radiation from distant Nebulae, absence of cosmic birefringence and non observation of ultra-high-energy cosmic photons can be accommodated within the aforementioned stringy space-time foam model. I argue that, at least within the currently available sets of astrophysical data, the stringy foam model can avoid all theses constraints in a natural range of the string coupling and mass scale. The key features are: (i) transparency of the foam to electrons and charged probes in general, (ii) absence of birefringence effects and (iii) a breakdown of the local effective lagrangian formalism.  However, in order to accommodate, in this theoretical framework, the data of the FERMI satellite on the delayed arrival of photons from the short intense Gamma Ray Burst GRB 090510, in a way consistent with the findings of the MAGIC telescope, a non uniform density of brane foam defects must be invoked.

\keywords{String Foam; Lorentz Violation; Gamma Ray Astronomy}
\end{abstract}

\ccode{PACS numbers: 11.25.-w, 11.25.Uv, 11.30.Cp, 04.60.Bc, 95.85.Pw}

\section{Introduction and Summary}

\begin{figure}[ht]
\begin{center}
  \includegraphics[width=5.5cm,angle=-90]{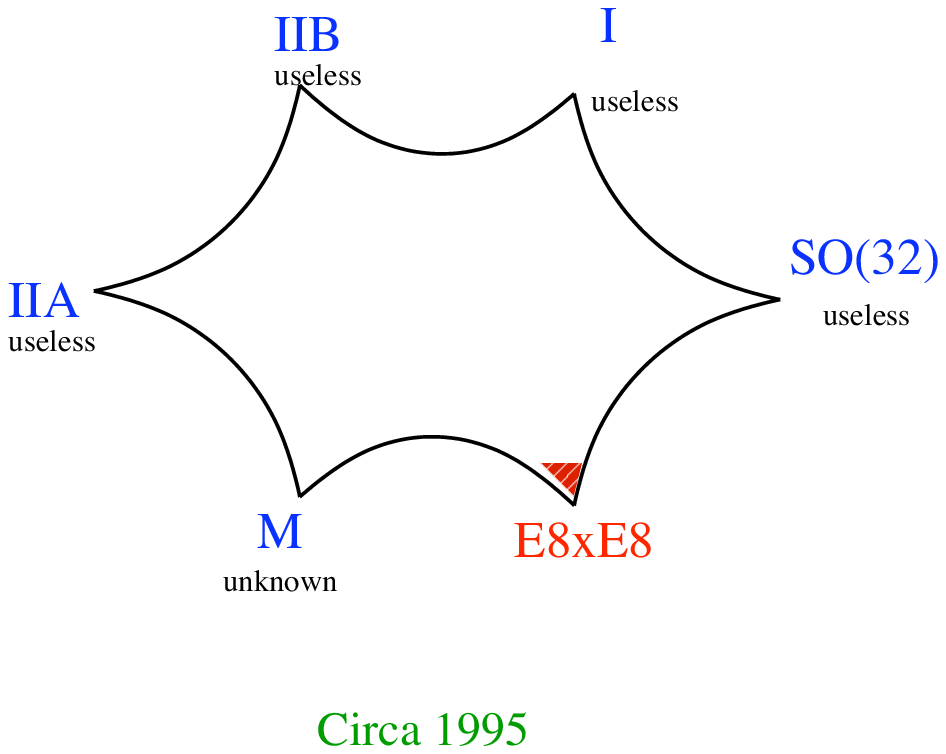} \hfill \includegraphics[width=5.5cm,angle=-90]{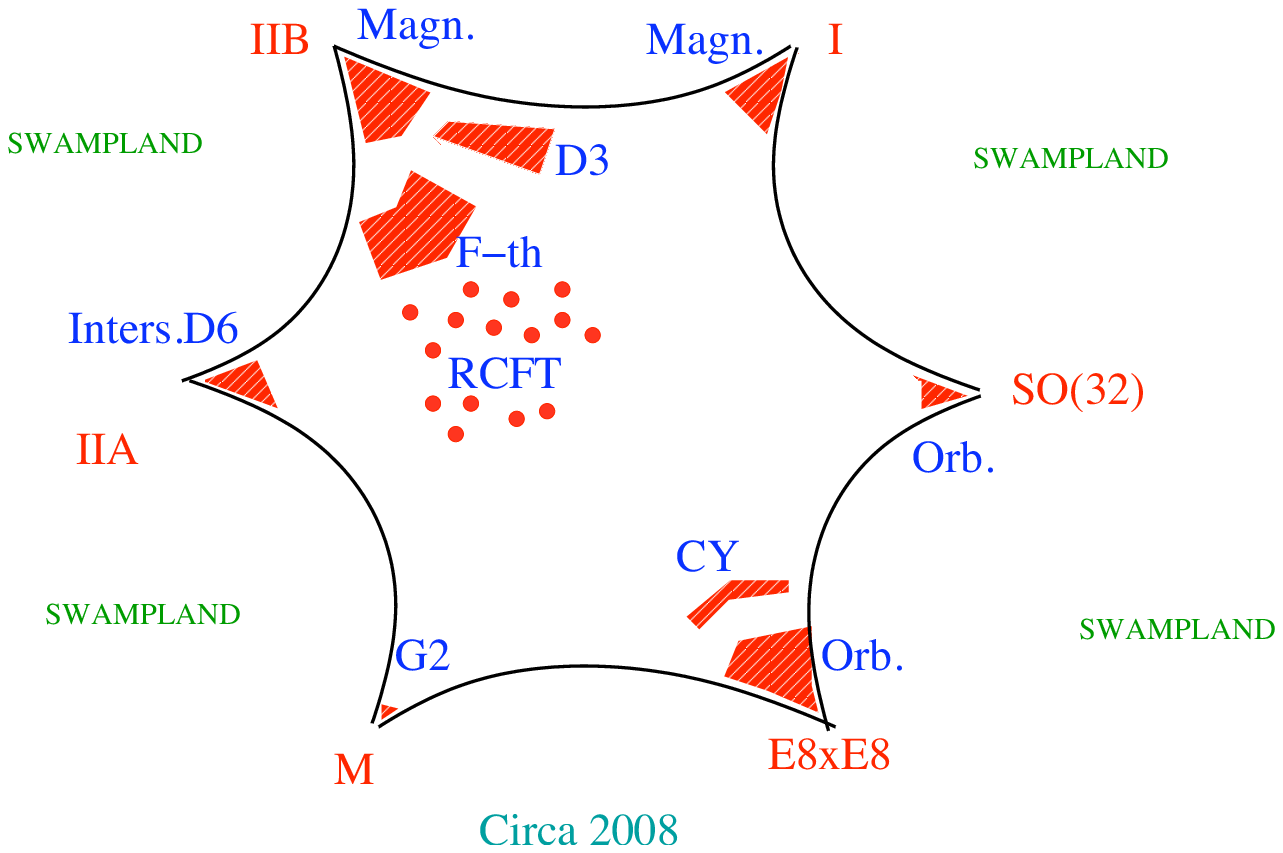}
\end{center}\caption{The diagram shows the current (quite limited) understanding of the M-theory (interior of polygon graphs) unifying all known string theories, which appear as its low-energy limits. The left diagram shows the understanding we had in the mid 90s and is placed here for comparison of the progress made. The characterisation \emph{useless} in the 1995 situation (upper panel)
indicates the ability of the corresponding string theory to include the standard model group.
The red-shaded interior regions lead to a partial understanding of M-theory issues, either
through F-theory (termed F-th) considerations or rational conformal field theory (RCFT) analysis.
The discovery (in 1996) of solitonic states in string theory, termed D(irichlet) branes changed the picture. Many of the previously thought as useless string theories can incorporate standard model groups in non-trivial ways, involving intersecting brane situations (termed Inters. on the lower panel) and compactification to magnetised manifolds (termed Magn. in the right picture), orbifold (termed Orb.) or Calabi-Yau spaces (CY) \emph{etc}.
The M low-energy theory on the lower right corner of the super-unification graphs is identified now with the 11-dimensional supergravity (Pictures \emph{taken from Ref.~\protect\cite{ibanez}.})}
\label{fig:mtheory}
\end{figure}

An alternative title for the review article could be \emph{MAGIC Strings}:
Usually the terminology \emph{M-theory}, with M standing for either \emph{Magical} or Marvelous or Mysterious, is attributed to the underlying (yet not completely understood) unifying theory of all known string theories~\cite{polch}, as a result of the many appealing duality and other symmetries it possesses, which result in the unification of the five known string theories, viewed as a low-energy limit of M-theory. This is a super-unification picture, which may prompt the way for a detailed understanding of the yet elusive theory of the quantum structure of space-time, otherwise termed as ``Quantum Gravity'' (QG). Our current knowledge/understanding of M-theory is limited. Schematically, this knowledge is restricted to a few small regions of the interior of the diagram of fig.~\ref{fig:mtheory}.
Nevertheless, for an analysis of some of the predictions of string theory that could have some relevance to observable low-energy physics this may not be an obstacle, as we shall attempt to discuss in this work. We shall put emphasis on astrophysical tests of some versions of string theory entailing Lorentz-Invariance Violating space-time foam structures
for the quantum gravity ground states.

\begin{figure}[ht]
\begin{center}
  \includegraphics[width=3cm, angle=0]{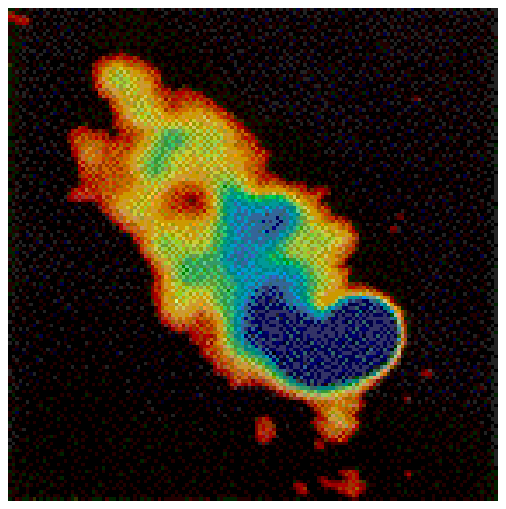} \hspace{2cm}
 \includegraphics[width=5cm,angle=0]{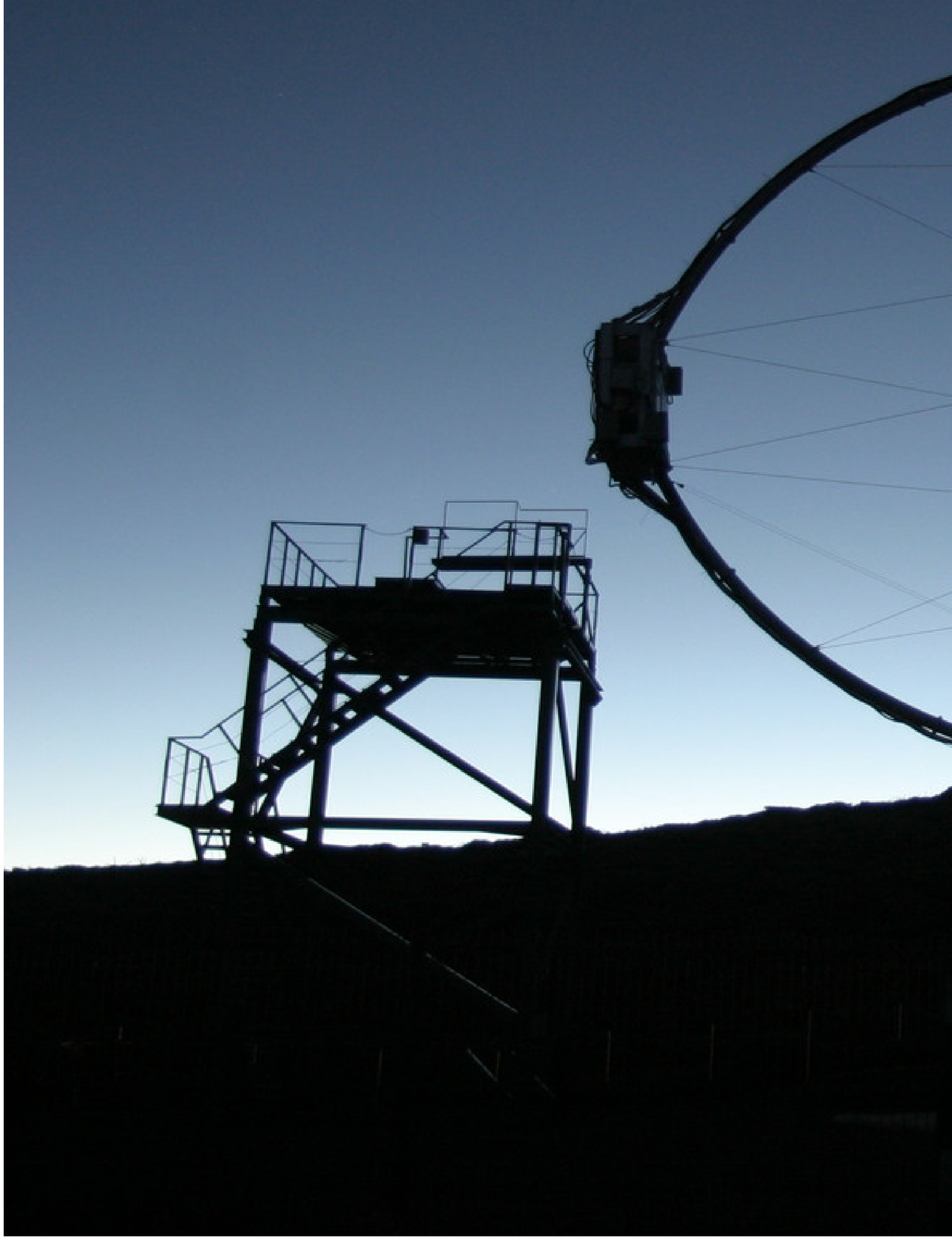} \vfill
\end{center}
\caption{Major Atmospheric Gamma-ray Imaging Cherenkov Telescope at the Canary Islands (Spain) Observatory (right panel). The telescope observed very high energy gamma rays, with energies up to the order of 10 TeV, from the active galactic nucleus Markarian 501 (radio image on left panel, by J.M. Wrobel and J.E. Konway,
picture taken from \texttt{http://www.vlba.nrao.edu/whatis/mark.html}).}
\label{fig:magic}
\end{figure}

This brings us to the meaning of the word \emph{MAGIC} used in this review.
Here \emph{MAGIC} is
an acronym pertaining to the initials describing the full name of a Physics Instrument
(\emph{M.A.G.I.C} = \emph{\textbf{M}}ajor \emph{\textbf{A}}tmospheric \emph{\textbf{G}}amma-ray \emph{\textbf{I}}maging \emph{\textbf{C}}herenkov telescope). Specifically, it refers to a Telescope based on the Canary Islands observatory (\emph{c.f.} fig.~\ref{fig:magic}), dedicated to the study of
Cherenkov radiation emitted by highly energetic cosmic particles as they enter our atmosphere.
From the study of the emitted Cherenkov radiation one can deduce several important conclusions on the nature of the initial particle and through this to try to understand the mechanisms of production of such energetic cosmic particles.

\begin{figure}[ht]
\begin{center}
\includegraphics[width=10cm]{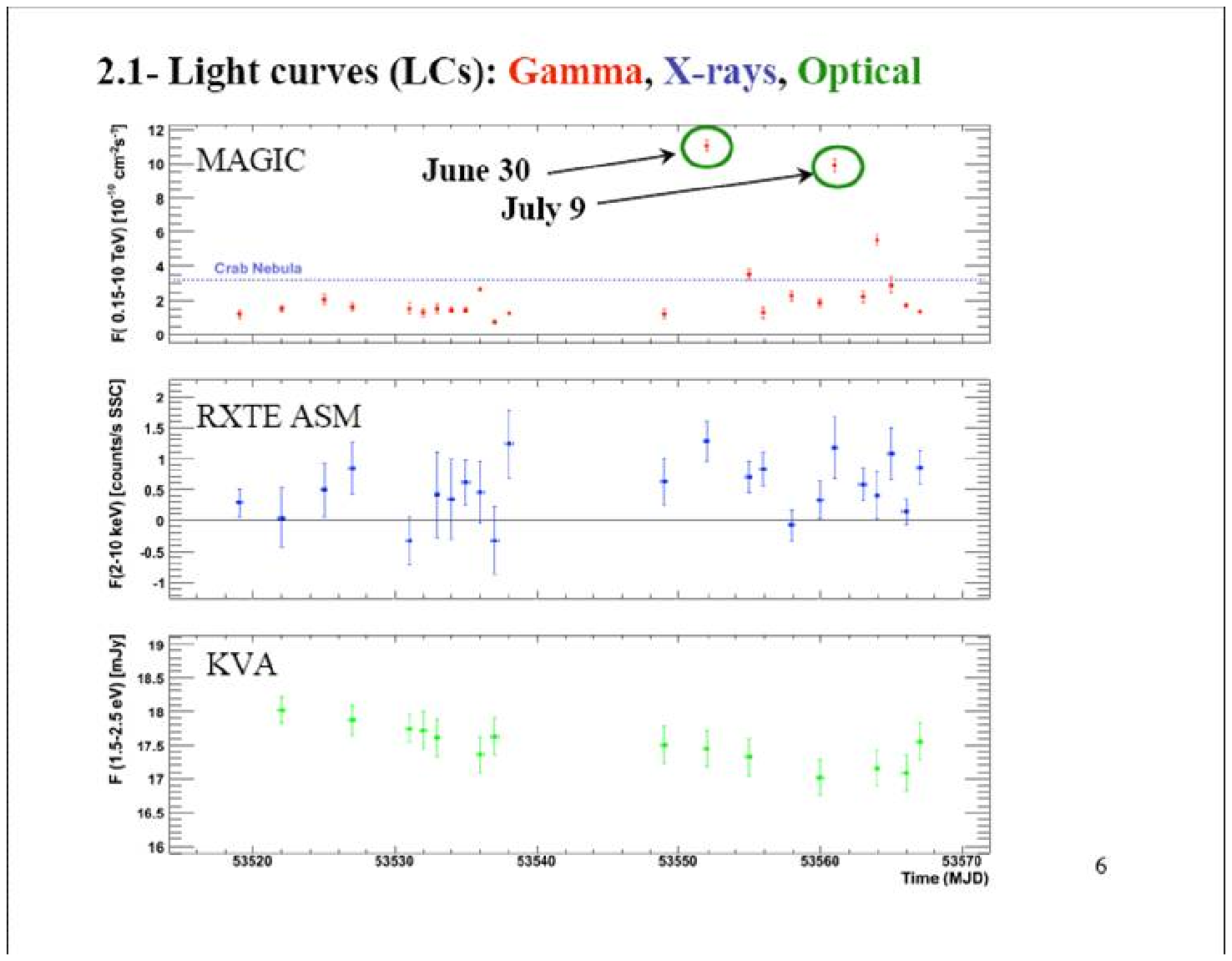}  \hfill \includegraphics[width=10cm]{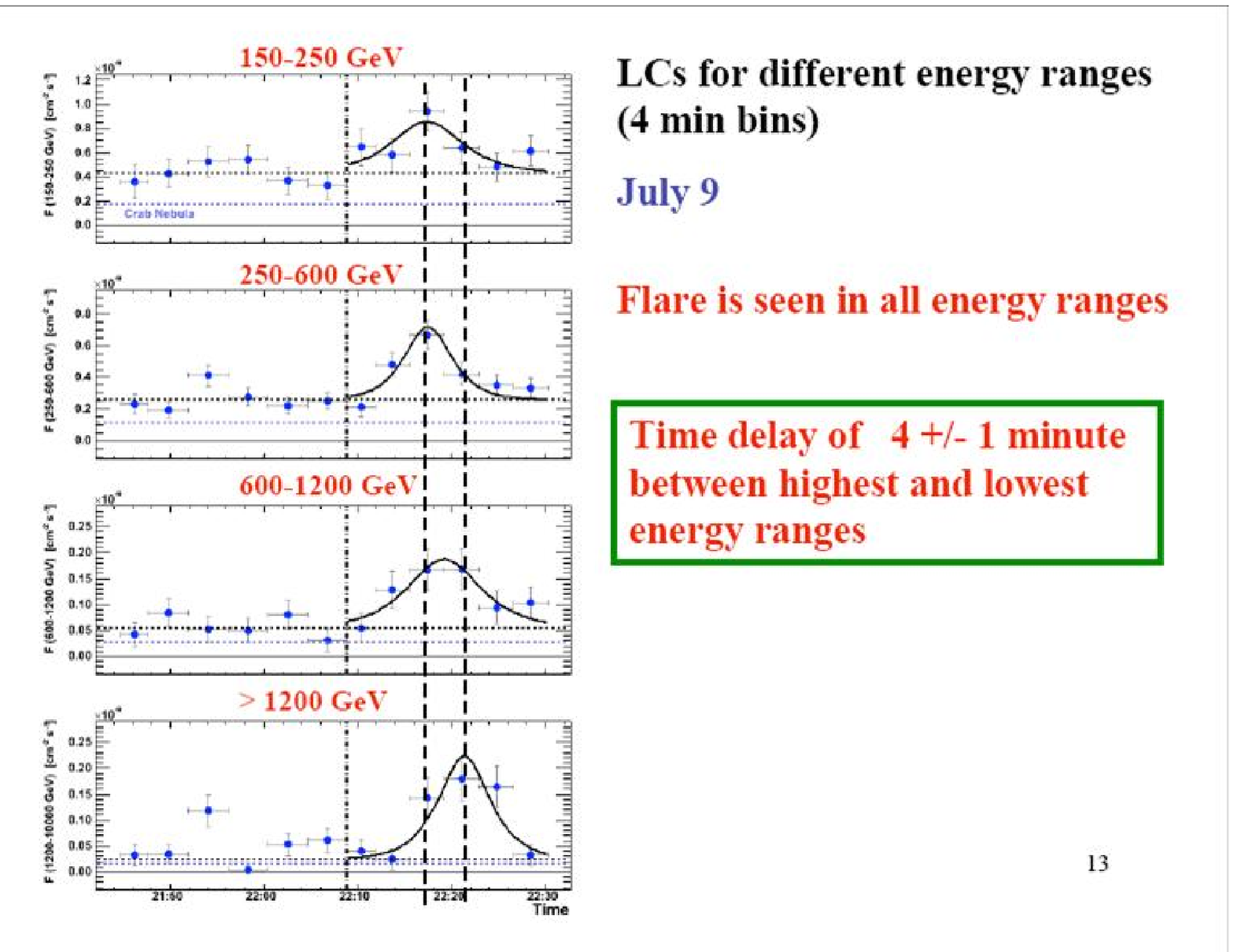}
\end{center}
\caption{ The observations of the MAGIC telescope~\protect\cite{MAGIC} regarding very high energy Gamma Rays (with energies in the TeV range) showed that the most energetic photons
were delayed up to four minutes as compared with their lower-energy counterparts (in the 0.6 TeV or lower range). The figures show light curves (LC), \emph{i.e}. the photon flux vs. time of arrival. Observations like this may be used to prompt new fundamental physics on the structure of space-time. The lower panel shows LC  at different energy ranges, demonstrating clearly the time delay (of order of 4 $\pm 1$ minutes) of the more energetic photons.}
\label{fig:magic2}
\end{figure}

On July 9th 2005, the telescope observed~\cite{MAGIC} (\emph{c.f.} fig.~\ref{fig:magic}) very high energy gamma rays from the active galactic nucleus Markarian 501 (Mkn 501),
which lies at red-shift $z=0.034$  (\emph{i.e}. about half a million light years away from Earth).
 The highest energy photons with energy of order 10 TeV  (1 TeV = $10^3$ GeV = $10^{12}$ eV), were delayed up to four minutes as compared with their lower-energy counterparts (in the 0.6 TeV or lower range) (\emph{c.f}. fig.~\ref{fig:magic2}). This was the first observation
of such a distinct delay.

Three years after the MAGIC observations, in September 2008, the FERMI (formerly known as GLAST) Satellite Telescope~\cite{glast}, also observed time delays of the higher-energy photons, from the distant Gamma Ray Burst (GRB) 080916c~\cite{grbglast}, at red-shifts $z = 4.35$,  and later on from GRB 090510~\cite{grb090510}, at red-shift $z=0.9$ and from GRB 09092B, at redhifts $z = 1.822$~\cite{grb09092b}. GRBs are cosmic explosions of titanic proportions, due to collapsing massive stars at distant parts of the universe. As we shall discuss in this review article, the pattern of these delays fits~\cite{emnfit} a string model of quantum-gravity-induced refractive index, with the pertinent quantum gravity energy scale being essentially the same as that inferred from the MAGIC observations (of order $10^{18}$~GeV). Viewed as a lower bound, this scale is also compatible with that obtained from other Gamma-Ray data of the H.E.S.S. Collaboration~\cite{hess2155,hessnew}.
However, as we shall discuss in detail in this article, in order to accommodate the findings of MAGIC with the recently observed time delays of photons from the extremely short Gamma-Ray burst GRB 090510~\cite{grb090510}, provided that the latter can be trusted~\footnote{Indeed, there are uncertainties in this measurement concerning, for instance, the precise emission time of photons due to the duration of pre-cursors to burst activities and other such issues, which need to be confirmed by other measurements of similarly short bursts, that presently are lacking.}, one needs
inhomogeneous densities of defects in the foam~\cite{emndvoid}, and in this sense such astrophysical observation are quite essential in falsifying models.

Of course, it goes without saying that the delay effects may be due to the conventional astrophysics of the active galactic nucleus or Gamma-Ray Burst (\emph{source} effect), which, however, as we shall discuss below, is not well understood at present. In fact, currently there seem to be no consensus among the astrophysicists on the appropriate mechanism for the production of such high-energy photons at the source.
These uncertainties prompted more ambitious, although admittedly far-fetched, explanations~\cite{MAGIC2},
pertaining to  new fundamental physics, affecting the photon \emph{propagation} due to space-time \emph{foamy} vacuum structures that lead to modified  dispersion relations for photons~\cite{robust,mitsou,JP}. If true, this would be a clear
departure from the Lorentz invariant energy (E)-momentum ($\vec p$) relations of Special Relativity, $E=|\vec p|c$.

The reader should bear in mind that
at small length scales, of the order of the Planck length, $\ell_P = \sqrt{\frac{\hbar G_N}{c^3}} = \frac{\hbar}{M_P c} \sim 10^{-35}$~m, which is the characteristic scale at which quantum gravity effects are expected to become dominant, the structure of space time may be quite different from what we perceive at our (low-energy) scales. It might be even discrete and non-commutative, that is the space-time coordinates (as we perceive them at present) might be average values of non commutative quantum operators. Moreover, one may have highly curved non-trivial fluctuations of the space-time metric, giving space time a ``\emph{foamy}'' structure~\cite{wheeler}. In such complicated Quantum Gravity (QG) vacua, the concept of Lorentz symmetry might break down at these short length scales, pointing towards the possibility of \emph{spontaneous Lorentz symmetry breaking} by the QG ground state, since if Lorentz symmetry is intact,
it strictly prohibits such modifications in the dispersion relations. Such departures from the
 standard Special Relativity form of dispersion relations
correspond to the generation of a momentum dependent mass gap for the photon, and hence a non trivial
refractive index, as the photon propagates through the \emph{medium} of quantum gravity.

It is the point of this review to touch upon such issues, through the description of
the most important physical consequences of such a breaking. As we shall see, surprisingly enough, many models of QG that entail such a breaking can already be falsified in current astrophysical experiments, which set very stringent bounds on Lorentz symmetry breaking. Some of the models we are going to discuss have experimental consequences that can be tested (or are already falsified(?)) in Nature, at least in principle. In this article we shall discuss in some detail only one class of theories of QG that can entail such a breaking, which is a subset of the modern version of string theory, including brane (domain-wall-like) defects in space time. Such defects will play the r\^ole of the non trivial space time structures that would be deemed responsible for the spontaneous breaking of the Lorentz symmetry  by the ground state of these systems.

As we shall discuss here, there are very stringent conditions for such
exotic explanations of the MAGIC observations to come into play in agreement with
the plethora of many other astrophysical tests of Lorentz symmetry, existing currently. Specifically,
the space-time foam must be transparent to electrically charged probes, such as electrons, while photons should exhibit non-trivial refractive indices in this theory, but with no birefringence effects.
Moreover, the local effective lagrangian description of the effects, that is a low-energy representation
of the QG medium effects in terms of local higher derivative operators in flat space-time backgrounds,
should break down.

Thus, although one cannot exclude the possibility that
\emph{both} effects, \emph{source} and \emph{propagation} due to quantum gravity, may be simultaneously responsible for the observed photon delays in the MAGIC experiment, nevertheless the available theoretical models that do the job are very limited.
The thesis of this article will be that only certain models~\cite{emnw,emnnewuncert,li} of (the modern version of) string theory, including space-time defects, whose dynamics breaks Lorentz symmetry and provides a ``foamy'' structure of space-time, can satisfy the above conditions and thus offer an explanation for the MAGIC photon-arrival-times anomaly, consistently  with all other current astrophysical constraints on Lorentz symmetry violations.
The string model(s) provide an explanation for the observed photon time delays, in a natural range of the string coupling and mass scale, and avoid all the other stringent constraints of Lorentz Invariance Violation coming from non observation of birefringence effects or of very high energy photons~\cite{sigl}.

Our point in this article is not to advocate string theory as a superior candidate to other quantum gravity models available to date, but rather to discuss situations in which string theory predictions can be falsified by experiment. And high energy photon astrophysics may be a useful arena for this purpose!
Of course, it goes without saying that, at present, we are very far from reaching any conclusions on such matters. Definite falsification of these stringy quantum gravity scenarios, if at all possible,  would require a plethora of further studies by means of other high-energy astrophysics or particle physics processes and observations.

The structure of the article is as follows: in the next section, \ref{sec:LV}, as a means of introduction, I discuss   issues pertaining to the violation of Lorentz invariance in media or
quantum electrodynamics vacua with non-trivial vacuum refractive index, \emph{e.g}. the case of thermal plasma. Such cases may be thought of as (simplified) analogues of Lorentz-Violating Quantum Gravity (QG) space-time foam vacua, but
they also characterize the source regions of the cosmic high energy Gamma Rays.
In this latter sense, it is important to understand photon propagation in those cases first, so as to disentangle possible source effects from the QG-induced ones. In section \ref{sec:mainstring}, I proceed to the main topic of this review, namely a discussion on the string/brany model of space-time foam. The model involves membrane-like defects in a higher-than-four dimensional space time,
with our world being viewed as a hyper-membrane (D(irichlet)-brane~\cite{polch}) embedded in this space time.
The model is of the kind of large-extra-dimension models to be tested at LHC and future colliders, with the important ingredient of having point-like space-time (D-particle) defects, responsible for the ``foamy'' structure of space time.
I describe first the cosmological features of this model, specifically in connection with modifications of the energy budget of the Dark Sector (Dark Matter and Dark energy), compared to the standard $\Lambda$CDM model. Then, I proceed to discuss its ``optical properties'', namely the induction of effective vacuum refractive indices, the absence of birefringence, and the break down of a local effective lagrangian formalism insofar as the description of the induced time delays of high energy photons (due to interactions with the foam) are concerned. I also discuss a novel ultra-violet ``cutoff'' that may characterize ultra high energy photon scattering for specifically stringy reasons.
In  section \ref{sec:3}, I discuss high energy Gamma Ray astrophysics as a potential probe of the stringy space-time foam model. First, I analyze possible interpretations of the MAGIC (and FERMI) observations, including exotic ones involving QG dispersive media. Then, I discuss bounds and sensitivities of various astrophysical experiments, and state carefully the stringent requirements that must be met by a theoretical model of QG, in order for the observed delays in the MAGIC and FERMI Telescopes to be attributed to effects due to a space-time-foam medium, in agreement with all other current tests. All these requirements are surprisingly respected by the string model discussed in section \ref{sec:mainstring}.  Finally, section \ref{sec:5} contains concluding remarks and an outlook.

A note is in order at this point concerning the \emph{units} used in this work: throughout the article, unless otherwise stated, we shall work in \emph{natural Planck units}, in which $\hbar = c = 1$. In these units, length and time are identified, and they are inversely proportional to mass or energy. The
 latter are also identified and expressed in units of multiples of eV (=$1.6 \times 10^{-19}$ Joules), in particular GeV (=$10^9$ eV) and TeV (=$10^{12}$ eV) in this work. From time to time, for concreteness, the speed of light \emph{in vacuo} $c$ and Planck's constant $h$ or $\hbar =h/2\pi$ may appear explicitly in some formulae.

\section{Lorentz Invariance and the Vacuum Structure of Quantum Fields \label{sec:LV}}

One of the cornerstones of Modern Physics is Einstein's theory of Special Relativity (SR), which is based on the assumption that the speed of light in vacuo $c$ is an invariant under all observers. In fact, this implies
invariance of the physical laws under the Lorentz transformations in flat space times, and the r\^ole of $c$ as a universal limiting velocity for \emph{all} particle species.

The generalization (by Einstein) of SR to include curved space times, that is the theory of General Relativity (GR), encompasses SR locally in the sense of the \emph{strong form of the equivalence principle}. According to it,  \emph{at every space-time point, in an arbitrary gravitational
field, it is possible to choose a locally
inertial (`free-float')  coordinate frame, such that within a sufficiently
small region of space and time around the point in question,
the laws of Nature are described by special relativity, \emph{i.e.} are of the
same form as in unaccelerated Cartesian coordinate frames in the absence
of Gravitation.} In other words, locally one can always make a coordinate transformation
such that the space time looks {\it flat}. This is not true globally, of course, and this is why GR is a more general theory to describe gravitation. The equivalence principle relies on another fundamental invariance of GR, that of general coordinate, that is the invariance of the gravitational action under arbitrary changes of coordinates. This allows GR to be expressed in a generally covariant form.

In such a locally \emph{Lorentz-invariant} vacuum,
the photon dispersion relation, that is a local in space-time relation between the photon's four-wavevector components $k^\mu =(\omega, \vec k)$ (where $\omega$ denotes the frequency, and $\vec k$ the momentum) reads in a covariant notation:
\begin{equation}
  k^\mu k^\nu \eta_{\mu\nu} = 0
\label{photonflat}
\end{equation}
where repeated indices $\mu, \nu = 0, 1, \dots 3$, with $0$ referring to temporal components, denote summation and $\eta_{\mu\nu}$ denotes the Minkowski space-time metric, with components $\eta_{00} = -1, ~\eta_{0i}=\eta_{i0}=0, ~ \eta_{ij}=\eta_{ji}=\delta_{ij}~, i = 1,2,3$ with $\delta_{ij}$ the Kronecker delta symbol.

The above relation (\ref{photonflat}) implies the equality of all three kinds of photon velocities \emph{in vacuo} that stem from its wave nature (due to the particle-wave duality relation):
\begin{eqnarray}
&& {\rm phase}: \qquad v_{\rm ph} = \frac{\omega}{|\vec k|} \equiv \frac{c}{n(\omega)} = c \nonumber \\
&& {\rm group}: \qquad v_{\rm gr} = \frac{\partial \omega }{\partial |\vec k|} \equiv \frac{c}{n_{\rm gr}(\omega)} = c~,~ \quad n_{\rm gr}(\omega) = n(\omega) + \omega \frac{\partial n(\omega)}{\partial \omega}  \nonumber \\
&& {\rm front}: \qquad v_{\rm front} = c/n(\infty) = c
\label{velocities}
\end{eqnarray}
since the phase and group \emph{refractive indices} of the trivial \emph{vacuum } equal unity $n(\omega) = n_{\rm gr}(\omega) = 1$. For brevity we shall work from now on in units where $c=1$.

\subsection{Photon Propagation in Conventional Media  and in Non-trivial field-theory vacua with a refractive index \label{sec:ntv}}

The above results (\ref{velocities}) change significantly when light propagates in a material \emph{medium}, in which its speed is different from $c$ \emph{in vacuo}. This is due to the non trivial refractive index the material has, as a result of the electromagnetic interactions of the photon with the electrons in the medium. As we shall discuss later on, this case seems to bear some quite instructive analogies with our string model of space time foam, which lies at the focus of our attention in this review.

The simple model of quantum oscillators has been adopted by Feynman~\cite{feynman} as a
simplified but well motivated analogue for describing the situation in  the case of
photons in ordinary media.
In that case, the electrons of the medium, of mass $m$, are represented by simple harmonic oscillators, with frequency $\omega_0$,
which provides the necessary \emph{restoring force} during the scattering of light off the electrons in order to keep the latter oscillating around their initial position.
In that problem the induced refractive index is obtained by the reduction of the phase velocity of the photon wave.

We consider~\cite{feynman} the electrons, of mass $m$, as forced simple-harmonic oscillators with a resonant frequency $\omega_0$, subject to the force $F$ exerted by an
oscillating external electric field of frequency $\omega$: $F= e E_0 e^{i \omega t}$, where $e$
is the electron charge.  The corresponding equation of motion is:
\begin{equation}
m \left(\frac{d^2}{dt^2} x + \omega_0^2 x \right) = e E_0 e^{i \omega t}
\end{equation}
Assuming for concreteness a plate of thickness $\Delta z$, representing the medium through which an electromagnetic wave travels, where $z $ is perpendicular to $x$,
one can compute in a standard way the electric field $E_a$ produced by the excited atoms:
\begin{equation}
 E_a = -\frac{e n_e}{\epsilon_0 c}i \frac{e E_0}{m (\omega^2 - \omega_0^2)}e^{i\omega (t - z)} ,
\label{elfield}
\end{equation}
where $\epsilon_0$ is the dielectric constant of the vacuum and $n_e$ is the area density of
electrons in the medium (plate), which is given by $n_e = \rho_e \Delta z$,
where $\rho_e$ is the volume density of electrons.

We next recall that light propagates through a medium with a refractive index $n$ with a
speed $c/n$, causing a delay $\Delta t$ in traversing the distance $\Delta z$, given by:
\begin{equation}\label{delayfeynman}
\Delta t = (n - 1)\Delta z/c~.
\end{equation}
Representing the electric field before and
after passing through the plate as
$E_{\rm before} = E_0 e^{i\omega (t - z/c)}$ and
$E_{\rm after} = E_0 e^{i\omega (t - z/c - (n-1)\Delta z/c)}$, in the case
of small deviations from the vacuum refractive index
we have: $E_{\rm after} \simeq E_0 e^{i \omega(t - z/c)} - i[\omega(n -1)\Delta z/c]
E_0 e^{i\omega(t-z/c)}$.
The last term on the right-hand-side of this relation is just the field $E_a$ produced
in the region of space after
the plate by the oscillating electrons.
We then obtain from (\ref{elfield}):
\begin{equation}
(n - 1)\Delta z = \frac{n_e e^2}{2\epsilon_0 m (\omega_0^2 - \omega^2)},
\end{equation}
and hence the following formula for the refractive index in a conventional medium:
\begin{equation}
   n = 1 + \frac{{\rho_e}_e e^2 }{2\epsilon_0 m (\omega_0^2 - \omega^2)}.
   \label{refrordinary}
   \end{equation}
We see in (\ref{refrordinary}) that the refractive index in an ordinary medium is inversely proportional to (the square of) the frequency $\omega$ of light, as long as it smaller than the oscillator
frequency, where the refractive index diverges.

If the couplings of the two polarizations of the photon to the electrons in the medium are different,
the phenomenon of birefringence emerges, namely different refractive indices for the two
polarizations. Moreover, we see from (\ref{refrordinary}) that the propagation of light is
subluminal if the frequency (energy) of the photon $\omega < \omega_0$, whereas it is
superluminal for higher frequencies (energies)~\cite{feynman}.
This reflects the fact that the phase shift induced for the scattered light can be either positive or negative, but there such a superluminal refractive index causes no issue with causality,
since the speed at which information may be sent is still subluminal.

As we shall see later on, this conventional model will provide us with a rather good analogue of what happens in some string models of quantum space time foam, also characterized by non-trivial refractive indices, which we shall analyse in section \ref{sec:mainstring}.
However, as we shall see there, contrary to the conventional situation discussed in this section, in the string case the refractive index is found proportional to the photon frequency, while the effective mass scale that suppresses the effect is the quantum gravity (string) scale and not the electron mass as in (\ref{refrordinary}).

A word of caution on this point concerns the disentanglement of the time delay due to a non-trivial refractive index (\ref{delayfeynman}) from a modified dispersion relation. In some cases the two are equivalent. In the string model, however, discussed in section \ref{sec:string}, the pertinent delays will be associated with stringy uncertainties of the foam, and will not be directly related to modified dispersion relations for photons~\cite{emnnewuncert,emncomment}.

\begin{figure}[ht]
 \begin{center} \includegraphics[width=9cm]{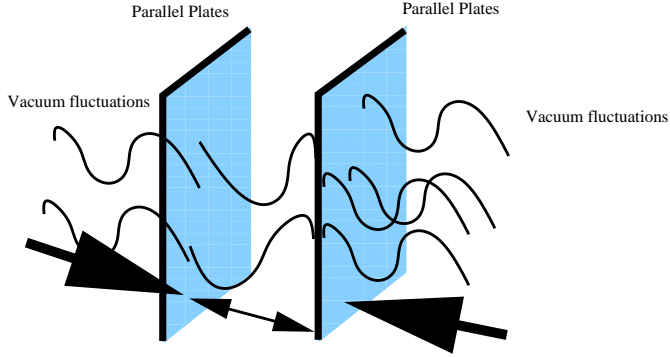} \end{center}
\caption{The Casimir Vacuum of Quantum Electrodynamics: in the compact space between the parallel plates (or more generally in other compact geometries), the quantum fluctuations of the electromagnetic field are responsible for the macroscopically measured
Casimir force between the plates, scaling with the size $L$ of the compact space as $L^{-4}$ in four space-time dimensions. The force can be either attractive or repulsive, depending
on the geometric set up. However, besides this force from ``nothing'', the Casimir Vacuum provides another interesting phenomenon in quantum physics. According to calculations by Scharnhorst and collaborators, the presence of the boundaries
breaks manifestly Lorentz invariance, and results in
modified dispersion relation, and hence a non-trivial refractive index, for the virtual photons
of the \emph{non-trivial Quantum Electrodynamics vacuum} in the compact region. From this latter perspective, the situation is entirely analogous to what happens in thermal plasmas, such as the ones occurring in the interior of stars or source regions of active galactic nuclei, which we are interested in in this work. There, the periodic boundary conditions of the Casimir case are replaced by similar ones due to the finite temperature, which also break Lorentz symmetry of the ground state.}
\label{fig:casimir}
\end{figure}

The r\^ole of a non-trivial refractive index material can be played under certain circumstances by
a \emph{non-trivial vacuum} in which photons propagate, such as quantum electrodynamics at finite temperature plasmas~\cite{latorre} or the Casimir vacuum between parallel capacitor plates~\cite{scharn} (or other geometries, as long as the space is bounded appropriately)
(\emph{c.f.} fig.~\ref{fig:casimir}). In such cases, the loop corrections due to \emph{vacuum polarization}, \emph{i.e.} creation and annihilation of virtual electron-positron pairs, in quantum electrodynamics (QED) result in a modified photon propagator, and a non-trivial group velocity and refractive index $n(\omega) \ne 1$. The reason for this is the breaking of Lorentz invariance, due to either the existence of spatial boundaries (Casimir vacuum) or finite temperature (thermal vacuum in case of plasmas). Moreover, one may consider quantum electrodynamics in a homogeneous and isotropic Friedman-Robertson-Walker (FRW) expanding-Universe background~\cite{hath} and examine the non-trivial effects of vacuum polarization on photon propagation there. This was in fact historically the first instance where the effects of curvature induced \emph{superluminal} propagation for low-frequency photon modes.

The situation concerning all the above cases can in fact be represented in a rather unified way by the following formula for the group velocity of (low-energy) photons, due to vacuum polarization in the context of four-dimensional QED in such non-trivial vacua~\cite{latorre}:
\begin{equation}
v_{\rm gr} = 1 - \frac{44}{135}\alpha^2 \frac{\rho}{m_{\rm e}^4}~.
\label{unified}
\end{equation}
This formula is valid in all cases except the Gravitational Background case of \cite{hath}, where $\alpha^2$ should be replaced by $\alpha m_{\rm e}^2 G_{\rm N}$, with $G_N$ the Newton gravitational constant.
Here,  $\rho$ denotes the energy density \emph{relative} to the standard vacuum, and can be negative or positive.
In the (four space-time dimensional) Casimir vacuum, for instance, the presence of boundaries in space (e.g. capacitor parallel plates at a distance $L$ in the simplest geometry), imply loss of low-energy photon modes, and as such the energy density of the vacuum is lowered relative to the standard one (without the boundaries). In this case $\rho = - \frac{\pi^2}{720 L^4} < 0$ in (\ref{unified}), and one has the phenomenon that the effective low-energy photon modes appear to propagate in a \emph{superluminal way}, $v_{\rm gr} > 1$ (we give here the formula for propagation \emph{perpendicular} to the plates for concreteness and ease of comparison with the plasma case later on):
\begin{equation}
v_{\rm gr}^{\rm Casimir} = 1 + \frac{11 \pi^2}{8100}\alpha^2 \frac{1}{L^4 m_{\rm e}^4} > 1
 \label{casimir}
 \end{equation}
 As explained in \cite{latorre}, there is no contradiction with relativity here, as this result applies only to low energy photons, with energies much lower than $m_{\rm e}$, and indicates the effective loss of degrees of freedom due to the spatial boundaries.

In the plasma case at finite temperature $T$, which characterizes the interior regions of stars or
active galactic nuclei, of interest to us in this work,
there is a formal analogy~\cite{latorre} between the r\^oles played by the temperature $T$ and the plate separation $L$ in the Casimir vacuum set up. They both break Lorentz invariance in a rather similar way. In fact, there is a correspondence between the Casimir and thermal-plasma vacuum formulae by replacing $2T $ by $L^{-1}$. In the plasma case, the low energy photon modes with momentum $k^2 \ll m_{\rm e}^2$ ($k \equiv |\vec k|$) have a group velocity:
\begin{equation}
v_{\rm gr}(kT \ll m_{\rm e}^2 ) = 1 + \frac{11 \pi^2}{8100}\alpha^2 \left(\frac{2T}{m_{\rm e}} \right)^4 ~> ~1
\end{equation}
which can be directly compared with the Casimir one (\ref{casimir}) upon the substitution $2T \to L^{-1}$.

For future reference we give the group velocity for the high-energy (as compared to the effective QED scale $m_{\rm e}$) photon modes~\cite{latorre} in the finite $T$ plasma case:
\begin{equation}
v_{\rm gr}(kT \gg m_{\rm e}^2 ) = 1 - \frac{\alpha^2 }{6}\left(\frac{T}{k}\right)^2 {\rm ln}^2\left(\frac{kT}{m_{\rm e}^2}\right) ~< ~1
\label{highenergy}
\end{equation}
which again can be compared with the corresponding Casimir vacuum case (\ref{casimir}), for photon modes \emph{perpendicular} to the plates~\cite{scharn}, upon the replacement $2t \to L^{-1}$.
The reader should notice here the momentum-$(k)$ dependence of the subluminal velocity for the high-energy photon modes.
In such non-trivial vacua, therefore, high energy photon modes will exhibit a momentum dependent subluminal refractive index, which will affect their arrival times at the observation point, if one considers simultaneous emission of modes within a certain energy range. From (\ref{highenergy}) we observe that the higher the momentum $k$ of the high-energy photon mode the higher the group velocity, since $\frac{\partial v_{\rm gr}(kT \gg m_{\rm e}^2)}{\partial k} = \frac{\alpha^2 T^2}{3k^3}\left({\rm ln}(\frac{kT}{m_{\rm e}^2})\right) > 0$, and thus fast modes will arrive first if emitted simultaneously in this vacuum.
Similar effects also characterise the curved background case of \cite{hath}, where again the high-momentum photon modes are subluminal.

We remark at this point that the relation (\ref{highenergy}) can characterise the source regions of active galactic nuclei, and thus
such effects can be responsible for inducing time delays of photons from these regions.
We shall come back to this point when we shall discuss the MAGIC delays in section \ref{sec:3magicqg} below.
As we shall see there, however, the plasma-induced time delays are much smaller than the observed delays in the MAGIC experiment, and hence cannot provide the dominant explanation for the effect.

At this point it is instructive to make some clarifications regarding the speed of light and causality, that is the fact that signals do not arrive before they occur.
The phase, front and group velocities
have all been found to exceed the value of the speed of light in vacuo. However, this is not in conflict with
causality.

Indeed, the various light velocities do not have to be subluminal, as they carry no information.  Information can be transmitted by (or not) sending pulses. The information then \emph{appears} at first sight to propagate with the group velocity, \emph{i.e.} the velocity of the peaks of the pulses. However,
as experimentally demonstrated in several instances since the 1980's, the group velocity of the photons can also be super-luminal. This is because the group velocity can be disentangled from the \emph{signal velocity}, \emph{i.e.} the speed by which information can be transferred, and thus it can exceed the speed of light \emph{in vacuo}, $c$, without contradicting the \emph{causality} requirements of Special Relativity. Group velocities larger than $c$ can occur \emph{e.g}. in tunneling experiments and appear to lead to superluminal transmission. However, there is no
contradiction with causality or Special Relativity in such cases. The error lies in identifying the peak of the pulse with the temporal position of the carried information. For example, a Gaussian-shaped pulse can be detected long before its peak due to the rise of intensity at earlier times. Therefore, a different kind of signal must be considered, where no information at all is sent out before a certain moment of time. For such signals, it can be proven that the earliest time at which that switching event can be observed is limited exactly by propagation with the vacuum velocity $c$. A so-called \emph{precursor} is traveling with that speed, but is normally too weak to be detected, except in certain circumstances.
For direct measurements of optical precursors in regions with anomalous dispersion the reader is referred, for instance, to ref.~\cite{precur}.

Super-luminal group photon velocities have been measured in laser pulses passing through specially prepared materials in ref. \cite{brunner}. Using special set up involving optical fibers, a super-luminal  group velocity of photons (in the fiber) has been measured and found different from the \emph{signal
velocity}, which is defined as the speed of the front of a square pulse. It is the signal speed that
cannot exceed $c$, due to \emph{causality}, as mentioned above. The precursors (or ``forerunners'') of the signal, mentioned above, which travel with sub-luminal speed, arrive first before the main front. This signal velocity
was measured for the first time directly in an experiment in \cite{brunner}, and indeed was found to be less than $c$.

This is an important issue that the reader should have in mind, especially when we discuss non-trivial vacuum refractive indices in several
non trivial ground states of quantum systems, including gravity.

In fact, as argued in ref.~\cite{libe} (specifically for the case of Casimir vacuum~\cite{scharn} but the discussion can be generalised), super-luminal group velocities are ``benign'' as far as causality and  compatibility with the kinematics of Special Relativity are concerned. In particular, the kinematics of Special Relativity requires \emph{only} an invariant speed \emph{not} actually a maximum one. Moreover, causality can be guaranteed in such super-luminal cases because the pertinent kinematics is equivalent to an ``effective'' metric in space-time, $g_{\rm eff}^{\mu\nu}$, $\mu,\nu =0,\dots 3$, different from the Minkowski one, that describes the kinematics in the non-trivial (Casimir) vacuum. The modified dispersion relations leading to super-luminal group velocities (\ref{casimir}) acquire the form
 \begin{equation}
  k_\mu k_\nu g_{\rm eff}^{\mu\nu} = 0~,
  \label{effmetr}
\end{equation}
where as usual repeated indices denote summation, $k_0 = -\omega$, $k_i = (\vec k)_i, i=1,2,3$,
\begin{equation}\label{effmetricsr}
g_{\rm eff}^{\mu\nu} = \eta^{\mu\nu} + \xi n^\mu n^\nu ~,
\end{equation}
with $\eta^{\mu\nu}$ the (inverse) Minkowski metric, $n^\mu$ a unit (space-like) vector
orthogonal to the plates of fig.~\ref{fig:casimir}, and $\xi = \frac{11 \pi^2 \alpha ^2}{4050 L^4 m_{\rm e}^4}$, with $L$ the distance between the Casimir plates.

The basic point of the discussion in \cite{libe}, and how causality is maintained, is that
the presence of the effective metric (\ref{effmetricsr}) in (\ref{effmetr}) widens slightly (due to the deviations of order $\xi \ll 1$ of the metric from the Minkowski one) the light cone in the direction orthogonal to the plates,
and hence light in that direction travels at a speed $c_{\rm light} > c$, while light in the direction parallel to the plates still travels with speed $c$. In a given inertial reference frame, moving with  four-velocity $u^\mu$ with respect to the rest frame of the apparatus of fig.~\ref{fig:casimir},
the photons inside the Casimir cavity travel, at a direction perpendicular to the plates, with a definite speed $c_{\rm light}^{(u)} > c$, which has \emph{only one value} for each observer and is not universal among observers.
In this way  super-luminal
group velocities (\ref{casimir}) are compatible with causality, since violation of the latter occurs \emph{only } if signals travel with the \emph{same} speed greater than $c$ in two different frames.

As a final comment on properties of the speed of light in non-trivial vacua we mention some recent work in \cite{shore} according to which the above-mentioned features of superluminal phase-, and hence front-, velocity of light
might have some effects on \emph{micro-causality},
which is a property related to the \emph{local} commutativity (or anti-commutativity in the case of fermions) of fields
(\emph{i.e.} the requirement that the vacuum expectation value of the commutator of two field operators at space like separations of $x,y$, $\langle 0|[A(x), A(y) |0\rangle $ be zero). This is a fundamental axiom of local quantum field theories. In fact, the authors of \cite{shore} argue that such correlators no longer vanish in the above-mentioned  non-trivial vacua.

Indeed, although the subluminal nature of the high frequency photon modes guarantees \emph{macro-causality}, that is the property that signals are not measured before they arrive, nevertheless the superluminal nature of the phase velocity of the low-energy photon modes might~\cite{shore} create some problems with the micro-causality in curved space time backgrounds. Microcausality is usually studied by examining the validity of
Kramers-Kronig (KK) dispersion relation relating the (complex in general) phase refractive index at high frequency to that at low frequency, namely
\begin{equation}
{\rm Re}n(\infty) - {\rm Re}n(0) = -\frac{2}{\pi}\int_0^\infty \frac{d\omega}{\omega} {\rm Im}n(\omega)
\label{kramers}
\end{equation}
It should be remarked at this point that in standard optics this relation is assumed to be valid, but this may not be the case of (quantum) gravitational backgrounds, for which the situation is still far from being understood in detail, due to the lack of a fundamental underlying theory.

The imaginary parts of the refractive index in (\ref{kramers}) refer to absorptive properties of the medium. In fact the KK dispersion relation follows from analyticity requirements of the (complex) refractive index in the upper half plane. If one assumes (\ref{kramers}) for classical gravity, then it becomes evident that, by assuming a positive ${\rm Im}n(\omega) > 0$, which is the standard property of an absorptive medium and follows from unitarity arguments of
a quantum field theory, one arrives at:
$ {\rm Re}n(\infty) \equiv 1/v_{\rm ph}(\infty) >  {\rm Re}n(0) = 1/v_{\rm ph}(0)$, \emph{i.e.} $v_{\rm ph}(\infty) > v_{\rm ph}(0)$. Hence, a superluminal low-frequency phase velocity, as seems to be the result of vacuum polarization in QED in curved space times~\cite{hath},
would imply a superluminal front velocity, with the consequent violation of causality. Assuming the KK dispersion relation, therefore, the only way out of this would be to assume a negative ${\rm Im}n(\omega) < 0$ for gravity, \emph{i.e.} classical gravity does not behave as an absorptive but rather as a \emph{gaining} medium. In nature one can meet situations like this, for instance in laser-atom interactions, which can induce gain, leading to superluminal low frequency phase velocities, but preserving a front velocity equal to the speed of light in vacuo $c$ and hence the KK dispersion relation.

However, as indicated in \cite{shore}, QED seems to violate the KK relation itself, by introducing \emph{non analyticities} in the corresponding refractive index.
Since this relation relies only on analyticity of
the refractive index in the upper half plane, a possible violation would indicate the breakdown of micro-causality.

However, the alert reader should bear in mind that the requirement of micro-causality actually requires
Lorentz invariance of the Scattering matrix, and hence it may not be valid in a generic curved space time theory, where Lorentz invariance may be violated. In fact, as argued in \cite{shore}, it appears that QED violates analyticity in the presence of curved space times, and in this way KK relation fails, and causality is maintained. The authors of \cite{shore} found that in the
high-frequency limit, the phase velocity always approaches $c$, so they  determined $v_{\rm front}/c =
1$. Moreover,  they have also shown that where the background gravitational field
induces pair creation, $\gamma \to  e^+e^-$ the ${\rm Im}n(\omega)$  is indeed positive as required by unitarity. However, the refractive index $n(\omega)$ is \emph{not analytic} in the upper half-plane, and the KK dispersion relation is modified accordingly. In particular, for the specific cases of space-time backgrounds of constant scalar curvatures $R$, they examined for concreteness, they found that there are branch cuts in the upper half plane of the refractive index, due to vacuum polarization (computed using the world-line formalism of QED in order to deal properly with the high frequency regime), which modify the KK relation such that, for conformally flat
backgrounds, for instance, one obtains:
\begin{equation}
{\rm Re}n(\infty) - {\rm Re}n(0) = \frac{\alpha R}{36\pi m_{\rm e}^2} > 0
\label{kramersII}
\end{equation}
One might think that this implies a
violation of micro-causality. In our opinion, however, the issue of the refractive index of background gravity is not the end of the story in a dynamical space time, where back reaction effects of the propagating matter have to be properly taken into account, in particular at high energies. In this respect, there may be problems with the KK dispersion relation, associated with the fact that at the high frequency branch, $\omega \to \infty$, one is actually dealing with modes with super-Planckian energies, \emph{i.e}. regimes where the \emph{full quantum gravity} theory may come into play. Hence, it may be misleading to attempt to define concepts from flat space time at such regimes. In particular, back reaction effects of \emph{transplanckian} modes may affect the unitarity properties of a quantum field theory. At such regimes, the structure of space time is highly curved, or even discrete, so any attempt to rely on analyticity properties of the scattering matrix may breakdown.

Above the energy scales $M_{\rm QG}$, where quantum gravity may set in, the structure of the theory may change completely. Concepts like those of a local effective lagrangian or quantum coherence may be in jeopardy~\cite{hawking,ehns}, leading to violations of fundamental principles of ordinary quantum field theories, such as invariance of the effective lagrangian or correlation functions under CPT, \emph{i.e.} the successive operations (at any order) of
Charge conjugation (C), Parity (reflexion) and Time reversal (T) symmetries.
Hence, there is no guarantee that the formal KK relation, which involves an integration all the way up to $\omega \to \infty$, is valid above such high scales, in the
complete situation, where back reaction effects are taken into account and gravity is no longer treated classically as a background.

The issue of the refractive index of gravity depends crucially also on the type of the underlying microscopic theory.
The reader should have these in mind when studying space-time foam theories of quantum gravity.
In this review we shall discuss in section \ref{sec:string} a particular type of string foam, leading to a \emph{subluminal} refractive index, whose anomalous effects are proportional to the photon frequency. This is a result of non-trivial quantum interactions of photon with space-time stringy defects. In this model, the above issues on possible violation of micro-causality do not arise. Nevertheless, the effects of the stringy foam on low-energy particles cannot be described within the framework of local effective Lagrangians.

\subsection{Non-trivial Optical properties of the Quantum-Gravity Vacuum ? }

A truly unspeakable feature on the speed of light may appear when considering the ground state of \emph{Quantum Gravity} \emph{per se } as a non-trivial vacuum with non-standard optical properties, leading to a \emph{non-trivial} refractive index. This may characterize, for instance, certain approaches in which path-integration over microscopic singular fluctuations of the metric field, such as Planck size black holes and other topologically non-trivial configurations, implies a sort of ``\emph{foamy}'' structure of the space time at small (Planck-size) length scales, over which photons can propagate. This
proposal was made initially by J.A. Wheeler~\cite{wheeler}.
In such situations, the concept of a \emph{local Effective Quantum-Field Theory} Lagrangian may \emph{break down}, and the situation resembles that of a quantum-decoherent motion of matter in open quantum mechanical systems interacting with an environment~\cite{hawking,ehns,banks}.

 The ground state of such QG foam situations
may behave as a (subluminal) \emph{refractive medium}, as suggested originally in \cite{aemn},
which to our knowledge, constitutes the first concrete attempt to consider the non-trivial optics effects effect of such vacua on massless particle (photon) propagation,
based on earlier works on non-critical strings in black hole backgrounds~\cite{emn}.
We note at this stage that similar ideas on modified dispersion relations for particles in a quantum gravity medium, but on a purely phenomenological basis, without any attempt to present concrete models, have also been advocated in ref.~\cite{mestres}. Subsequent to the suggestion of \cite{aemn}, modified dispersion relations have also been suggested to characterize \emph{some} Loop quantum gravity ground states~\cite{gambini}.

In the quantum gravity case, the effective modified dispersion relations of the low-energy matter theory, assumes the generic form:
\begin{equation}
     E^2 = |\vec p|^2 + m^2 + \sum_{n=1}^{\infty} c_n |\vec p|^2\left(\frac{|\vec p|}{M_{\rm QG}}\right)^n
\label{mdr}
\end{equation}
where $m$ is the rest mass of the probe, and $c_n$ are constant coefficients, with signature and values that depend crucially on the type of theory considered. It is not clear whether the series converges, or is resummable, as this information depends crucially on the details of the underlying microscopic theory.
The important point to notice in (\ref{mdr}) is that the natural suppression scale of the (Lorentz-symmetry-violating) correction terms of the standard Special Relativity dispersion relation is that of Quantum Gravity itself, $M_{\rm QG}$. According to what was mentioned earlier, there is no fundamental issue with violation of (micro-)causality~\cite{libe} or contradiction with the principles of Special Relativity, at least for some
cases where super-luminal group velocities arise from QG anomalous dispersion relations (\ref{mdr}). The important point to observe is that, as in the Casimir
vacuum case (\ref{effmetr}), effective metrics can be found, at least in some cases, which describe the QG anomalous dispersion (\ref{mdr}). Thus, causality can be saved for those QG cases by applying
the same logic as for the Casimir vacuum, discussed previously.
Nevertheless, as we shall discuss later on in subsection \ref{sec:biref}, in cases where super-luminal modes exist, one has QG-induced birefringence phenomena, which are severely constrained by the current phenomenology (\emph{i.e} by the absence of the relevant signals).

Before closing this subsection we also mention that there is another approach towards modified dispersion relations of particles, of the type (\ref{mdr}), based on the so-called doubly special or deformed special relativity  (DSR) theories~\cite{dsr,smolin}. In such approaches, which are formulated on flat space times, the modification to the dispersion relation arises by the postulate that the local symmetry group of space time is no longer the Lorentz group but a different one. In their original version~\cite{dsr}, DSR modified dispersion relations were obtained by the requirement that the length scale of quantum gravity (``Planck'') remain invariant under transformations, which thus leads to deviations from the Lorentz group. There is no unique prescription to achieve this, however, and in this way one may even arrive at models where there is no upper limit in momentum. In their subsequent version~\cite{smolin}, DSR models postulated the existence of upper limits in velocities of species, and in this way the modified local group was determined by appropriate combinations of dilatations and translations. However \emph{not all} DSR theories exhibit modified dispersion~\cite{glikman}, and the issue there lies on what one defines as the physical momenta.

It is unclear to us whether such theories are fundamental or effective, and in the former case, whether they can be quantized consistently. For instance, a known problem is the behaviour of multi-particle states in such DSR fundamental theories. Another problem for quantizing DSR theories with modified dispersion relations is associated with their \emph{lack} of \emph{locality}, as pointed out recently in \cite{hossen}, although currently there is an ongoing debate as to whether there is a resolution to this problem.

We shall not discuss these theories further here, apart from mentioning that, as in the stringy model we shall consider in section \ref{sec:mainstring}~\cite{aemn,horizons,recoil,szabo,emnnewuncert,li}, they are not characterised by birefringence effects~\cite{dsrreview}, leading only to subluminal propagation (defined appropriately~\cite{dsr,smolin}). There is, however, an important difference from the stringy model: the action of gravity is universal among particle species in DSR theories, and therefore the resulting modification in the dispersion relation pertain to all species. In contrast, for specifically stringy reasons to be outlined in section \ref{sec:string}, in the string model only photons and at most electrically neutral particles~\cite{ems} are allowed to interact non-trivially with the stringy defects, and are thus subjected to non-trivial modification of their dispersion relations. Moreover, in our string foam models, the modified dispersion relations arise as a result of \emph{local} distortions of space time due to the recoil movement of space time defects, inducing metrics that depend on momenta (Finsler-like~\cite{finsler}). The underlying string model is perfectly well defined as a quantum theory, and the associated effects are perfectly consistent with stringy uncertainty principles, as we shall discuss below. Hence,  criticisms of the type appearing in \cite{hossen} for DSR are not relevant for our string foam models. Nevertheless, as we shall also discuss here, there seems to be a breakdown of a local effective field theory description of the effects of string foam on the propagation of particles, which in fact proves crucial for the model to be able to evade~\cite{emncomment} stringent bounds on Lorentz Violation from ultra high energy cosmic ray physics~\cite{sigl}.

\section{Lorentz-Invariance Violating Stringy Space-time ``Foam'' \label{sec:mainstring}}

In string theory, the simplest way to obtain Lorentz-Symmetry Violation, consistently with the conformal properties
of a $\sigma$-model vacuum, as appropriate for a first-quantized version of strings, is to consider
constant electric and/or magnetic background fields. The Lorentz symmetry violation is associated with the preferred reference frame specified by the direction of the background field. In a $\sigma$-model language, the presence of a constant electric or magnetic background field, in which a string propagates, is described by deforming the world-sheet action by appropriate antisymmetric tensor $B_{\mu\nu} = - B_{\nu\mu}$ backgrounds.
Constant Magnetic fields are associated with the \emph{spatial} components of the B-field, $B_{ij} \ne 0$, while a constant electric field is associated with the $0i$-component $B_{0i} \ne 0$, with $0$ the temporal component and $i$ the spatial component in the direction of the electric field.

The presence of a B-field background leads also to
\emph{non commutativity} (NC) of the target-space string coordinates~\cite{seibergwitten}. For \emph{spatial-coordinate} NC one needs only spatial components of the $B$-field to be non zero (corresponding to constant ``magnetic'' fields in the simplest case), while for time-space non commutativity one needs~\cite{sussk1} $B_{0i} \ne 0$, which is equivalent to constant ``electric'' background fields, in its simplest version.

\subsection{Strings in Constant Background Fields: the simplest case of Lorentz Violation \label{sec:livstring}}

To understand better the above points, we should note that, in general, the $B$-field $\sigma$-model vertex operator
on the world sheet $\Sigma$ in which open strings propagate has the form
\begin{equation}
V_B = \int_{\Sigma}  \epsilon^{\alpha\beta} B_{\mu\nu} \partial_\alpha X^\mu \partial_\beta X^\nu
\label{bfield}
\end{equation}
where $B_{\mu\nu} = - B_{\mu\nu}$, $\alpha, \beta =1,2$ are world-sheet indices, $\mu, \nu = 0, 1, \dots D$ are target space indices and $\epsilon^{\alpha\beta}$ is the two-dimensional antisymmetric (Levi-Civita) symbol. In the modern version of strings, including branes~\cite{polch}, the dimensionality $D$ may represent the target-space dimensionality of brane worlds, which the ends of the open strings are attached to. In the case of a constant background gauge field, with constant field strength $F_{\mu\nu}^{(0)}$ one identifies $B_{\mu\nu} = -\frac{1}{2}F_{\mu\nu}^{(0)}$, as follows by the application of Stokes theorem to the vertex operator of the corresponding gauge potential
$$ \oint_{\partial \Sigma} A_\mu \partial_\tau X^\mu = \int_{\Sigma} \epsilon^{\alpha\beta} \partial_\alpha \left(A_\mu \partial_\beta X^\mu\right) = -\frac{1}{2}\int_{\Sigma} \epsilon^{\alpha\beta} F_{\mu\nu} \partial_\alpha X^\mu \partial_\beta X^\nu ~,$$
where $\partial \Sigma$ the boundary of the world-sheet and $F_{\mu\nu} = \partial_\mu A_\nu - \partial_\nu A_\mu$~.

The presence of the B-field leads to mixed-type boundary conditions for open strings on the boundary $\partial \mathcal{D}$  of world-sheet surfaces with the topology of a disc, we consider here for concreteness and relevance to our discussion below:
\begin{equation}
      g_{\mu\nu}\partial_n X^\nu + B_{\mu\nu}\partial_\tau X^\nu |_{\partial \mathcal{D}} = 0~,
\label{bc}
\end{equation}
where $g_{\mu\nu} $ denotes the metric of target space-time.

In what follows, we shall concentrate for concreteness to the case of constant electric background fields $\textbf{E}$.
This will be relevant for our subsequent discussion on space-time foam.
Considering commutation relations among the coordinates of the first quantized $\sigma$-model in the above background, one obtains a non-commutative (NC) time-space relation~\cite{sussk1}, with the pertinent spatial coordinates along the direction of the electric field, say $X^1$ for concreteness and brevity. The NC relations assume the form:
\begin{equation}
[ X^1, t ] = i \theta^{10} ~, \qquad \theta^{01} (= - \theta^{10}) \equiv \theta =  \frac{1}{E_{\rm c}}\frac{\tilde E}{1 - \tilde{E}^2}
\label{stnc}
\end{equation}
where $t$ is the target time. The quantity $\tilde{E}_i \equiv \frac{E_i}{E_{\rm c}}$ and  $E_{\rm c} = \frac{1}{2\pi \alpha '}$ is the Born-Infeld \emph{critical} field. The space-time NC relations (\ref{stnc}) are consistent with the space-time string uncertainty principle~\cite{yoneya}
\begin{equation}
   \Delta X \Delta t \ge \alpha '
\label{stringyunc}
\end{equation}

As discussed in detail in refs.~\cite{sussk1,seibergwitten},
 there is also an induced open-string \emph{effective target-space-time metric}.
To find it,
one should consider the world-sheet propagator on the disc $\langle X^\mu(z,{\overline z})X^\nu(0,0)\rangle$, with the boundary conditions (\ref{bc}).
Upon using a conformal mapping of the disc onto the upper half plane
with the real axis (parametrized by $\tau \in R$) as its boundary~\cite{seibergwitten},
one then obtains:
\begin{equation}
     \langle X^\mu(\tau)X^\nu(0)\rangle = -\alpha ' g^{\mu\nu}_{\rm open,~electric}{\rm ln}\tau^2 + i\frac{\theta^{\mu\nu}}{2}\epsilon(\tau)
\label{propdisc}
\end{equation}
where $\epsilon (\tau)$ is the function that is $1$ ($-1$) for positive (negative) $\tau$. The non-commutativity parameters  $\theta^{\mu\nu}$ are given by (\ref{stnc}),
 while the effective open-string metric, due to the presence of the electric field $\vec{E}$, whose direction breaks target-space Lorentz invariance, is given by:
\begin{eqnarray}
           g_{\mu\nu}^{\rm open,electric} &=& \left(1 - {\tilde E}_i^2\right)\eta_{\mu\nu}~, \qquad \mu,\nu = 0,1 \nonumber \\
           g_{\mu\nu}^{\rm open,electric} &=& \eta_{\mu\nu}~, \mu,\nu ={\rm all~other~values}~,
\label{opsmetric}
\end{eqnarray}
Moreover, there is a modified effective string coupling~\cite{seibergwitten,sussk1}:
\begin{equation}
   g_s^{\rm eff} = g_s \left(1 - \tilde{E}^2\right)^{1/2}
\label{effstringcoupl}
\end{equation}

Notice that the presence of the \emph{critical} ``electric'' field is associated with a \emph{singularity} of both the effective metric and the non commutativity parameter, while the effective string coupling vanishes in that limit. This reflects the \emph{destabilization of the vacuum} when the ``electric'' field intensity approaches the \emph{critical value}, which was noted in \cite{burgess}.

\subsection{Non-commutative Effective Field Theory \label{sec:effact}}

Once non-commutativity is established, one may proceed to write down a generic (low-energy) effective field theory, by following the rather generic approach of \cite{carroll}. For constant electric fields, which we are restricting our attention to for the purposes of this review, the resulting effective field theory is of a type that can be accommodated within the so-called Standard Model Extension (SME) framework~\cite{kostelecky}.

For brevity, and relevance to our phenomenological discussion in this work, we shall restrict ourselves here to a discussion of Non-commutative Quantum Electrodynamics (NCQED), following \cite{carroll}, for a generic non-commutativity among coordinates of the form
\begin{equation}
[ X^\mu, \, X^\nu ] = i \theta^{\mu\nu}~, \qquad \theta^{\mu\nu} \theta_{\mu\nu} > 0 ~,
\end{equation}
which is certainly of the type induced by a constant electric field (\emph{c.f}. (\ref{stnc})). The condition $\theta^{\mu\nu} \theta_{\mu\nu} > 0$ guarantees \emph{perturbative unitarity} of the effective theory~\cite{carroll}.

One approach to constructing a
non-commutative (NC) quantum field theory
is to promote an established ordinary theory
to a noncommutative one
by replacing ordinary fields with noncommutative fields
and ordinary products with Moyal $\star$ products,
defined by
\begin{equation}
f\star g(x) \equiv
\exp(\frac{1}{2} i\theta^{\mu\nu}\partial_{x^\mu}\partial_{y^\nu}) f(x) g(y)\big|_{x=y}.
\label{moyal}
\end{equation}
For gauge theories, one can define appropriate extensions of the pertinent gauge tansformations, and the covariant derivatives acting on the NC fields. Denoting by a caret symbol over a field the corresponding NC extension,
we write for the gauge potential, $\widehat{A_\mu}$, the fermion field $\widehat \psi$, the Maxwell tensor
$\widehat{F}_{\mu\nu} = \partial_\mu \widehat{A}_\nu - \partial_\nu \widehat{A}_\mu - i [ \widehat{A}_\mu, \widehat{A}_\nu ]$ and the corresponding NC extension of the covariant derivative:
$\widehat{D}_\mu \widehat{\psi} = \partial_\mu \widehat{\psi} - i \widehat{A}_\mu \star \widehat{\psi} $.
The gauge invariant NCQED Lagrangian then, proposed in \cite{carroll}, reads:
\begin{equation}
\mathcal{L}_{NCQED} =
\half i \overline{\widehat{\ps}} \star
\ga^\mu \lrhatDmu \widehat \ps
- m \overline{\widehat{\ps}} \star \widehat \ps
- \fr 1 {4q^2} \widehat F_{\mu\nu} \star \widehat F^{\mu\nu}~,
\label{ncqed}
\end{equation}
with
$\widehat f \star \lrhatDmu \widehat g \equiv
\widehat f \star \widehat D_\mu \widehat g
- \widehat D_\mu \widehat f \star \widehat g$.
Any NC field theory, and hence (\ref{ncqed}), violates Lorentz symmetry, since the NC parameter $\theta^{\mu\nu}$ carries space-time indices. In the case of strings in external background fields the violation of Lorentz symmetry by the direction of the field is obvious.

An important point is in order here, concerning the charge assignments of (\ref{ncqed}). From the definition of the NC gauge transformations and the associated gauge covariant NC derivatives only particles with charges $q= 0, \pm 1$ are allowed, which certainly seems problematic from the point of view of associating (\ref{ncqed}) to an effective action of string theory in constant external electric field backgrounds, which in general incorporates the Standard Model.

This issue is resolved~\cite{carroll} by the so-called Seiberg-Witten map~\cite{seibergwitten}, which in the case of NCQED and for small NC parameter, we assume here, defines the physical excitations associated with the caret fields in (\ref{ncqed}) as follows:
\begin{eqnarray}
\widehat A_\mu &=&
A_\mu - \half \th^{\al\be} A_\al
(\prt_\be A_\mu + F_{\be\mu}),
\nonumber\\
\widehat\ps &=& \ps - \half \th^{\al\be}
A_\al \prt_\be \ps.
\label{swmap}
\end{eqnarray}
This leading-order form suffices for many purposes,
since any physical non-commutativity in nature must be small. In fact,
as we shall discuss later on in section \ref{sec:ncrecoil}, in our stringy space-time foam model
the r\^ole of the non commutativity parameter is played by the recoil velocity of the heavy D-particle defects in the space-time foam, which is small.

Plaguing the above expressions (\ref{swmap}) to the action (\ref{ncqed}) and expanding in the small NC parameter $\theta^{\mu\nu}$ leads to the following effective action that is physically equivalent to noncommutative QED
to leading order in $\theta^{\mu\nu}$~\cite{carroll}:
\begin{eqnarray}
\cL &=&
\half i \overline{\ps} \ga^\mu \lrDmu \ps
- m \overline{\ps} \ps
- \frac 1 4 F_{\mu\nu} F^{\mu\nu}
\nonumber\\
&&
- \frac 1 8 i q\th^{\al\be} F_{\al\be}
\overline{\ps} \ga^\mu \lrDmu \ps
+ \frac 1 4 i q\th^{\al\be} F_{\al\mu}
\overline{\ps} \ga^\mu \lrDbe \ps
\nonumber\\
&&
+ \frac 1 4 m q \th^{\al\be}F_{\al\be} \overline{\ps} \ps
\nonumber\\
&&
- \frac 1 2 q \th^{\al\be} F_{\al\mu} F_{\be\nu} F^{\mu\nu}
+ \frac 1 8 q \th^{\al\be} F_{\al\be} F_{\mu\nu} F^{\mu\nu}.
\label{effactionncqed}
\end{eqnarray}
In this equation,  one has redefined the gauge field $A_\mu \to q A_\mu$
to display the charge coupling of the physical fermion,
and $D_\mu \ps = \prt_\mu\ps - i q A_\mu\ps$ as usual.
This action is manifestly CPT invariant, and as such it constitutes only part
of the SME.

Let us now connect these results to those coming from strings in external electric field backgrounds, which is our main topic of discussion here. In the context of strings in external background fields the above action can be obtained from the standard procedure
of considering the equations of motions for the various background fields, obtained from the requirement of
world-sheet conformal invariance. Let us ignore the fermions for simpplicty.

It is a standard result in open string theory~\cite{tseytlin} propagating in Abelian background gauge $A_\mu$ and antisymmetric tensor fields $B_{\mu\nu}$ that the effective target-space action of strings with their ends attached to a D3-brane, of interest to us here, is of Born-Infeld type:
\begin{equation}\label{bi}
\mathcal{L}_{BI-D3} = \frac{1}{2\pi \alpha '}\int d^3 x \sqrt{-{\rm Det}_4\left(g_{\mu\nu} + B_{\mu\nu} + \alpha' F_{\mu\nu}\right)}~,
\end{equation}
where we assume for generality that there is an induced metric on the D3 brane $g_{\mu\nu}$, and the D3 brane tension is determined by the Regge slope $T_3 = \frac{1}{2\pi \alpha '}$.
The reader should first notice that in string theory, the background fields are the physical fields, connectred via the inverse Seiberg-Witten map (\ref{swmap}) to the NC fields in this case. The $B$-field represents the constant electric field background, and hence the source of NC.

The Born-Infeld action (\ref{bi}) is invariant under CPT, despite the Lorentz violation entailed in the direction of the constant field, and this invariance
 was to be expected from the manifest validity of this symmetry in the underlying string theory.

The expansion of the square root to first order in the NC parameter, \emph{i.e}. the $B$-field in this case, reproduces the corresponding terms in (\ref{effactionncqed}). To see this easily, we may redefine
$\alpha ' F_{\mu\nu} + B_{\mu\nu} \rightarrow \mathcal{F}^{\mu\nu}$. In four space-time dimensions, we have the identity~\cite{tseytlin}:
\begin{equation}
-{\rm Det}_4 \left( g_{\mu\nu} + \mathcal{F}_{\mu\nu} \right) = \sqrt{-g}\left( 1 +\frac{1}{2}\mathcal{F}_{\mu\nu}\mathcal{F}^{\mu\nu} - \frac{1}{16} \left(\mathcal{F}_{\mu\nu}\mathcal{F}^{\star\mu\nu}\right)^2  \right)~, \quad \mathcal{F}^{\star\mu\nu} \equiv \frac{1}{2}\epsilon^{\mu\nu\rho\sigma}\mathcal{F}_{\rho\sigma}~.
\label{detbi}
\end{equation}
Substituting $\mathcal{F}_{\mu\nu} = B_{\mu\nu} + \alpha ' F_{\mu\nu}$ in (\ref{detbi}), taking the square root,
considering  the well-known formula for the product of two $n$-dimensional ($n=4$ in our case) antisymmetric symbols
$$ \epsilon_{i_1 i_2 \dots i_n} \epsilon_{j_1 j_2 \dots j_n} = {\rm Det}\begin{pmatrix} \delta_{i_1 j_1} & \delta_{i_1 j_2} & \dots &\delta_{i_1 j_n} \\ \delta_{i_2 j_1} & \delta_{i_2 j_2} &\dots & \delta_{i_2 j_n} \\
\cdot & \cdot & \cdot & \cdot \\\cdot & \cdot & \cdot & \cdot \\\cdot & \cdot & \cdot & \cdot \\
\delta_{i_n j_1} & \delta_{i_n j_2} &\dots & \delta_{i_n j_n} \end{pmatrix}~, $$
and expanding to first order in the $B$-field, we can easily verify the appearance of the two terms involving $\theta^{\alpha\beta} \propto B^{\alpha\beta}$ in  (\ref{effactionncqed}). The terms involving the dual of the Maxwell tensor had not been considered in ref. \cite{carroll}, as they are quartic in derivatives and the construction of that work was to leading order in derivatives only.  Neverhtless these terms, along with an infinity of higher derivative terms appear in the effective action obtained from open strings in the background of external fields. The inclusion of fermion backgrounds leads to the other terms in (\ref{effactionncqed}), to leading order in a target-space derivative expansion.

By applying the reverse Seiberg-Witten map (\ref{swmap}), then, to the fields $A$, $\psi$ one can construct in principle the NC fields and the corresponding NC action from the Born Infeld action, or extensions thereof~\cite{furtherborninfeld} involving higher orders in a target-space derivative expansion.

\subsection{Non-commutativity in String theory and Causality: photon-photon scattering and causal time delays \label{sec:photon_photon}}

\begin{figure}[ht]
\begin{center}
  \includegraphics[width=5.5cm]{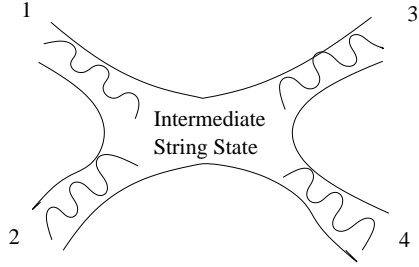}
\end{center}\caption{Photon-photon scattering in open string theory (four point Veneziano amplitude).}
\label{fig:veneziano}
\end{figure}

In the above context of open strings in external electric field backgrounds, one can consider photon-photon scattering and study the effects of the induced non-commutativity
and its connection with causality~\cite{sussk1}. To this end, let one consider the Veneziano four-point amplitude, describing the scattering of two vector (spin one) particles, representing photon excitations, in open first-quantized string theory (\emph{c.f.} fig.~\ref{fig:veneziano}).
In the presence of the external electric field there are non-commutative phases in the amplitude. The result for a world-sheet with the topology of a disc, where we restrict our attention here, is~\cite{seibergwitten,sussk1}:
\begin{eqnarray}
&& A_4 \sim g_s^{\rm eff}\, \left( K_{st} e^{i(p_1\wedge p_2 + p_3 \wedge p_4 )} + K'_{st} e^{i(p_1\wedge p_4 + p_3 \wedge p_2 )}\right)\, \frac{\Gamma (-2 s \alpha ') \, \Gamma(-2 t \alpha ')}{\Gamma ( 1 + 2 u \alpha ')} + \nonumber \\
&& \sim g_s^{\rm eff}\, \left( K_{su} e^{i(p_1\wedge p_2 + p_4 \wedge p_3 )} + K'_{su} e^{i(p_1\wedge p_4 + p_2 \wedge p_3 )}\right)\, \frac{\Gamma (-2 s \alpha ') \, \Gamma(-2 u \alpha ')}{\Gamma ( 1 + 2 t \alpha ')} + \nonumber \\
&&\sim g_s^{\rm eff}\, \left( K_{tu} e^{i(p_1\wedge p_3 + p_2 \wedge p_4 )} + K'_{tu} e^{i(p_1\wedge p_3 + p_4 \wedge p_2 )}\right)\, \frac{\Gamma (-2 t \alpha ') \, \Gamma(-2 u \alpha ')}{\Gamma ( 1 + 2 s \alpha ')}~,
\label{venezianoampl}
\end{eqnarray}
with $s=2 p_1 \cdot p_2$, $t = 2 p_1 \cdot p_4$ and $u = 2 p_1 \cdot p_3$ the Mandelstam variables for massless particle states ($s + t + u =0 $) in the non-trivial induced space-time metric (\ref{opsmetric}), which should be used for contraction of target-space-time indices,  and we use the notation $p \wedge k = \theta^{01} \left( p_0 k_1 - k_0 p_1 \right)$. The quantities $K_{st}, K'_{st} \dots $ denote the standard tree-level kinematic factors of the amplitude, involving the photon polarization vectors~\cite{polch}. We thus observe that the presence of the external field induce non-commutative phases in the amplitude.
Moreover, the perturbative amplitude expression is valid for values of the electric field below the critical value, as mentioned previously, since above that value the space-time is destabilized.

If we look at backward scattering ($u =0$)~\cite{sussk1}, the amplitude (\ref{venezianoampl}) reduces to:
\begin{equation}
A_{st} \sim g_s^{\rm eff}\, \left( K_{st} e^{2\pi i \tilde{E} s \alpha '} + K'_{st}e^{-2\pi i \tilde{E} s \alpha '} \right) \, \Gamma (-2 s \alpha') \Gamma (2 s\alpha ')~,
\end{equation}
where in this case $K_{st} = a_1 s^2, \, K'_{st} = a_2 s^2$, with $a_i, i=1,2$  independent of $s$.
Using the identity $y \Gamma (y) \Gamma (-y) = -\pi/{\rm sin}(\pi y)$, we can write:
\begin{equation}
A_{st} \sim g_s^{\rm eff} \left(a_1 e^{2\pi i \tilde{E} s \alpha '} + a_2 e^{-2\pi i \tilde{E} s \alpha '}\right)\frac{1}{{\rm sin}(2\pi s \alpha')}~,
\end{equation}
which, upon replacing $s \rightarrow s + i \epsilon$, $\epsilon \to 0^+$, in order to treat the \emph{poles} of the amplitude (at $s=n/(2\alpha')$ with $n$=integer) appropriately, and expanding the sinusoidal function,
leads to:
\begin{equation}
\label{poles}
A_{st} \sim g_s^{\rm eff} \sum_{n>0, \, {\rm odd} } \left(a_1 e^{2\pi i(n + \tilde{E})s \alpha ' } +
a_2 e^{2\pi i(n - \tilde{E})s \alpha ' }\right) + \mathcal{O}(\epsilon)
\end{equation}
It is important to observe that the phases are \emph{causal}, in the sense that only
\emph{retarded} outgoing waves are emitted. This is to be contrasted with the non-commutative local field theory case, where a corresponding calculation leads to both advanced and retarded waves, and thus violates causality~\cite{sussk1}.
This is an important feature of the string calculation.

For our purposes in this work, we are mostly interested in the fact that there is a series of retarded outgoing waves, with attenuating amplitudes. The first of this waves is emitted with a
\emph{time delay} of order
\begin{equation}\label{timedelaystring}
\Delta t = \frac{\alpha ' p^0}{1 - \tilde{E}^2}~,
\end{equation}
where $p^0$ denotes the total incident energy (of the two photon states in this case).
We notice that the induced time delays survive the limit of zero electric external field.

This result can be physically interpreted as follows~\cite{sussk1}: as the two open strings in the diagram of fig.~\ref{fig:veneziano} come together, an intermediate string state is created, which acquires $N$ internal oscillator excitations, growing in size from zero length to a maximum length $L$ and back to zero size (oscillating $N$ times).  The whole process is physically possible because the intermediate string absorbs the incident energy $p^0$ (conservation of energy). Since the string has a tension $1/\alpha '$ one has (let us ignore the external field for brevity):
\begin{equation}
p^0 = \frac{L}{\alpha'} + \frac{N}{L}
\label{consenergy}
\end{equation}
Minimizing the right-hand-side with respect to $L$ and taking into account that the ends of the intermediate string state move with the \emph{speed of light in vacuo}, $c=1$ (in our units), we arrive at an estimate of the time delay
which is in agreement with the detailed (microscopic) string amplitude calculation above (in the absence of the external field for simplicity):
\begin{equation}\label{timedelayezero}
\Delta t \sim \alpha ' p^0~.
\end{equation}
The result (\ref{timedelayezero}) is consistent with the time-space uncertainty relation (\ref{stringyunc}) as well as the standard position-momentum string corrected uncertainties~\cite{venheisenberg} $\Delta P \Delta X \ge 1 + \alpha ' (\Delta p)^2 + \dots $, (in our units $\hbar = c =1 $).

An interesting question that arises at this point is whether we can measure experimentally such delays, and thus probe the string scale by considering photon-photon scattering in dense regions, such as central regions of galaxies or active galactic nuclei. The reason why we need dense regions is because the causal time delays (\ref{timedelaystring}) or (\ref{timedelayezero}) are \emph{additive} for multiple photon-photon scatterings, and hence the total delay is proportional to the number of scatterings a photon undergoes in such galactic regions before observation. Since the total delay grows proportional to the photon energy, the higher energy photons will be delayed more compared to their lower-energy counterparts, and if there is a significant number of scatterings in such galactic regions this may result in observable time delays of the higher energy photons (provided the photons are emitted ``simultaneously'' (within the experimental sensitivities)). A natural question that arises at this stage is whether this ``stringy'' cumulative delays from multiple photon-photon scatterings at the source of the emission of such cosmic photons could account for the observed delays in MAGIC~\cite{MAGIC2} and/or FERMI observations~\cite{grbglast,grb090510}. This would imply that such celestial sources could probe stringy uncertainties and thus play the r\^ole of ``Heisenberg microscopes'.

Unfortunately, for standard brane world or stringy scenarios,
with conventional string scales $\alpha ' \sim 10^{-34}$ m (\emph{i.e}. string mass scales of order $M_s \sim 10^{18}$ GeV), in which the intermediate space between such sources and the observation point is defect free, one would need unphysically high concentrations of photons in the source region (whose typical extent is some hundreds of Kpc (1 pc = 10$^{16}$ m)) in order to produce total time delays of the observed order. Even for low string scales of the order of TeV, the associated time delay from a string photon-photon scattering with total incident energy of a few TeV, is of order
$\Delta t \sim 10^{-27}$ sec, which again requires enormous concentrations of scattering events at the source regions in order to reproduce the observed high energy photon delays in MAGIC or FERMI observations.

Hence, unfortunately, for standard brane world scenarios, the possibility of celestial sources serving as stringy uncertainty probes is not realized physically.

\subsection{A stringy model for space time medium with non-trivial ``optical'' properties: D--Particle ``foam''\label{sec:string}}

However, as we shall argue now, such a possibility of \emph{amplification} of the \emph{stringy uncertainty effects} may be provided in cases where the bulk space between brane worlds is punctured by appropriate brane defects which provide the seeds for ``foam'' structures.
In such a case, photons interact with (brany) space-time ``defects'' as they traverse the enormous distances, of cosmological size, between the emission regions and observation points on Earth or at satellites nearby. The cumulative (causal) time delays from such interactions, which, as we shall discuss below, assume the same form as the time delays of photon-photon scattering, as far as the energy dependence is concerned, can then become of the order of
the observed delays in the MAGIC or FERMI Telescopes for sufficiently high concentration of space-time foamy defects.
This is the model of the so-called ``D-particle'' foam, proposed in \cite{emnw,emnnewuncert,li}, which we now come to discuss in some detail before moving into its phenomenology.
In such models, the sourced regions (Active Galactic Nuclei or Gamma Ray Bursts)
can thus play the r\^ole of ``\emph{Heisenberg microscopes}'', probing stringy uncertainties but also the space-time quantum foam structure itself. As we shall point out, in fact, it is rather easy to falsify a large class of such models from the already existing experimental data in astrophysics.

\begin{figure}[ht]\begin{center}
  \includegraphics[width=6cm]{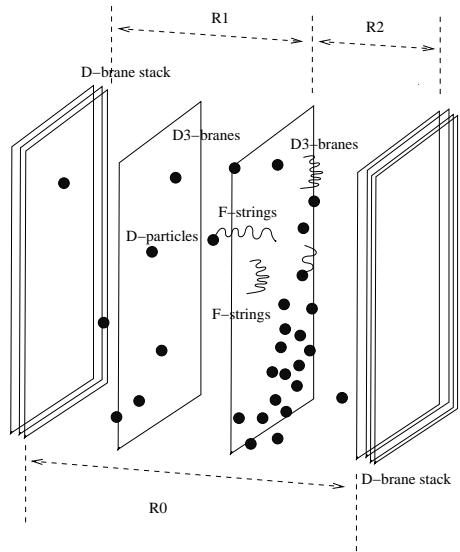}
\end{center}\caption{A type IIA string theory model of D-particle ``foam''. The model consists of appropriate stucks of parallel D(irichlet)branes, some of which are moving in a higher-dimensional Bulk space time, punctured by point-like D-brane defects (D-particles). The observable Universe is represented by one of such moving branes, compactified to three spatial dimensions (D3-brane). As the D3-brane world moves, D-particles from the bulk cross it and, thus, to an observer on the brane, they appear as ``flashing on and off'' space-time foam defects (``D-particle foam''). Photons, represented by open strings with their ends attached on the D3 brane, interact with these defects via capture/recoil, and this leads to non-trivial refractive indices. The effect is therefore ``classical'' from the bulk space time point of view, but appears as an effective ``quantum foam'' from the D3-brane observer effective viewpoint.}
\label{dfoam:fig}
\end{figure}

In \cite{emnw}
we have attempted to construct a brane/string-inspired
model of space time foam which could have realistic cosmological properties.
For this purpose we exploited the modern approach to string theory~\cite{polch}, involving
membrane hypersurfaces (D(irichlet)-branes). Such structures are responsible for reconciliating (often via duality symmetries) certain string theories (like type I), which before were discarded as physically uninteresting, with Standard-Model phenomenology in the low-energy limit.

In particular, we
considered (\emph{c.f}. figure \ref{dfoam:fig}) a ten-dimensional bulk bounded
by two eight-dimensional orientifold planes, which contains two stacks of
eight-dimensional branes, compactified to three spatial
dimensions. Owing to special reflective properties, the orbifolds act as
boundaries of the ninth-dimension. The bulk space is punctured by point-like D0-branes
(D-particles), which are allowed in type IIA string theory (a T-dual of type I strings~\cite{schwarz}) we consider in \cite{emnw} and here~\footnote{One can extend the construction to phenomenologically realistic models of type IIB strings~\cite{li}, as we shall discuss in section~\ref{sec:typeIIB}.}. These are massive objects in string theory~\cite{polch}, with masses $M_s/g_s$, where $M_s$ is the string mass scale (playing the r\^ole of the quantum gravity scale in string theory), and $g_s < 1$ is the string coupling, assumed weak for our purposes. These
objects are viewed as space-time \emph{defects}, analogous, \emph{e.g.} to cosmic strings, but these are point-like and electrically neutral. I have to stress at this point that, according to modern ideas in string theory~\cite{polch}, the scale $M_s$ is in general different from the four-dimensional Planck-mass scale $M_P = 1.2 \times 10^{19}$ GeV/$c^2$, and in fact it is a free parameter in string theory to be constrained by experiment.
The energy scale $M_s c^2$ can be as low as a few TeV; it cannot be lower than this, though, since in such a case we should have already seen fundamental string structures experimentally, which is not the case.

Supersymmetry dictates the number of
D8-branes in each stack in the model, but does not restrict the number of D0-branes in the bulk.
For definiteness, we restrict our attention for now to the type-IIA model, in which the
bulk space is restricted to a finite range by two appropriate stacks of D8-branes, each
stack being supplemented by an appropriate orientifold eight-plane with specific
reflecting properties, so that the bulk space-time is effectively compactified to a finite
region, as illustrated in Fig.~\ref{dfoam:fig}. We then postulate that two of the D8-branes
have been detached from their respective stacks, and are propagating in the bulk.
As discussed above, the bulk region is punctured by D0-branes (D-particles),
whose density may be \emph{inhomogeneous}.
When there are no relative motions of the D3-branes or D-particles,
it was shown in~\cite{emnw} that the ground-state energy vanishes, as decreed by
the supersymmetries of the configuration. Thus, such static configurations constitute appropriate
ground states of string/brane theory. On the other hand, relative motions of the branes break
target-space supersymmetry explicitly, contributing to the dark energy density.

Our model assumes a collision between two branes from the original stack of branes (\emph{c.f.} figure \ref{dfoam:fig}) at an early
epoch of the Universe, resulting in an initial cosmically catastrophic
Big-Bang type event in such non-equilibrium cosmologies~\cite{brany}. After
the collision, the branes bounce back.

It is natural to assume that, during the current (late) era of the Universe, the D3-brane
representing our Universe is moving slowly
towards the stack of branes from which they
emanated and the configuration is evolving
adiabatically. Hence populations of bulk D-particles cross the brane worlds and
interact with the stringy matter on them. To an observer on the brane the
space-time defects will appear to be {\em \textquotedblleft
flashing\textquotedblright } \emph{on} and \emph{off}. The model we are using involves
eight-dimensional branes and so requires an appropriate \emph{compactification}
scheme to three spatial dimensions e.g. by using manifolds with non-trivial
\emph{fluxes} (unrelated to real magnetic fields). Different coupling of fermions
and bosons to such external fields breaks target space supersymmetry, in a way independent of that induced by brane motion, which could be the dominant one in phenomenologically realistic models. The
consequent induced mass splitting~\cite{bachas,gravanis} between fermionic
and bosonic excitations on the brane world is proportional to the intensity
of the flux field (a string generalization of the well-known Zeeman effect of ordinary quantum mechanics, whereby the presence of an external field leads to energy splittings, which are however different  between (charged) fermions and bosons).
In this way one may obtain phenomenologically realistic mass
splittings in the excitation spectrum (at TeV or higher energy scales) owing
to \emph{supersymmetry obstruction} rather than spontaneous breaking (this terminology, which is due to E.~Witten~\cite{obstr}, means that,
although the ground state could still be characterised by zero vacuum energy, the masses of fermion and boson excitations differ and thus supersymmetry is broken at the level of the excitation-spectrum).
The assumption
of a population concentration of
massive D-particle defects in the haloes of galaxies can lead to modified galactic dynamics~%
\cite{recoil2}. Hence, we have an alternative scenario to standard cold dark
matter, using vector instabilities arising from the splitting of strings and
attachment of their free ends to a D-particle defect (\emph{c.f.} fig.~\ref{dfoam:fig}).

One can calculate the vacuum energy induced on the brane world
in such an adiabatic situation by considering its interaction with the D-particles as well as the
other branes in the construction. This calculation was presented in~\cite{emnw},
where we refer the interested reader for further details. Here we mention only the results
relevant for the present discussion.

We concentrate first on D0-particle/D8-brane interactions in the type-IIA
model of~\cite{emnw}. During the late era of the
Universe when the approximation of adiabatic motion is valid, we use a weak-string-coupling
approximation $g_s \ll 1$. In such a case, the D-particle masses
$\sim M_s/g_s$ are large, \emph{i.e.}, these masses could be of the Planck size:
$M_s/g_s \sim M_P = 1.22 \cdot 10^{19}$~GeV or higher.
In the adiabatic approximation for the relative motion, these
interactions may be represented by a string stretched between the
D0-particle and the D8-brane, as shown in Fig.~\ref{dfoam:fig}. The world-sheet amplitude of
such a string yields the appropriate potential energy between the D-particle and the D-brane,
which in turn determines the relevant contribution to the vacuum energy of the brane.
As is well known, parallel relative motion does not generate any potential, and the only
non-trivial contributions to the brane vacuum energy come from motion transverse
to the D-brane. Neglecting a velocity-independent term in the D0-particle/D8-brane
potential that is cancelled for a D8-brane in the presence of orientifold $O_8$
planes~\cite{polch}~\footnote{This cancellation is crucial for obtaining
an appropriate supersymmetric string ground state with zero ground-state energy.}, we find:
\begin{eqnarray}
\mathcal{V}^{short}_{D0-D8} & = & - \frac{\pi \alpha '}{12}\frac{v^2}{r^3}~~{\rm for} ~~
r \ll \sqrt{\alpha '}~,
\label{short}\\
\mathcal{V}^{long}_{D0-D8} & = & + \frac{r\, v^2}{8\,\pi \,\alpha '}~~~{\rm for} ~~
r \gg \sqrt{\alpha '}~.
\label{long}
\end{eqnarray}
where $v \ll 1$ is the relative velocity between the D-particle and the brane world, which is
assumed to be non-relativistic.
We note that the sign of the effective potential changes between short distances (\ref{short})
and long distances (\ref{long}). We also note that there is a minimum distance given by:
\begin{equation}
r_{\rm min} \simeq \sqrt{v \,\alpha '}~, \qquad v \ll 1~,
\label{newmin}
\end{equation}
which guarantees that (\ref{short}) is less than $r/\alpha'$, rendering
the effective low-energy field theory well-defined. Below this minimum distance,
the D0-particle/D8-brane string amplitude diverges
when expanded in powers of $(\alpha')^2 v^2 /r^4$. When they are separated from
a D-brane by a distance smaller than $r_{\rm min}$, D-particles should be considered as lying on
the D-brane world, and two D-branes separated by less than $r_{\rm min}$
should be considered as coincident.

We now consider a configuration with a moving D8-brane located at distances $R_{i}(t)$
from the orientifold end-planes, where $ R_1(t) + R_2(t) = R_0$ the fixed extent of the ninth bulk dimension, and the 9-density of the D-particles in the bulk is  denoted by $n^\star (r) $: see
Fig.~\ref{dfoam:fig}. The total D8-vacuum-energy density $\rho^8$ due to the relative motions
is~\cite{emnw}:
\begin{eqnarray}
&&  \mathcal{\rho}^{D8-D0}_{\rm total} = -  \int_{r_{\rm min}}^{\ell_s}  n^\star (r) \, \frac{\pi \alpha '}{12}\frac{v^2}{r^3}\, dr -  \int_{-r_{\rm min}}^{-\ell_s}  n^\star (r) \, \frac{\pi \alpha '}{12}\frac{v^2}{r^3}\, dr
   +  \nonumber \\ &&  \int_{-\ell_s} ^{-R_{1}(t)} \,n^\star (r) \frac{r\, v^2}{8\,\pi \,\alpha '}\, dr
+ \int_{\ell_s} ^{R_{2}(t)} \,n^\star (r) \frac{r\, v^2}{8\,\pi \,\alpha '}\, dr + \rho_0
\label{total}
\end{eqnarray}
where the origin of the $r$ coordinate is placed on the 8-brane world and $\rho_0$
combines the contributions to the vacuum energy density from inside the band
$-r_{\rm min} \le r \le r_{\rm min}$, which include the brane tension.
When the D8-brane is moving in a uniform bulk distribution of D-particles,
we may set $n^*\star(r) = n_0$, a constant, and the
dark energy density $\mathcal{\rho}^{D8-D0}_{\rm total}$ on the D8-brane is also
(approximately) constant for a long period of time:
\begin{equation}
\mathcal{\rho}^{D8-D0}_{\rm total} =  -n_0 \frac{\pi }{12} v(1 - v) + n_0 v^2 \frac{1}{16\pi \alpha '} (R_1(t)^2 + R_2(t)^2 - 2 \alpha ')
+ \rho_0~.
\label{total2}
\end{equation}
Because of the adiabatic motion of the D-brane, the time dependence of $R_i(t)$ is
weak, so that there is only a weak time dependence of the D-brane vacuum energy density:
it is positive if $\rho_0 > 0$, which can be arranged by considering branes with positive tension.

However, one can also consider the possibility that the D-particle density $n^\star (r)$
is inhomogeneous, perhaps because of some prior catastrophic cosmic collision,
or some subsequent disturbance. If there is a region depleted by D-particles - a {\it D-void} -
the relative importance of the terms in (\ref{total2}) may be changed.
In such a case, the first term on the right-hand side of (\ref{total2}) may
become significantly  smaller than the term proportional to $R_i^2(t)/\alpha'$.
As an illustration, consider for simplicity and concreteness a situation in which there
are different densities of D-particles close to the D8-brane ($n_{\rm local}$) and at long distances to
the left and right ($n_{\rm left}, n_{\rm right}$). In this case, one obtains  from (\ref{total}):
 \begin{eqnarray}
&& \mathcal{\rho}^{D8-D0}_{\rm total} \simeq  -n_{\rm local} \frac{\pi }{12} v(1 - v) + n_{\rm left} v^2 \frac{1}{16 \pi \alpha '} \left(R_1(t)^2 - \alpha '\right)+  \nonumber \\ && n_{\rm right} v^2 \frac{1}{16 \pi \alpha '} \left( R_2(t)^2 - \alpha '\right)
+ \rho_0~.
\label{total3}
\end{eqnarray}
for the induced energy density on the D8-brane. The first term can be significantly smaller in
magnitude than the corresponding term in the uniform case, if the local density of D-particles
is suppressed. Overall, the D-particle-induced energy density on the D-brane world
{\it increases} as the brane enters a region where the D-particle density is {\it depleted}.
This could cause the onset of an accelerating phase in the expansion of the Universe. It is
intriguing that the red-shift of GRB 090510~\cite{grb090510} is in the ballpark of the redshift range where the
expansion of the Universe apparently made a transition from deceleration to acceleration~\cite{decel}.
According to the above discussion, then, within our string foam model this \emph{may not} be a coincidence~\cite{emndvoid}.

The result (\ref{total3}) was derived in  an oversimplified case, where the possible
effects of other branes and orientifolds were not taken into account. However, as we now show,
the ideas emerging from this simple example persist in more realistic structures.
As argued in~\cite{emnw}, the presence of orientifolds, whose reflecting properties cause a
D-brane on one side of the orientifold plane to interact non-trivially with its image on the
other side, and the appropriate stacks of D8-branes
are such that the \emph{long-range contributions} of the D-particles to the D8-brane
energy density \emph{vanish}. What remain are the short-range D0-D8 brane
contributions and the contributions from the other D8-branes and O8 orientifold planes.
The latter are proportional to the fourth power of the relative velocity of the moving D8-brane world:
\begin{equation}
\mathcal{V}_{\rm{long} \,, \,D8-D8, , D8-O8} = V_8\frac{\left(aR_0 - b R_2(t)\right)v^4}{2^{13}\pi^9 {\alpha '}^5}~,
 \label{vlongo8d8}
 \end{equation}
where $V_8$ is the eight-brane volume, $R_0$ is the size of the orientifold-compactified
ninth dimension in the arrangement shown in Fig.~\ref{dfoam:fig}, and the numerical coefficients in
(\ref{vlongo8d8}) result from the relevant factors in the appropriate amplitudes of strings in
a nine-dimensional space-time.
The constants $a > 0$ and $b >0$ depend on the number of moving D8-branes in the
configuration shown in Fig.~\ref{dfoam:fig}.
If there are just two moving D8-branes that have collided in the past, as in the model we consider
here, then $a=30$ and $b = 64$~\cite{emnw}.
The potential (\ref{vlongo8d8}) is positive during late eras of the Universe
as long as $R_2(t) < 15R_0/32$. One must add to (\ref{vlongo8d8}) the negative
contributions due to the D-particles near the D8-brane world, so the total energy
density on the 8-brane world becomes:
\begin{equation}
\mathcal{\rho}_{8} \equiv \frac{\left(aR_0 - b R_2(t)\right)v^4}{2^{13}\pi^9 {\alpha '}^5} -n_{\rm short} \frac{\pi }{12} v(1 - v)~.
 \label{vtotal2}
 \end{equation}
As above, $n_{\rm short}$ denotes the nine-dimensional bulk density of D-particles near
the (moving) brane world.

As in the previous oversimplified example, the transition of the D8-brane world from a region
densely populated with D-particles to a depleted D-void causes these negative contributions
to the total energy density to diminish, leading potentially to an
acceleration of the expansion of the Universe. The latter lasts as long as the energy density
remains positive and overcomes matter contributions. In the particular example shown in
Fig.~\ref{dfoam:fig}, $R_2(t)$ diminishes as time elapses and the D8-brane moves towards
the right-hand stack of D-branes, so the net long-distance
contribution to the energy density (\ref{total2}) (the first term) increases, tending further to
increase the acceleration.

It is the linear density of the D-particle defects $n(z)$ encountered by a propagating photon
that determines the amount of refraction. The density of D-particles crossing the D-brane
world cannot be determined from first principles,
and so may be regarded as a parameter in phenomenological models.
The flux of D-particles is proportional also to the velocity $v$ of the D8-brane in the bulk,
if the relative motion of the population of D-particles is ignored.

In order to make some phenomenological headway,
we adopt some simplifying assumptions. For example, we may assume that between
a redshift $z < 1$ and today ($z=0$), the energy density (\ref{vtotal2}) has
remained approximately constant, as suggested by the available cosmological data.
This assumption corresponds to~\footnote{We recall that, for the case of two D8 branes
moving in the bulk in the model of~\cite{emnw}, we have $b=64 = 2^6$.}:
\begin{equation}\label{zero}
0 \simeq \frac{d\mathcal{\rho}_8}{d t } = H(z) (1 + z) \frac{d\mathcal{\rho}_8}{d z}  =
\frac{v^5}{2^{7}\pi^9 {\alpha '}^5} -\frac{d n_{\rm short}}{d t} \frac{\pi }{12} v(1 - v)~.
\end{equation}
where $t$ denotes the Robertson-Walker time on the brane world, for which
 $d/dt = -H(z) (1 + z) d/d z$, where $z$ is the redshift and $H(z)$ the Hubble parameter of the Universe,
and we take into account the fact that $dR_2(t)/dt = -v$ with $v > 0$,
due to the motion of the brane world towards the right-hand stack of branes in the model
illustrated in Fig.~\ref{dfoam:fig}. Using $a(t) = a_0/(1 + z)$ for the cosmic scale factor,
equation (\ref{zero}) yields:
\begin{equation}\label{nchange}
\frac{v^5}{c^4\,2^{7}\pi^9 {\alpha '}^5} = -H(z) (1 + z) \frac{d n_{\rm short}}{d z} \frac{\pi }{12} \frac{v}{c}(1 - \frac{v}{c})~,
\end{equation}
where we have incorporated explicitly the speed of light \emph{in vacuo}, $c$.
Equation (\ref{nchange}) may be then integrated to yield:
\begin{equation}
\label{nofz}
n_{\rm short} (z) = n_{\rm short}(0) - \frac{12}{2^{7}\pi^{10}{\alpha '}^5} \frac{1}{\ell_s^9}\,\frac{c}{\ell_s}\, \frac{(v/c)^4}{(1 - \frac{v}{c})}\int_0^z \frac{d z'}{H(z') \, (1 + z')} \; \; {\rm where} \; \;  \ell_s \equiv \sqrt{\alpha '}~.
\end{equation}
We shall make use of this result later on, when we fit the available data from MAGIC and FERMI telescopes to this model.

Before closing this section  we would like to make a comment regarding the modification of (thermal) Dark Matter (DM) relic abundances in the foam model. As discussed in \cite{vergou}, the quantum fluctuations of the D-particles act as sources of particle production and thus affect the respective Boltzmann equation determining the relic abundance of DM particles. Additional modifications to this equation are due to the Finsler-like~\cite{finsler} character of the induced space-time metric during the interaction of neutral particles (like the DM ones) with the foam, due to its dependence on D-particle recoil velocities (and hence momentum transfer) (\emph{c.f}. (\ref{opsmetric2}) below).
The modifications are suppressed by the square of the string scale (actually the D-particle mass $M_s/g_s$) and hence for relatively high string scales (much higher than TeV, of interest to us here) are small and do not lead to significant constraints on the density of foam.

It is important in this latter respect to make a comment regarding the nature of our foam: the D-particles in our approach~\cite{emnw,emnnewuncert,li} are viewed as \emph{background configurations} and \emph{not} as \emph{excitations} of the string vacuum. In other string/brane models, some authors have viewed the D-particles as localized excitations of the vacuum~\cite{dmatter}. In those cases, the D-particles  may be considered as dark matter candidates themselves and their density would be constrained by the cosmological observations on the DM sector to avoid overclosure of the Universe. As such they could not contribute to the refractive index.

Indeed, for the latter property to occur, light must interact coherently with the D-particles,
rather than scattering on them individually. Otherwise, there
would not just be a time delay  and thus an index of refraction, but the
light would be incoherently deflected at arbitrarily large angles.
Coherent scattering can only occur if the wavelength of light is
much greater than the mean separation of the scatterers. In
case the D0 particles behave like dark matter~\cite{dmatter}, with masses
near the Planck scale (to account for the MAGIC delays~\cite{MAGIC2}), their number density must be less than
$10^{-20}$ m$^{-3}$ to avoid overclosure of the Universe. Thus their mean
separation would be greater than $100$ m. The gamma rays that are of interest to us here
have a much shorter wavelength (smaller than cm) and therefore
cannot experience refraction due to these D-particles~\footnote{At any rate, such super-heavy DM would have been washed out by inflation in any realistic cosmology, so the scenario of \cite{dmatter} for D-particle excitations to play the r\^ole of DM pertains to much lighter D-particles in theories with low string mass scales.}.

In our D-foam model~\cite{emnw,emnnewuncert}, where the D0-branes (or the compactified D3 branes around 3-cycles in the model of \cite{li}) are viewed only as background defects, such an issue does not arise, and the density of the foam cannot be constrained by overclosure of the Universe issues. As we have discussed above, in this case the D-particles contribute to the dark energy sector (when in motion) consistently with current cosmological data. The cancellation between attractive (gravitational) and repulsive flux forces guarantees a supersymmetric vacuum, with zero vacuum energy,  if no relative motion of foam occurs.

\subsection{Time Delays in D-particle foam: a matter of Uncertainty with...strings attached \label{sec:uncert}}

\begin{figure}[H]
 \begin{center} \includegraphics[width=7cm]{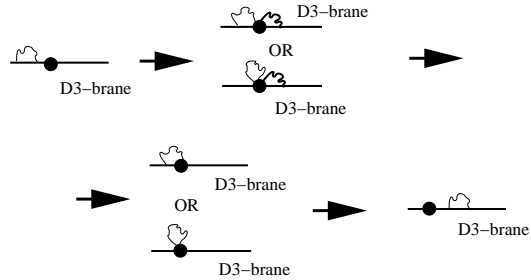}
\end{center}\caption{Schematic view of the capture process of an open string state, representing a photon propagating on a D3-brane world, by a D-particle on this world. The intermediate string state, indicated by thick wave lines, which is created on capture of the end(s) of the photon by the D-particle, stretches between the D-particle and the brane world, oscillates in size between $0$ and a maximal length of order $\alpha ' p^0$, where $p^0$ is the energy of the incident photon, and thus produces a series of outgoing photon waves, with attenuating amplitudes, constituting the re-emission process. The intermediate string state provides also the restoring force, necessary for keeping the D-particle roughly in its position after scattering.}
\label{fig:restoring}
\end{figure}

After these global considerations, we can now proceed to discuss a possible origin of time delays induced in the arrival time of photons, emitted simultaneously from an astrophysical object, as a result of their propagation
in the above-described D-particle space-time foam model. This comes by considering local interactions of photons, viewed as open string states in the model, with the D-particle defects.
As we shall argue, the above model is in principle capable of reproducing photon arrival time delays proportional to the photon energies, of the kind observed in the MAGIC experiment~\cite{MAGIC,MAGIC2}. This microscopic phenomenon, which is essentially stringy and does not \emph{characterize} local field theories, contributes to a sub-luminal non-trivial
refractive index \emph{in vacuo}, induced by the capture of photons or electrically neutral probes by the D-particle
foam~\cite{emnnewuncert}.
The capture process is described schematically in figure \ref{fig:restoring} and we next proceed to describe the underlying physics, which is essentially \emph{stringy}.

An important feature of the model is that, on account of electric charge conservation, only electrically neutral excitations are subjected to capture by the D-particles. To the charged matter the D-particle foam looks \emph{transparent}. This is because the capture process of fig.~\ref{fig:restoring} entails a splitting of the open string matter state. Charged excitations are characterized by an electric flux flowing across the string, and when the latter is cut in two pieces as a result of its
capture by the D-particle defect, the flux should go somewhere because charge is conserved.
The D-particle, being neutral, cannot support this conservation, and hence only \emph{electrically neutral} excitations, such as photons, are subjected to this splitting and the associated delays. The reader should bear in mind of course that the D-particles carry other kinds of fluxes, unrelated to electromagnetism~\cite{polch}. These are conserved separately, and it is for this reason that isolated D-particles cannot exist, but there must always be in the company of other D-branes, as in our model above, so that the relevant fluxes are carried by the stretched strings between the latter and the D-particles.

For our purposes in this work it is also important to remark that the D-particles are treated as static when compared to photons. This is because the ends of the open string representing the photon move on the D3 brane world with the speed of light in (normal) vacuo, while the relative velocities of the D-particles with respect to the brane world (which propagates in the bulk space) are much lower than this. For instance, as discussed in \cite{brany}, to reproduce cosmological observations in this model, in particular the spectrum of primordial density fluctuations, which are affected by the relative motion of the brane world, the speed of propagation of the D3-brane Universe should be smaller than $10^{-4}c$.

When the end(s) of the open-string photon state are attached to the D-particle, there is an intermediate string state formed, \emph{stretched} between the D-particle and the D3-brane, which absorbs the incident energy $p^0$ of the photon state to grow in size from zero  to a maximum length $L$ that is determined by the requirement of energy minimization as follows: one assumes~\cite{sussk1} that the intermediate string state needs $N$ oscillations to achieve its maximal length $L$, as in the standard string-string scattering case, examined in section \ref{sec:photon_photon} above. Following the same logic, we first observe that
energy conservation during the capture process of fig.~\ref{fig:restoring} leads to relations of the form (\ref{consenergy}), from which one
arrives at causal time delays for the outgoing photon waves of order~\cite{emnnewuncert}:
\begin{equation}
  \Delta t \sim  \alpha ' p^0~.
\label{delayonecapture}
\end{equation}
This delay is causal and can also be obtained by considering scattering of open string states off D-particle backgrounds. For type IIA strings, which admit point-like D-particles, the situation is a straightforward extension of techniques
applied to the open string-string scattering case of \cite{sussk1}, discussed in section \ref{sec:photon_photon}.
One assumes that the incident open string state splits into two open string states upon interaction with the D-particle defect, which after their scattering in the D-particle background, the presence of which is implemented through the appropriate Dirichlet world-sheet boundary conditions of the open strings, re-join to produce the outgoing state.
The intermediate stage involves two open strings that scatter off each other.
The (quantum) oscillations of the intermediate string state will produce a series of outgoing wave-packets, with attenuating amplitudes, which will correspond to the re-emission process of the photon after capture.
The presence of the stretched string state, which carries the characteristic flux of the D-brane interactions, provides the \emph{restoring force}, necessary to keep the D-particle in its position after scattering with the photon.

The situation may be thought of as the stringy/brany analogue of the restoring force in situations in local field theories of photons propagating in media with non trivial refractive indices, as discussed by Feynman~\cite{feynman} and reviewed above in section \ref{sec:ntv}. The r\^ole of the electrons in that case (represented as harmonic oscillators) is played here by the D-particle defects of the space-time. The stringy situation, however, is more complicated, since the D-particles have an infinite number of oscillatory excitations, represented by the various modes of open strings with their ends attached to them. Moreover, contrary to the conventional medium case, in the string model the refractive index is found proportional to the photon frequency, while the effective mass scale that suppresses the effect (\ref{delayonecapture}) is the quantum gravity (string) scale $M_s$ and not the electron mass, as in (\ref{refrordinary}). The latter property can be understood qualitatively by the fact that the mass of these D-particle defects is of order~\cite{polch} $1/(g_s~\sqrt{\alpha '}) = M_s/g_s$, where $g_s$ is the string coupling (in units of $\hbar = c = 1$).

We remark at this stage that the above time delays are a direct consequence of the \emph{stringy uncertainty principles}. Indeed, strings are characterized by two kinds of uncertainty relations:
the phase-space Heisenberg uncertainty, modified by higher order terms in $\alpha '$~\cite{venheisenberg} as a result of the existence of the minimal string length $\ell_s \equiv \sqrt{\alpha '}$ in target space-time,
\begin{equation}
  \Delta X \Delta P \ge \hbar  + \alpha ' (\Delta P)^2 + \dots
\label{heisenberg}
\end{equation}
and the time-space uncertainty relation~\cite{yoneya} (\ref{stringyunc}).
Since the momentum uncertainty $\Delta P < p^0$, we have from (\ref{heisenberg}), to leading order in $\alpha '$ (in units $\hbar = 1$):
\begin{equation}
\Delta X \ge \frac{1}{\Delta P} > \frac{1}{p^0}
\end{equation}
In view of (\ref{delayonecapture}), we then arrive at consistency with the space-time un certainty (\ref{stringyunc}),
\begin{equation}
\Delta X \ge \frac{\alpha'}{\alpha ' p^0} \sim \frac{\alpha '}{\Delta t}
\label{timespacenew}\end{equation}
As in the conventional string theory photon-photon scattering, reviewed in section \ref{sec:photon_photon}, these delays are \emph{causal}, \emph{i.e.} consistent with the fact that signals never arrive before they occur.
Hence they are \emph{additive} for multiple scatterings of photons by the foam defects from emission till observation.
 As we shall discuss in section~\ref{sec:totaldelay},  this provides
the necessary \emph{amplification}, so that the total delay of the more energetic photons can be~\cite{emnnewuncert} of the observed order in MAGIC and FERMI experiments.

\subsection{D-particle Recoil as External Lorentz-Violating $\sigma$-model Background and Non-Commutativity in D-particle foam \label{sec:ncrecoil}}

In addition to this leading refractive index effect (\ref{delayonecapture}),
 there are corrections induced by the recoil of the D-particle itself, which contribute to space-time distortions that we now proceed to discuss.
From a world-sheet view point, the presence of $D$-particle recoil may be represented by adding to a fixed-point (conformal) $\sigma$-model action, the following deformation~\cite{recoil,szabo}:
\begin{equation}
\mathcal{V}_{\rm{D}}^{imp}=\frac{1}{2\pi\alpha '}
\sum_{i=1}^{D}\int_{\partial D}d\tau\,u_{i}%
X^{0}\Theta\left(  X^{0}\right)  \partial_{n}X^{i}. \label{fullrec}%
\end{equation}
where $D$ in the sum denotes the appropriate  number of spatial target-space dimensions.
For a recoiling D-particle confined on a D3 brane, $D=3$.

There is a specific type of conformal algebra, termed logarithmic conformal algebra~\cite{lcft}, that the recoil operators satisfy~\cite{recoil,szabo}.
This algebra is the limiting case of world-sheet algebras that can still be classified by conformal blocks. The impulse operator
$\Theta(X^0)$ is regularized so that the logarithmic conformal field theory algebra is respected~\footnote{This can be done by using the world-sheet scale, $\varepsilon^{-2} \equiv {\rm ln}\left(L/a\right)^2$, with $a$ an Ultra-Violet scale and $L$ the world-sheet area, as a regulator~\cite{recoil,szabo}: $\Theta_\varepsilon (X^0) = -i\,\int_{-\infty}^\infty \frac{d\omega}{\omega- i\varepsilon} e^{i\omega X^0}$. The quantity $\varepsilon \to 0^+$ at the end of the calculations.}.
The conformal algebra is consistent with momentum conservation during recoil~\cite{recoil,szabo}, which allows for the expression of the recoil velocity $u_i$ in terms of momentum transfer during the scattering
\begin{equation}
u_i = g_s\frac{p_1 - p_2}{M_s}~,
\label{recvel}
\end{equation}
with $\frac{M_s}{g_s}$ being the D-particle ``mass'' and $\Delta p \equiv p_1 - p_2$ the associated momentum transfer of a string state during its scattering  with the D-particle.

We next note that one can write the boundary recoil/capture operator (\ref{fullrec}) as a total derivative over the bulk of the world-sheet, by means of the two-dimensional version of Stokes theorem. Omitting from now on the explicit summation over repeated $i$-index, which is understood to be over the spatial indices of the D3-brane world, we write then:
\begin{eqnarray}\label{stokes}
&& \mathcal{V}_{\rm{D}}^{imp}=\frac{1}{2\pi\alpha '}
\int_{D}d^{2}z\,\epsilon_{\alpha\beta} \partial^\beta
\left(  \left[  u_{i}X^{0}\right]  \Theta_\varepsilon \left(  X^{0}\right)  \partial^{\alpha}X^{i}\right) = \nonumber \\
&& \frac{1}{4\pi\alpha '}\int_{D}d^{2}z\, (2u_{i})\,\epsilon_{\alpha\beta}
 \partial^{\beta
}X^{0} \Bigg[\Theta_\varepsilon \left(X^{0}\right) + X^0 \delta_\varepsilon \left(  X^{0}\right) \Bigg] \partial
^{\alpha}X^{i}
\end{eqnarray}
where $\delta_\varepsilon (X^0)$ is an $\varepsilon$-regularized $\delta$-function.
For relatively large times after the impact at $X^0=0$ (which we assume for our phenomenological purposes in this work), this is equivalent to a deformation describing an open string propagating in an antisymmetric  $B_{\mu\nu}$-background corresponding to an external constant in target-space ``electric'' field,
\begin{equation}
B_{0i}\sim u_i ~, \quad B_{ij}=0~, \quad (X^0 > 0)
\label{constelectric}
\end{equation}
where the $X^0\delta (X^0)$ terms in the argument of the electric field yield vanishing contributions in the large time limit, and hence are ignored from now on.

To discuss the space time effects of a recoiling D-particle on an open  string state propagating on a D3 brane world, we should consider a $\sigma$-model in the presence of the B-field (\ref{constelectric}), which leads to mixed-type boundary conditions (\ref{bc}) for open strings on the boundary $\partial \mathcal{D}$ of world-sheet surfaces with the topology of a disc.
 Absence of a recoil-velocity $u_i$-field leads to the usual Neumann boundary conditions, while the limit where $g_{\mu\nu} \to 0$, with $u_i \ne 0$, leads to Dirichlet boundary conditions.

In analogy with the standard string case in background electric fields, discussed in section \ref{sec:photon_photon},
one obtains a \emph{non-commutative space-time} if recoil of the D-particle is taken into account. This can be seen
upon considering commutation relations among the coordinates of the first quantised $\sigma$-model in the background
(\ref{constelectric}). As in the standard case of a constant electric background field, in the presence of a recoiling D-particle, the pertinent non commutativity is between time and the spatial coordinate along the direction of the recoil velocity field (for large times $t$ after the impact at $t=0$):
\begin{equation}
[ X^1, t ] = i \theta^{10} ~, \qquad \theta^{01} (= - \theta^{10}) \equiv \theta =  \frac{1}{u_{\rm c}}\frac{\tilde u}{1 - \tilde{u}^2}
\label{stnc2}
\end{equation}
where, for simplicity and concreteness, we assume recoil along the spatial $X^1$ direction. Thus, the induced non commutativity is consistent with the breaking of the Lorentz symmetry of the ground state by the D-particle recoil. The quantity $\tilde{u}_i \equiv \frac{u_i}{u_{\rm c}}$ and  $u_{\rm c} = \frac{1}{2\pi \alpha '}$ is the Born-Infeld \emph{critical} field.
The space-time uncertainty relations (\ref{stnc2}) are consistent with the corresponding space-time string uncertainty principle (\ref{stringyunc}).

Of crucial interest in our case is the form of the induced open-string \emph{effective target-space-time metric}.
The situation parallels that of open strings in external electric field backgrounds, discussed  in refs.~\cite{sussk1,seibergwitten}, and reviewed in section~\ref{sec:livstring}. Hence,
the effective open-string metric, $g_{\mu\nu}^{\rm open,electric}$, which is now due to the presence of the recoil-velocity field $\vec{u}$, whose direction breaks target-space Lorentz invariance, is obtained by
extending appropriately the result (\ref{opsmetric}) to the background (\ref{constelectric}):
\begin{eqnarray}
           g_{\mu\nu}^{\rm open,electric} &=& \left(1 - {\tilde u}_i^2\right)\eta_{\mu\nu}~, \qquad \mu,\nu = 0,1 \nonumber \\
           g_{\mu\nu}^{\rm open,electric} &=& \eta_{\mu\nu}~, \mu,\nu ={\rm all~other~values}~.
\label{opsmetric2}
\end{eqnarray}
For concreteness and simplicity, we considered a frame of reference where the matter particle
has momentum only across the spatial direction $X^1$, \emph{i.e.} $0 \ne p_1 \equiv p \parallel u_1~, p_2=p_3 =0$.
Moreover, as in the standard case of strings in an electric field background, there is a modified effective string coupling~\cite{seibergwitten,sussk1} (\emph{c.f.} (\ref{effstringcoupl})):
\begin{equation}
   g_s^{\rm eff} = g_s \left(1 - \tilde{u}^2\right)^{1/2}
\label{effstringcoupl2}
\end{equation}
The fact that the metric in our recoil case depends on momentum transfer variables implies that D-particle recoil induces Finsler-type metrics~\cite{finsler}, \emph{i.e}. metric functions that depend on phase-space coordinates of the matter (photon) state.

As already mentioned, the presence of the critical ``background field'' $u_c$  is associated with the \emph{destabilization of the vacuum} when the field intensity approaches the \emph{critical value}. Since in our D-particle foam case, the r\^ole of the `electric' field is played by the recoil velocity of the
D-particle defect, the critical field corresponds to the relativistic speed of light, in accordance with special relativistic kinematics, which is respected in string theory, by construction.
On account of (\ref{recvel}), then, this implies an upper bound on the induced momentum transfer, and hence on the available momenta, for the effective field theory limit to be valid. Indeed, if we represent $\Delta p$ in (\ref{recvel}) as a fraction of the incident momentum $p_1$,  $\Delta p = r p_1$, $r < 1$, then the condition that the recoil velocity of the D-particle is below the speed of light in vacuo, as required by the underlying consistency of strings with the relativity principle, implies~\cite{emncomment}
\begin{equation}\label{novelgzk}
\Delta p \equiv r p_1 < \frac{M_s}{g_s} \Rightarrow p_1 < \frac{M_s}{r\, g_s}~.
\end{equation}
When the incident momentum approaches the order of this cutoff, the effective string coupling (\ref{effstringcoupl2}) vanishes, while above that value the coupling becomes imaginary, indicating complete absorption of the string state by the D-particle. The space-time distortion due to recoil is so strong in such a case that there is no possibility of re-emergence of the string state, the defect behaves like a black hole, capturing permanently the string state.

It is important to notice that in modern string theory the quantities $M_s, g_s$ are completely phenomenological.
In fact, it is possible to construct phenomenologically realistic string/brane-Universe models, in the sense of being capable to incorporate the Standard Model particles at low energies, with string scales in such a way that
$M_s/g_s$ is significantly lower than the Planck scale. For instance, there are constructions~\cite{pioline} in the large extra dimension framework, for which
$M_s/g_s$ is of the order of $10^{19}$ eV, \emph{i.e}. of order of the conventional GZK cutoff in Lorentz invariant particle physics models. Thus, by embedding the D-particle foam to such models, one may have the appearance of a \emph{novel type} Gretisen Zatsepin Kusmin (GZK) cutoff~\cite{GZK}, of similar order to the one obtained from conventional Lorentz invariance arguments, but of quite different origin: here it is the \emph{subluminal} nature of the \emph{recoil velocity} of the foam that sets the new upper bound in momentum transfer. This is not fine-tuning in our opinion, but indicates the appearance of a new upper bound in momenta, related implicity to the underlying Lorentz invariance of the string theory, which is broken spontaneously by the recoiling D-particle background. This point will become of importance later on, when we discuss
constraints of our model coming from ultra-high-energy/infrared photon/photon scattering~\cite{sigl}.

Before closing this subsection we make an important remark. The induced metric (\ref{opsmetric2}) will affect the dispersion relations of the photon state:
\begin{equation}\label{drps}
p^\mu p^\nu g_{\mu\nu}^{\rm open,electric} = 0~.
\end{equation}
However, because the corrections on the recoil velocity $u_i$ are quadratic, such modifications will be suppressed by the square of the D-particle mass scale.
On the other hand, the presence of the D-particle recoil velocity will affect the induced time delays (\ref{delayonecapture})
 by higher-order corrections of the form, as follows by direct analogy of our case with that of open strings in  a constant electric field~\cite{sussk1}:
 \begin{equation}\label{timerecoil}
 \Delta t_{\rm with~D-foam~recoil~velocity} = \alpha ' \,\frac{p^0}{1 - {\tilde u}_i^2}~.
\end{equation}
Thus, the D-particle recoil effects are quadratically suppressed by the D-particle mass scales, since (\emph{c.f.} (\ref{recvel})) ${\tilde u}_i \propto g_s \Delta p_i /M_s$, with $\Delta p_i$ the relevant string-state momentum transfer.
However, the leading order delay effect (\ref{timerecoil}), obtained formally by considering the limit of vanishing recoil velocity, is linearly suppressed by the string scale, and thus the induced time delays are disentangled in the string foam model from the modified Finsler dispersion relations (\ref{drps}).
This is important for the phenomenology of the model, as we shall discuss below.

\subsection{Lack of a Local Effective Field Theory Formalism in recoiling D-particle Foam \label{sec:nonlocaleft}}

An important comment arises at this stage concerning the construction of a possible local effective field theory action in the case of D-particle foam, when recoil of the D-particle defect is considered.
In view of the formal analogy of the problem with that of an external electric background field, one is tempted to apply the same considerations as those in section \ref{sec:effact} leading to the non-commutative effective action (\ref{effactionncqed}) for the case of quantum electrodynamics in the presence of D-foam background, with the $\theta^{\alpha\beta}$ parameter being replaced by $\theta^{0i} = u_i = g_s \frac{\Delta p_i}{M_s}$, the D-particle recoil velocity.

However, this is not correct. As a result of the momentum-transfer dependent nature of the non-commutativity parameter in the D-foam case, it is not possible to write down the effective action in target space as a power series of local quantum operators. The momentum transfer is not represented by such operators when taking the Fourier transform.

Thus writing down a local effective action for the recoiling D-particle in interaction  with, say, a photon, is not possible. This has important consequences for phenomenology. Indeed, the time delays (\ref{timerecoil}), which are associated with the stringy uncertainty, are found proportional to the incident energy of the (split) photon state
and are thus linearly suppressed by the string scale. The associated refractive index is therefore linearly suppressed, but this cannot be interpreted as an average propagation of photons in the context of some local effective action. As we have seen above, the associated anomalous photon dispersion terms in this case, induced by the Finsler metric (\ref{opsmetric2}), are  \emph{quadratically} suppressed by the string scale.

Therefore, any analysis, such as those involving high energy cosmic rays~\cite{sigl} using an effective linear dispersion relation, obtained from a local effective action, does not apply to our problem. Moreover, it is a generic feature of any local effective theory, that is a theory on flat space times involving higher derivatives operators, to yield birefringence, since the corresponding modified dispersion relations for photons reduce at the end of the day to a solution of the photon frequency as a function of the wave number $k$ which is obtained from a quadratic equation. The latter admits two solutions with different propagation for the two photon polarizations.

This is \emph{not} the case of the uncertainty related time delay (\ref{timerecoil}), which as we have seen is independent of the photon polarization~\cite{emnnewuncert}.

\subsection{D-particle ``foam'' in Type IIB strings \label{sec:typeIIB}}

The discussions in the previous subsections referred formally to D-brane defects of zero spatial size (point-like D-particles), which are allowed in type IIA string theory.
One can do a similar analysis for (the phenomenologically more realistic, from the point of view of incorporating Standard Model Physics in the low-energy limit) type IIB strings~\cite{li}, which, though, do not contain D-particles (D0 branes). In such a case the ``foamy'' defects can be constructed by wrapping up D3 branes around three cycles. The r\^ole of brane worlds in such type IIB string models is played by appropriately compactified D7 branes.
There are some technical subtleties in this case which are worthy of discussing, and which lead to some important differences from the type IIA D-particle foam model.

Let us consider the Type IIB string theory with D3-branes
and D7-branes where the D3-branes are inside the D7-branes.
The D3-branes wrap a three-cycle, and the D7-branes wrap a four-cycle.
Thus, the D3-branes can be considered as point
particles in the Universe, {\it i.e.}, the D-particles, while
the Standard Model (SM)
 particles are on the world-volume of the D7-branes.

\begin{figure}[ht]
\centering
\includegraphics[width=0.6\textwidth]{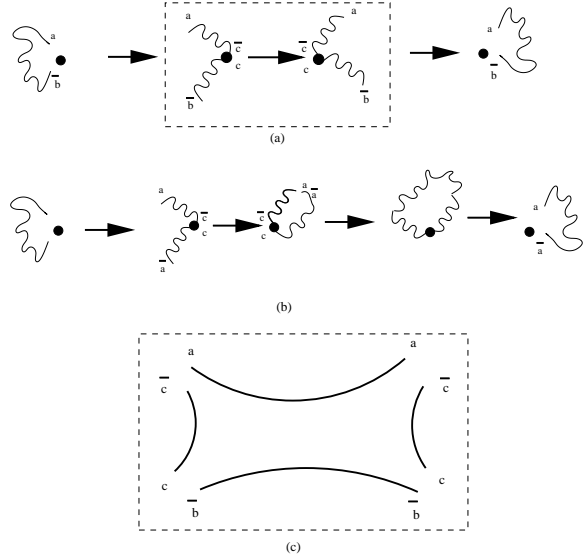}
\caption{\textbf{(a):} the splitting/capture/re-emission process of a (generic)
matter string by a D-particle from a target-space point of view, in the type IIB D-foam model. \textbf{(b)} The
same process but for photons (or in general particles in the Cartan subalgebra of
the Gauge group in this (intersecting) brane world scenario.
\textbf{(c):} The four-point string scattering amplitude (corresponding to the parts
inside the dashed box of (a)) between the constituent open ND strings of the splitting
process. Latin indices at the end-points of the open string refer to the
brane worlds these strings are attached to.}
\label{fig:scatt}
\end{figure}

For simplicity, we assume that the three internal space dimensions,
which the D3-brane wraps around, are cycles $S^1\times S^1\times S^1$, and we denote
the radius of the $i-$th cycle $S^1$ as $R_i$.
The mass of the D3-brane is~\cite{Johnson:2000ch}
\begin{equation}
M_{\rm D3} ~=~ {{R_1 R_2 R_3}\over {g_s  \ell_s^4}}~,~\,
\label{D3-Mass}
\end{equation}
whereby $g_s$ is the string coupling, and
 $\ell_s$ is the string length, {\it i.e.},
 the square root of the Regge
slope $\ell_s = \sqrt{\alpha '}$. If $R_i \le  \ell_s$, we can perform a
T-duality transformation along the $i-$th $S^1$ as follows
\begin{equation}
R_i \longrightarrow {{\ell_s^2}\over {R_i}}~,~\,
\end{equation}
and then we obtain $R_i \ge  \ell_s$. In string theories with compact internal space dimension(s),
there are Kaluza-Klein (KK) modes and string winding modes. Under
T-duality, the KK modes and winding modes are interchanged.
In particular, these two theories are physically identical~\cite{Johnson:2000ch}.
Therefore, without loss of generality, we can  assume
$R_i \ge  \ell_s$. Choosing $R_i = 10  \ell_s$ and $g_s \sim 0.5$,
we obtain $M_{\rm D3} \sim 2000/\ell_s$, and then the mass of the
D3-brane can be much larger than the string scale.
Therefore, anticipating the result (\ref{timerecoil}),
we shall ignore to a first approximation
the recoil of the D-particles. Their inclusion (as small,
 perturbative corrections) is straightforward,
 according to the discussion in the previous section,
and we shall come back briefly to this issue at the
end of the article. In this setting,
the particles (called ND particles) arising
from the open strings between the D7-branes and D3-branes
which satisfy the Neumann (N) and Dirichlet (D) boundary
conditions respectively on the D7-branes and D3-branes, have
 gauge couplings (with the gauge fields on the D7-branes) of the form:
\begin{equation}
\frac{1}{g_{37}^2} = \frac{V}{g_7^2}~,~\,
 \label{coupl}
\end{equation}
where $g_7$ are the gauge couplings on the D7-branes, and
$V$ denotes the volume of the extra four spatial dimensions of the D7 branes
transverse to the D3-branes~\cite{kutasov}.
Because the Minkowski space dimensions are non-compact,
$V$ is infinity and then $g_{37}$ is zero. Thus, the SM particles have no
interactions with the ND particles on the
D3-brane or D-particle.

To have non-trivial interactions between the particles on D7-branes
and the ND particles, we consider
a D3-brane \emph{foam}, {\it i.e.}, the D3-branes are distributed \emph{uniformly}
in the Universe, as a \emph{background}. We assume that the $V_{A3}$ is the average three-dimensional
volume that has a D3-brane in the Minkowski space dimensions, and $R'$ is the
radius for the fourth space dimension transverse to the D3-branes.
In addition, in the conformal field theory description, a D-brane is
an object with a well defined position. In contrast, in the string field theory,
a D-brane is a ``fat'' object with thickness of the order of the string
scale. In particular, the widths of the D-brane along the transverse
dimensions are about $1.55\ell_s$, as follows from an analysis of
the tachyonic lump solution in the string field theory
which may be considered as a D-brane~\cite{Moeller:2000jy}.
Thus, our ansatz for the gauge couplings between
 the gauge fields on the D7-branes and the ND particles is~\cite{li}:
\begin{equation}
\frac{1}{g_{37}^2} = \frac{V_{A3} R'}{(1.55\ell_s)^4}
\frac{\ell_s^4}{g_7^2} = \frac{V_{A3} R'}{(1.55)^4} \frac{1}{g_{7}^2}~.~\,
 \label{coupl-N}
\end{equation}

We denote a generic SM particle as an open string $a\bar b$ with both ends
on the D7-branes (\emph{c.f}. fig.~\ref{fig:scatt}). For $a=b$,
we obtain the gauge fields related to the
Cartan subalgebras of the SM gauge groups, and their supersymmetric
partners (gauginos), for example, the
photon, $Z^0$ gauge boson and the gluons associated with the $\lambda^{3}$
and $\lambda^{8} $ Gell-Mann matrices of the $SU(3)_C$ group.
For $a\not= b$, we obtain the other particles, for example, the electron,
neutrinos, $W^{\pm}_{\mu}$ boson, etc.
As indicated in fig.~\ref{fig:scatt}(a), when the open string
$a \bar b$ passes through the  D3-brane, it splits and becomes
two open strings (corresponding to the ND particles)
 $a \bar c$ and $c \bar b$ with one end on the D7-brane
($a$ or $\bar b$) and one end on the D3-brane ($c$ or $\bar c$).
Then, we can have the
 two-to-two process and have two outgoing particles
arising from the open strings  $a \bar c$ and $c \bar b$. Finally,
we can have an outgoing particle denoted as open string $a\bar b$.
In particular, for $a=b$, at leading order we can have the $s$-channel process
depicted in fig.~\ref{fig:scatt}(b).

The time delays arise from the two-to-two process in the box of
 fig.~\ref{fig:scatt}(a), with the corresponding string diagram give in fig.~\ref{fig:scatt}(c).
To calculate the time delays, we consider the four-fermion scattering
amplitude and use the results of ref.~\cite{benakli} for simplicity.
We can discuss the other scattering amplitudes similarly, for example,
the four-scalar scattering amplitude. the results for the delay are the same.
The total four-fermion scattering amplitude is obtained by summing up
 the various orderings~\cite{benakli}:
\begin{eqnarray}
\mathcal{A}_{\rm total } \equiv \mathcal{A}(1,2,3,4)
+ \mathcal{A}(1,3,2,4) +
 \mathcal{A}(1,2,4,3)~,~\,
\end{eqnarray}
where
\begin{eqnarray}
 \mathcal{A}(1,2,3,4) \equiv A(1,2,3,4) + A(4,3,2,1)~.~\,
\end{eqnarray}
$A(1,2,3,4)$ is the standard four-point ordered scattering amplitude
\begin{eqnarray}
&& \! \!(\!2\pi\!)^4 \! \delta^{(4)}(\sum_a k_a)\! A(1,2,3,4)\!
=  \! \frac{-i}{g_s l_s^4}\!
\int_{0}^{1}\!\! dx\!
\nonumber \\
&&
\left<\! {\cal V}^{(1)} (0,\!k_1) {\cal
V}^{(2)} (x,\!k_2) {\cal V}^{(3)} (1,\!k_3) {\cal V}^{(4)} (\infty,\!k_4)\!\right> ~,~\,
\end{eqnarray}
where $k_i$ are the space-time momenta, and we used the SL(2,R)
symmetry to fix three out of the four $x_i$ positions on the boundary of the upper
half plane, representing the
insertions of the open string vertex fermionic ND operators ${\cal V}^{(i)}$,
$i=1,\dots 4$, describing the emission
of a massless fermion originating from a string stretched between the D7 brane
and the D3 brane~\cite{benakli}.

The amplitudes depend on kinematical invariants expressible in terms of the
Mandelstam variables: $s=-(k_1 + k_2)^2$, $t=-(k_2 + k_3)^2$ and $u=-(k_1 + k_3)^2$,
for which $s + t + u =0$ for massless particles.
The ordered four-point amplitude $\mathcal{A}(1,2,3,4)$ is given by
\begin{eqnarray}
&& \mathcal{A} (1_{j_1 I_1},2_{j_2 I_2},3_{j_3 I_3},4_{j_4 I_4})= \nonumber \\
&&- { g_s} l_s^2 \int_0^1 dx \, \, x^{-1 -s\, l_s^2}\, \, \,
(1-x)^{-1 -t\, l_s^2} \, \, \,  \frac {1}{ [F (x)]^2 } \, \times  \nonumber \\
 &&  \left[   {\bar u}^{(1)} \gamma_{\mu} u^{(2)}
{\bar u}^{(4)} \gamma^{\mu} u^{(3)} (1-x) + {\bar u}^{(1)} \gamma_{\mu}
u^{(4)} {\bar u}^{(2)} \gamma^{\mu} u^{(3)}  x \right ] \,  \nonumber \\
&&  \times \{ \eta
\delta_{I_1,{\bar I_2}} \delta_{I_3,{\bar I_4}}
\delta_{{\bar j_1}, j_4} \delta_{j_2,{\bar j_3}}
\sum_{m\in {\bf Z}}  \, \,
e^{ - {\pi} {\tau}\,
   m ^2 \, \ell_s^2 /R^{\prime 2}   }
\nonumber \\
&& +  \delta_{j_1,{\bar j_2}}
\delta_{j_3,{\bar j_4}}
\delta_{{\bar I_1}, I_4} \delta_{I_2,{\bar I_3}}
\sum_{n\in {\bf Z}}  e^{-  {\pi \tau}   n^2
\,  R^{\prime 2} / \ell_s^{2} } \}~,~\,
\label{4ampl}
\end{eqnarray}
where $F(x)\equiv F(1/2; 1/2; 1; x)$ is the hypergeometric function,
$\tau (x) = F(1-x)/F(x)$,
$j_i$ and $I_i$ with $i=1, ~2, ~3, ~4$
are indices on the D7-branes and D3-branes, respectively, and $\eta$ is
\begin{eqnarray}
\eta={{(1.55\ell_s)^4} \over {V_{A3} R'}}~,~\,
\label{ETA-P}
\end{eqnarray}
in the notation of \cite{benakli},  $u$ is a fermion polarization spinor,
 and the dependence of the
appropriate Chan-Paton factors has been made explicit. Thus,
taking $F(x)\simeq 1$ we obtain
\begin{eqnarray}
\label{4ampldetail}
&&\mathcal{A}(1,2,3,4) \propto  g_s\ell_s^2 \left( t\ell_s^2
{\overline u}^{(1)}\gamma_\mu u^{(2)}{\overline u}^{(4)}\gamma^\mu u^{(3)}
\right. \nonumber\\ && \left.
+ s\ell_s^2{\overline u}^{(1)}\gamma_\mu u^{(4)}{\overline u}^{(2)}
\gamma^\mu u^{(3)}\right)
 \times \frac{\Gamma(-s\ell_s^2)\Gamma(-t\ell_s^2)}{\Gamma(1 + u\ell_s^2)}~,
\nonumber \\
&&\mathcal{A}(1,3,2,4) \propto
 g_s\ell_s^2 \left( t\ell_s^2
{\overline u}^{(1)}\gamma_\mu u^{(3)}{\overline u}^{(4)}\gamma^\mu u^{(2)}
\right. \nonumber\\ && \left.
+ u\ell_s^2{\overline u}^{(1)}\gamma_\mu u^{(4)}{\overline u}^{(3)}
\gamma^\mu u^{(2)}\right)  \times
\frac{\Gamma(-u\ell_s^2)\Gamma(-t\ell_s^2)}{\Gamma(1 + s\ell_s^2)}~, \nonumber \\
&&\mathcal{A}(1,2,4,3) \propto  g_s\ell_s^2 \left( u\ell_s^2
{\overline u}^{(1)}\gamma_\mu u^{(2)}{\overline u}^{(3)}\gamma^\mu u^{(4)}
\right. \nonumber\\ && \left.
+ s\ell_s^2{\overline u}^{(1)}\gamma_\mu u^{(3)}{\overline u}^{(2)}
\gamma^\mu u^{(4)}\right)
 \times \frac{\Gamma(-s\ell_s^2)\Gamma(-u\ell_s^2)}{\Gamma(1 + t\ell_s^2)}~,~\,
\end{eqnarray}
where the proportionality symbols incorporate Kaluza-Klein
or winding mode contributions,
which do not contribute to the time delays. Technically, it should be noted that the novelty of our results above, as compared with those of \cite{benakli}, lies on the specific compactification procedure we adopted, and the existence of a uniformly distributed population of D-particles (foam), leading to (\ref{coupl-N}).
It is this feature that leads to the replacement of the simple string coupling $g_s$ in the Veneziano amplitude by the effective D3-D7 effective coupling $g_{37}^2$~\footnote{We stress once again that this is only part of the process. The initial splitting of a photon into two open
string states and their subsequent re-joining to form the re-emitted photon both depend on the
couplings $g_{37}^2$, so there are extra factors proportional to $\eta$ in a complete treatment,
which we do not discuss here, as they are not relevant for the estimation of the time delays.}.

Similarly to the discussion in Ref.~\cite{sussk1}, time delays arise from  the
amplitude $\mathcal{A}(1,2,3,4)$ by considering
backward scattering $u=0$. Noting that $s+t+u=0$ for massless particles,
the first term in $\mathcal{A}(1,2,3,4)$
in Eq. (\ref{4ampldetail})  for $u=0$ is proportional to
\begin{eqnarray}
t\ell_s^2 \Gamma(-s\ell_s^2) \Gamma(-t \ell_s^2) &=& -s \ell_s^2 \Gamma(-s\ell_s^2)
\Gamma(s \ell_s^2)
\nonumber\\ &=&
{{\pi} \over {\sin(\pi s\ell_s^2)}}~.~\,
\end{eqnarray}
It has poles at $s=n/\ell_s^2$, with $n$ an integer. The divergence of the amplitude at the
poles is an essential physical feature of the amplitude,
a resonance corresponding
to the propagation of an intermediate string state over long space-time
distances. To define the poles we use the correct $\epsilon$ prescription
replacing $s \to s+ i\epsilon$, which shift the poles off the real axis.
Thus, the functions $1/\sin(\pi s\ell_s^2)$ can be expanded as a power
series in $y$:
\begin{eqnarray}
y~=~ e^{i\pi s \ell_s^2 - \epsilon} ~.~\,
\end{eqnarray}
On noting that $s = E^2$, where $E$ is the total energy of the incident string state (\emph{c.f}. fig.~\ref{fig:scatt}), we obtain the time delay at the leading order
\begin{eqnarray}
\Delta t = E \ell_s^2~.~\,
\label{t-delay}
\end{eqnarray}

Let us discuss the time delays
for concrete particles.
We will assume that $\eta $ is a small number (\emph{c.f}. below). Then, for the gauge fields (and their corresponding
gauginos) which are
related to the Cartan subalgebras of the SM gauge groups,
all the amplitudes $\mathcal{A}(1,2,3,4)$, $\mathcal{A}(1,3, 2,4)$,
and  $\mathcal{A}(1,2,4,3)$ will give the dominant contributions
to the total amplitude due to $j_1={\bar j_2}$. Thus,
they will have time delays as given
in Eq. (\ref{t-delay}). The resulting delay for photon
is independent of its polarization, and thus there
is \emph{no birefringence}, thereby leading to the evasion
of the relevant stringent astrophysical constraints~\cite{uv,grb,macio}, as we shall see later on.

However, for the other particles, we have $j_1\not= {\bar j_2}$, and
then only the amplitude $\mathcal{A}(1,3, 2,4)$ gives dominant
contribution. Considering backward scattering~\cite{sussk1} $u=0$ and
$s+t+u=0$, we obtain
\begin{eqnarray}
\mathcal{A}(1,3,2,4) & \propto &
 g_s\ell_s^2 \left({1\over {u\ell_s^2}}
{\overline u}^{(1)}\gamma_\mu u^{(3)}{\overline u}^{(4)}\gamma^\mu u^{(2)}
\right. \nonumber\\ && \left.
- {1\over {s\ell_s^2}}
u\ell_s^2{\overline u}^{(1)}\gamma_\mu u^{(4)}{\overline u}^{(3)}
\gamma^\mu u^{(2)}\right)  ~.~\,
\end{eqnarray}
Because they are just the pole terms,  we do not have time delays for
 other particles with $j_1\not= {\bar j_2}$ at the leading order, for example,
$W^{\pm}_{\mu}$ boson, electron, and neutrinos, etc.
At order $\eta$ (${\cal O} (\eta)$), we have time delays for
these particles, which arise from the forth line in Eq. (\ref{4ampl}).

These time delays are suppressed by terms of order $\eta$ compared to the corresponding time delays for photons or particles in the Cartan subalgebra of the gauge group of the model.
For some characteristic values of this parameter we note that, quite easily, and within natural ranges of the parameters of the model, $\eta$ can be of order
\begin{eqnarray}\label{rangeeta}
\eta  \leq 10^{-6}~.
\end{eqnarray}
For example, by taking the following
$V_{A3}$ and $R'$ in Eq. (\ref{ETA-P})
\begin{eqnarray}
V_{A3} \sim (10 \ell_s)^3~,~~~R'\sim 338 \ell_s~.~\,
\end{eqnarray}
one achieves the above range.

Although, as we have already mentioned, the time delays calculated from string amplitudes in the above way do not necessarily correspond to modified dispersion relations for the probe, nevertheless we note that
if one attempts to interpret the associated time delays for the electron in this model as modified dispersion relations, one would write $E^2 ~=~ p^2 +m_e^2 - \eta p^3/M_{\rm St}~,$ with $M_{\rm St}$ an effective quantum gravity scale. For $M_{\rm St}$ of the order of a string scale in the range of $10^{18}$~GeV, as required by the MAGIC fit of the observed photon delays~\cite{MAGIC2}, and $\eta$ in the range (\ref{rangeeta}), the constraints on the electron dispersion from the Crab Nebula synchrotron radiation observations~\cite{crab,crab2} would be easily satisfied.
However, as already mentioned, the ordered amplitude (\ref{4ampl})
describes only part of the process, and there are extra factors of $\eta$ coming from the initial
splitting of the photon into two open string states. In such a case, the
scale $M_{\rm St}$ would be inversely proportional to them, and so
the constraints from Crab Nebula can be satisfied for much larger values of $\eta$ than (\ref{rangeeta}).
Hence, as in the case of type IIA string foam, synchrotron radiation does not provide stringent constraints for type IIB D-foam.

\subsection{Multiple Photon-D-particle Scatterings in the Foam and Total (observed) Time Delays \label{sec:totaldelay}}

The above-discussed time delays (\ref{delayonecapture}) (or (\ref{t-delay})) pertain to a single encounter of a photon with a D-particle. In case of a foam, with a linear density of defects $n^*/\sqrt{\alpha '}$, \emph{i.e.} $n^*$ defects per string length, the overall delay encountered in the propagation of the photon from the source to observation, corresponding to a traversed distance $D$, is:
\begin{equation}
\Delta t_{\rm total} = \alpha ' p^0 n^* \frac{D}{\sqrt{\alpha '}} = \frac{p^0}{M_s} n^* D~,
\label{totaldelay}
\end{equation}
where $p^0$ denotes an average photon energy.
When the Universe's expansion is taken into account, one has to consider the appropriate red-shift-$z$ dependent stretching factors which affect the measured delay in the propagation of two photons with different energies as well as the Hubble expansion rate $H(z)$~\cite{robust,JP}.
Specifically, in a Roberston-Walker cosmology the delay due to any single scattering
event is affected by: (i) a time dilation factor~\cite{JP} $(1 + z)$ and (ii) the redshifting~\cite{robust}
of the photon energy which implies that the observed energy of a photon
with initial energy $E$ is reduced to $E_{\rm obs} = E_0/(1 + z)$. Thus, the observed
delay associated with (\ref{delayonecapture}) is~\cite{robust,JP}:
\begin{equation}\label{obsdelay}
\delta t_{\rm obs} = (1 + z) \delta t_0 = (1 + z)^2 \sqrt{\alpha '} E_{\rm obs} .
\end{equation}
For a line density of D-particles $n(z)$ at redshift $z$, we have $n(z) d\ell =  n(z) dt$ defects
per co-moving length, where $dt$ denotes the infinitesimal Robertson-Walker time interval
of a co-moving observer. Hence, the total delay of an energetic photon in a co-moving
time interval $dt$ is given by $n(z) (1 + z)^2 C \sqrt{\alpha '} E_{\rm obs} \, dt $. The
time interval $dt$ is related to the Hubble rate $H(z)$ in the standard way:
$dt = - dz/[(1 + z) H(z)]$.  Thus, from (\ref{obsdelay}) we obtain a total delay in the arrival times of photons with energy difference $\Delta E$,  which has the form considered in \cite{robust,mitsou}, namely it is proportional to $\Delta E$ and is suppressed linearly by the quantum gravity (string) scale, $M_s$:
\begin{equation}
(\Delta t)_{\rm obs} =  \frac{ \Delta E}{M_s} {\rm H}_0^{-1}\int_0^z n^*(z) \frac{(1 + z)dz}{\sqrt{\Omega_\Lambda + \Omega_m (1 + z)^3}}
\label{redshift}
\end{equation}
where $z$ is the red-shift, ${\rm H}_0$ is the (current-era) Hubble expansion rate, and we have
assumed for concreteness the $\Lambda$CDM standard model of cosmology, with
$\Omega_i \equiv \frac{\rho_{i(0)}}{\rho_c}$ representing the present-epoch energy densities ($\rho_i$) of matter (including dark matter), $\Omega_m$, and dark vacuum energy, $\Omega_\Lambda$, in units of the critical density $\rho_c \equiv \frac{3H_0^2}{8\pi G_N}$ of the Universe ($G_N$ is the Newton's gravitational constant), that is the density required so that the Universe is spatially flat. The current astrophysical measurements of the
acceleration of the Universe are all consistent with a non zero Cosmological-Constant Universe with Cold-Dark-Matter ($\Lambda$CDM Model), with $\Omega_\Lambda \sim 73 \%$ and $\Omega_m \sim 27\% $.

Notice in (\ref{redshift}) that the essentially stringy nature of the delay implies that the characteristic suppression scale is the string scale $M_s$, which plays the r\^ole of the quantum gravity scale in this case. The scale $M_s$ is a free parameter in the modern version of string theory, and thus it can be constrained by experiment. As we have discussed in this article, the observations of delays of energetic (TeV) photons from AGN by the MAGIC telescope~\cite{MAGIC} can provide such an experimental way of constraining $n^*/M_s$ in (\ref{redshift}). For $\Delta E \sim 10$ TeV, for instance, the delay (\ref{redshift}) can lead to the observed one of order of minutes, provided $M_s/n^* \sim 10^{18}$ GeV (in natural units with $c=1$)~\cite{MAGIC2}. This implies natural values for both $n^*$ and $M_s$, although it must be noted that $n^*$ is another free parameter of the bulk string cosmology model of \cite{emnw}, considered here. In general, $n^*(z)$ is affected by the expansion of the Universe, as it is diluted by it. This depends on the bulk model and the interactions among the D-particles themselves. For redshifts of relevance to the MAGIC experiments, $z=0.034 \ll 1$, one may ignore the $z$-dependence of $n^*$ to a good approximation.

The total delay (\ref{redshift}) may be thought of as implying~\cite{feynman} an effective \emph{subluminal} refractive index $n(E)$ of light propagating in this space time, since one may assume that the delay is equivalent to light being slowed down due to the medium effects.
 On account of the theoretical uncertainties in the source mechanism, however, the result of the
 AGN Mkn 501 observations of the MAGIC Telescope translate to \emph{upper} bounds for the quantity $n^*/M_s$ in (\ref{redshift}), which determines the strength of the anomalous photon dispersion in the string/D-particle foam model.

In view of the above discussion, if the time delays observed by MAGIC can finally be attributed partly or wholly to this type of stringy space-time foam, and therefore to the stringy uncertainties, then the AGN Mkn 501, and other such celestial sources of very high energy photons, may be viewed as playing the r\^ole of Heisenberg microscopes and amplifiers for the stringy space-time foam effects.

We next proceed to discuss the phenomenology of the D-foam model, and see whether it can survive the stringent constraints on Lorentz Violation imposed by a plethora of astrophysical measurements. To understand better the situation, and demonstrate why the exotic interpretation put forward in this article could be a viable hypothesis, we review briefly first the conventional Astrophysics mechanisms for cosmic particle acceleration, for which currently there is no consensus among the astrophysicists.

\section{High-Energy Gamma Ray Astronomy \& Fundamental Physics \label{sec:3}}

In this part of the review we shall discuss experimental tests of Lorentz symmetry using high energy cosmic probes.
Then we shall try to see what these observations imply for the Lorentz-violating model of stringy (D-particle) space-time foam discussed in previous sections and how the latter compare with other Lorentz violating field theories available to date. The important data for our purposes are delayed arrivals of high energy photons (compared to lower energy ones)
from distant cosmic sources. In particular, currently there are the following  ``anomalous'' photon events:
\begin{itemize}
\item{} MAGIC Telescope (source: Active Galactic Nucleus (AGN) Mkn 501, redshift z=0.034),
Highest Energy 1.2 TeV Photons,  Observed Delays of O(min)~\cite{MAGIC,MAGIC2}

\item{} HESS Telescope (source AGN PKS 2155-304, redshift   z=0.116), Highest Energy 10 TeV photons,                                                         Originally claim no observed time lags~\cite{hessnew}.

\item{} FERMI Satellite (source Gamma Ray Burst (GRB) 090816C, redshift z=4.35),
Highest Energy Photon 13.2 GeV, 4.5 s time-lag between E $>$ 100 MeV and E $<$ 100 KeV
Observed Time Delay  16.5 sec~\cite{grbglast}.

\item{} FERMI Satellite (source GRB 090510, redshift z=0.9), Highest Energy Photon 31 GeV,
several photons in range 1-10 GeV~\cite{grb090510}; this is a short and intense GRB. Observed Time Delays $<$ 1 sec~\footnote{However, in our opinion there are some ambiguities on the exact time emission of the highest energy photons, since the measurements on the precursor to the GRB are not yet conclusive as far as we understand. The assumption in this review, though, is that this measurement stands and we try to interpret it using our model in conjunction with the rest of the data.}. We also notice, out of curiosity, that the location of this GRB is in the ball park of the ranges of redshift where the cosmic deceleration/acceleration transition for our Universe occurs.

\item{} FERMI Satellite (source GRB 09092B, redshift z=1.822), Highest Energy Photon 33.4 GeV, Observed Time Delay 82 sec after GMB trigger, 50 sec after end of emission~\cite{grb09092b}.

\end{itemize}

To understand in detail the possible explanations of the above observations, let us first recapitulate the up to date knowledge on the production of very high energy gamma rays in the Universe~\cite{deangelis} and ways of cosmic acceleration. I must stress once again that there is no consensus among the astrophysical community as regards the various ways of particle acceleration at various regions in the Universe, and in fact it is most likely that there are several mechanisms taking place, depending on the source. The MAGIC and FERMI observations added to this puzzle, in particular why the more energetic photons from the respective cosmic sources arrived later than the lower energy ones with the observed time delays. Is there any correlation between the time delays and the photon energies ? Our understanding of the source mechanism for the production of high energy particles from cosmic sources will improve only by making more and more precision measurements from a variety of celestial sources, a process currently under way through the operation (or construction) of several terrestrial and extra-terrestrial facilities of high-energy astrophysics.

It is these uncertainties in the conventional astrophysics of the sources that allow for speculations that
fundamental physics, such as photon propagation in a quantum gravity ``medium'', might play a significant r\^ole on the MAGIC effect. However, if the latter has a chance of being true it must respect all the other stringent astrophysical constraints on Lorentz invariance that are currently available. As we have already discussed in section \ref{sec:string}, it seems that, at present, only a specific string theory model of quantum foam, in which only photons are not transparent to the foam effects,  could stand up to this chance. We shall come back to a detailed phenomenology of the model, later on. At present, we proceed to a brief review of the basic astrophysical mechanisms for cosmic particle acceleration available today in order to appreciate the difficulties in using them as possible explanations of the above-mentioned observed time delays of the more energetic photons.

\subsection{Conventional Astrophysics mechanisms for cosmic Very-High-Energy (VHE) Gamma-Ray production}

Gamma rays constitute the most interesting part of the spectrum of active galactic nuclei (AGN).
An AGNs is a compact region at the centre of a galaxy, with much higher than normal luminosity over some or all of the electromagnetic spectrum (in the radio, infrared, optical, ultra-violet, X-ray and/or gamma ray wavebands). The radiation from AGN is believed to be a result of accretion on to the supermassive black hole at the centre of the host galaxy. AGN are the most luminous persistent sources of electromagnetic radiation in the universe, and as such can be used as a means of discovering distant objects; their evolution as a function of cosmic time also provides constraints on cosmological models.

Gamma Rays with energies higher than 20 MeV and up to TeV have been observed today from such AGNs.
It is customary (although somewhat arbitrary) to classify these Gamma Rays according to their energies as follows:
\begin{itemize}

\item{(i)} \emph{High-Energy} Gamma rays: with energies from 20 MeV - 100 GeV~,
\item{(ii)} \emph{Very High-Energy} Gamma rays: with energies from 100 GeV - 30 TeV~,
\item{(iii)} \emph{Ultra High-Energy} Gamma rays: with energies from 30 GeV - 30 PeV~,
\item{(iv)} \emph{Extremely High-Energy} Gamma rays: with energies from 30 PeV - ?.
\end{itemize}
Theoretically, the last category incorporate energies up to the ultraviolet cutoff energy scale
(Planck-scale energies $10^{19}$ GeV) that defines the structure of low-energy field theories as we know them.

The production of very high energy gamma rays is still not understood well, and constitutes a forefront of research on galactic and/or extragalactic astrophysics.

\begin{figure}[H]
\begin{center}
  \includegraphics[width=5cm]{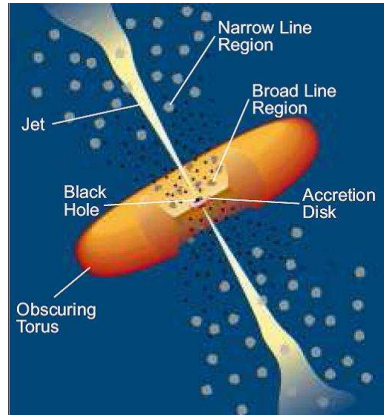}
\end{center}
\caption{The basics of a cosmic accelerator model. Very high energy Photon production from Gravitational Energy conversion (Penrose process) in AGNs, believed to take place in AGN Mkn501.
Relativistic matter, such as electrons, are beam ejected as a result of the enormous gravitational energy available during the collapse process forming a black hole at the center of the galaxy.
Such electrons,  undergo subsequent synchrotron radiation due to their interaction with the magnetic fields existing in the galactic region, and eventually inverse Compton scattering (IC), \emph{i.e.} interactions of these very high energy electrons with low-energy photons (say of eV-keV  energies), to produce
TeV photons from Mkn501 observed by MAGIC (see fig.~\ref{fig:ssc}). This combined process is called Synchrotron-self-Compton process (Figure taken from Ref.~\protect\cite{urry}).}
\label{fig:crproduction}
\end{figure}

At present there are three major categories believed responsible for the production of very- and ultra- high-energy Gamma rays:
\begin{itemize}

\item{(i)} Photons from conversion of gravitational energy in Active Galactic Nuclei (AGN) (\emph{c.f.} fig.~\ref{fig:crproduction}),

\item{(ii)} From self-annihilation of Dark Matter and

\item{(iii)} From decays of exotic massive particles (with masses of order $10^{15}-10^{16}$ GeV/c$^2$), appearing in Grand Unifying Models Beyond the Standard Model (such as string-theory inspired models), in the very Early Universe.

\end{itemize}
\begin{figure}[H]
\begin{center}
  \includegraphics[width=7cm]{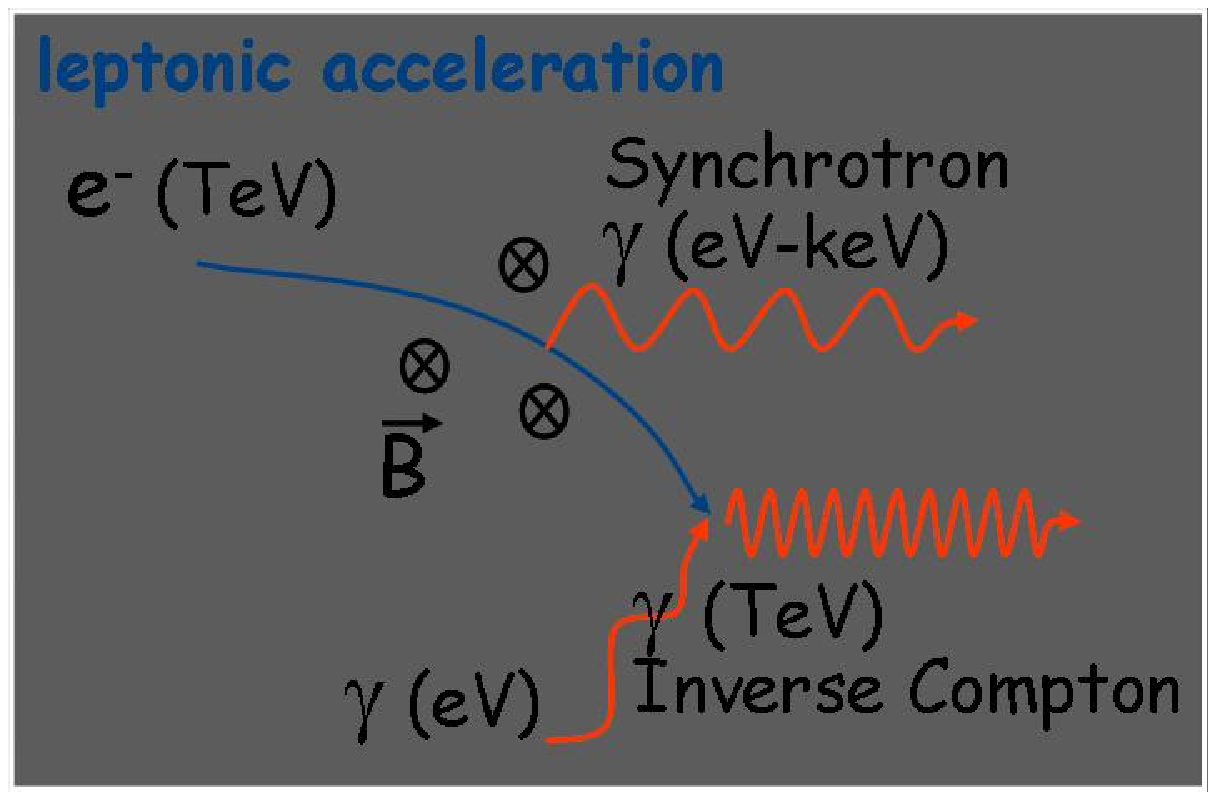} \hfill \includegraphics[width=7cm]{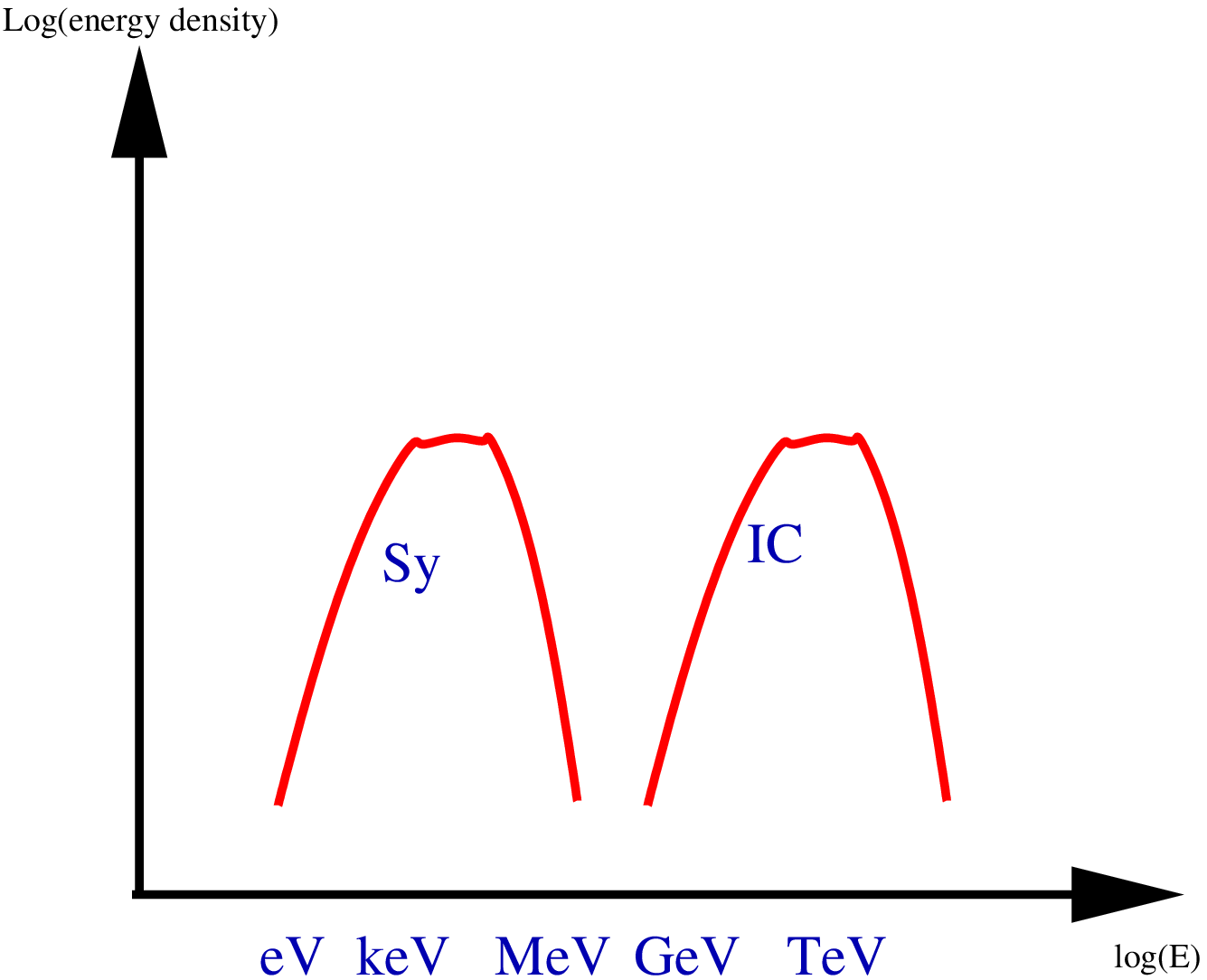}
  \vspace{0.2cm} \includegraphics[width=7cm]{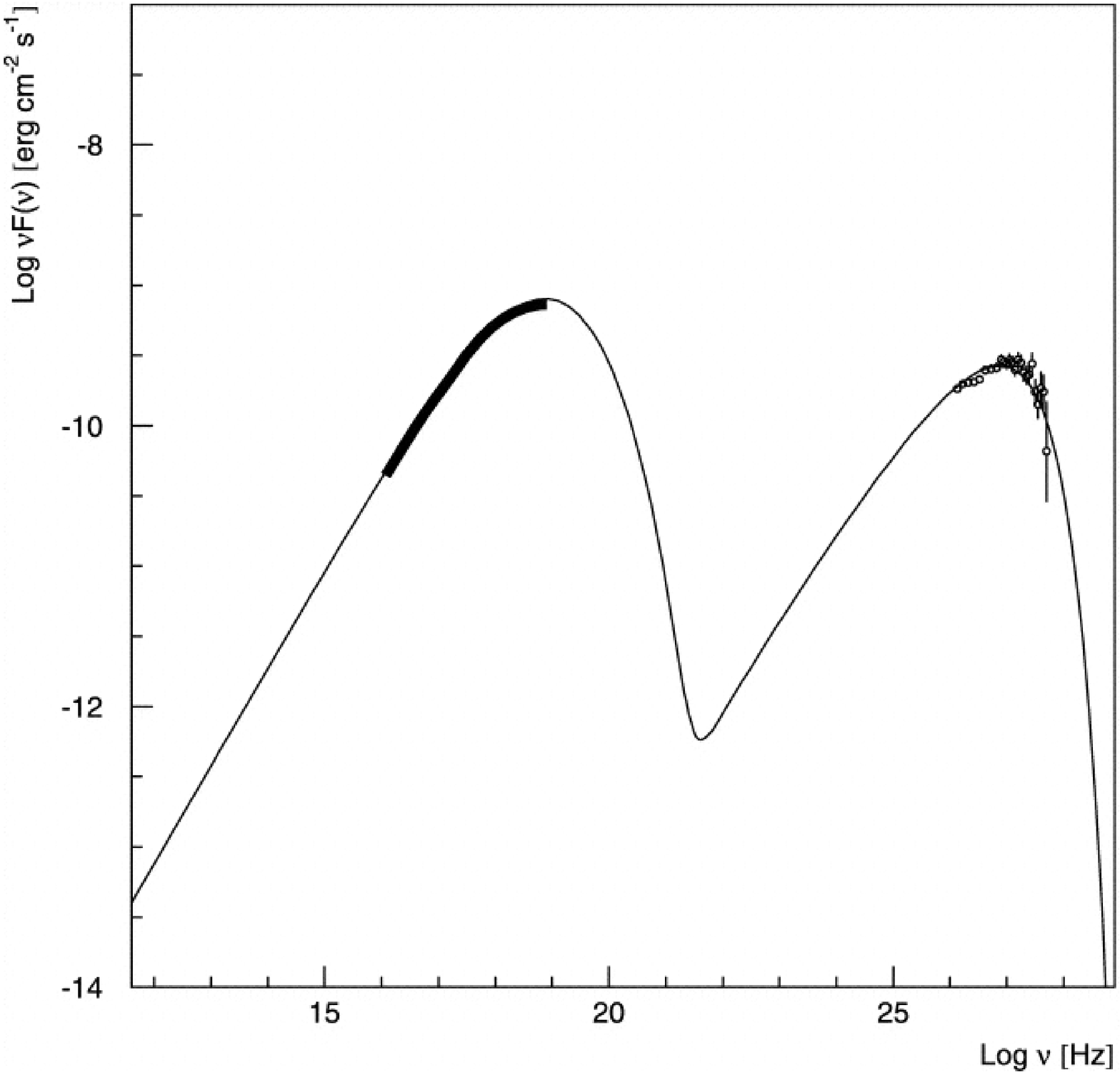}
  \end{center}
\caption{One of the suggested models (leptonic acceleration) for the production of very high energy gamma rays: Synchrotron-self-Compton (SSC), involving synchrotron (Sy) and inverse Compton (IC) scattering for the same electron, provides a mechanism for the production of photons with energies up to several TeV. In the middle figure the typical energy spectrum is indicated, with the characteristic
double peak of the SSC mechanism, one of the synchrotron radiation and the other for the IC scattering. The Inverse Compton spectrum has a peak at TeV energies, as those observed by MAGIC when looking at the AGN Mkn 501 (lower picture, from Ref.~\protect\cite{kono}).}
\label{fig:ssc}
\end{figure}

In what follows we shall restrict ourselves to (i), which most likely pertains to the production of very high energy Gamma rays observed in AGN such as Mkn 501 observed by the MAGIC Telescope. It is believed today that the centres of galaxies contain massive black holes due to matter collapse~\cite{blackholes}, with typical masses in the range $10^{6}-10^{9}$ solar masses. AGN therefore are celestial systems with very high mass density, and it has long been assumed that they consist of a massive black hole, of say $10^8$ solar masses or more, accreting the gas and dust at the center of the galaxy. The gravitational energy liberated during accretion onto a black hole is 10\% of the rest mass energy of that matter and is the most efficient mass–-energy conversion process known involving normal matter.
Collapsing matter towards this massive central galactic object releases gravitational energy (Gravitational Energy Conversion) and results in spectacular relativistic material jet emissions.
Since the accreting matter has in general a non-trivial angular momentum, angular momentum conservation is responsible for the matter orbiting the black hole and, through energy dissipation, the formation of a material (flat) accretion disk. This also results in material jets with ultra-relativistic particles outflowing the accretion plane (fig.~\ref{fig:crproduction}).

In addition, since the black holes at the centres of the AGN are probably rotating (as a result
of having a non-zero angular momentum (Kerr type)~\cite{blackholes}, due to angular momentum conservation in the formation (collapse) process), one might also speculate that part of the
relativistic jet might be due to the so-called Penrose process, which allows the extraction of energy from a rotating black hole~\cite{penrose}. The extraction of energy from a rotating black hole is made possible by the existence of a region of the Kerr spacetime called the ergoregion, in which a particle is necessarily propelled in locomotive concurrence with the rotating spacetime. In the process, a lump of matter enters into the ergoregion of the black hole and splits into two pieces, the momentum of which can be arranged so that one piece escapes to infinity, whilst the other falls past the outer event horizon into the hole (a rotating black hole has two event horizons, an outer and an inner one). The escaping piece of matter can possibly have greater mass-energy than the original infalling piece of matter. In summary, the process results in a decrease in the angular momentum of the black hole, leading to a transference of energy, whereby the momentum lost is converted to energy extracted. The process obeys the laws of black hole mechanics. A consequence of these laws is that if the process is performed repeatedly, the black hole can eventually lose all of its angular momentum, becoming rotationally stationary.

The particles in the relativistic jet undergo acceleration, but currently
the pertinent mechanism is a matter of debate and active research. Most likely it depends on the source.
In general there are two generic ways of cosmic acceleration.
\bigskip
\begin{center}

\emph{Conventional Astrophysics Mechanisms for Cosmic Acceleration}

\end{center}

\bigskip

\emph{Leptonic Acceleration:} Among the particles in the jet are charged electrons, whose paths are curved as a result of the existing magnetic fields in the galactic regions, which accelerate the electrons. The curved path of a charged object implies synchrotron radiation, as a result of energy conservation. Moreover, in AGN's like Mkn 501, the same (high-energy) electron can also undergo inverse-Compton scattering with low-energy photons (with energies of order eV, e.g. photons of the cosmic microwave background radiation that populate the Universe as remnants of the Big-Bang). The terminology inverse-Compton (IC) scattering refers here to the fact that, contrary to the conventional Compton photon-electron scattering, here it is the electron which is the
high energy particle, and whose loss of energy is converted to outgoing radiation (\emph{c.f.} fig.~\ref{fig:ssc}). This IC outgoing radiation can have very high energies. In fact, the Compton spectrum peak can be at several TeV energies, as observed for the AGN Mkn 501. This combined process, whereby the \emph{same electron} that is responsible for \emph{synchrotron} radiation in AGN also undergoes \emph{IC scattering} to produce high energy photons is known as Synchrotron-self-Compton (SSC) mechanism~\cite{ssc}, and is believed~\cite{deangelis}
-with some variations to be discussed below - that is responsible for the production of the very high energy photons observed in the AGN Mkn 501 (\emph{c.f.} fig.~\ref{fig:ssc}).
This is the so-called Leptonic acceleration mechanism, to distinguish it from a different type of acceleration, the \emph{hadronic} one, to be discussed below, and which is believed by many as being the main mechanism for extragalactic ultra-high energy cosmic rays. Notably, AGNs are also extragalactic objects and hence hadronic acceleration mechanism may be relevant for the production of very high energy Gamma rays, as alternative scenarios to (or co-existing with ) SSC-leptonic acceleration mechanism described above.

\begin{figure}[H]
\begin{center}
  \includegraphics[width=7cm, angle=-90]{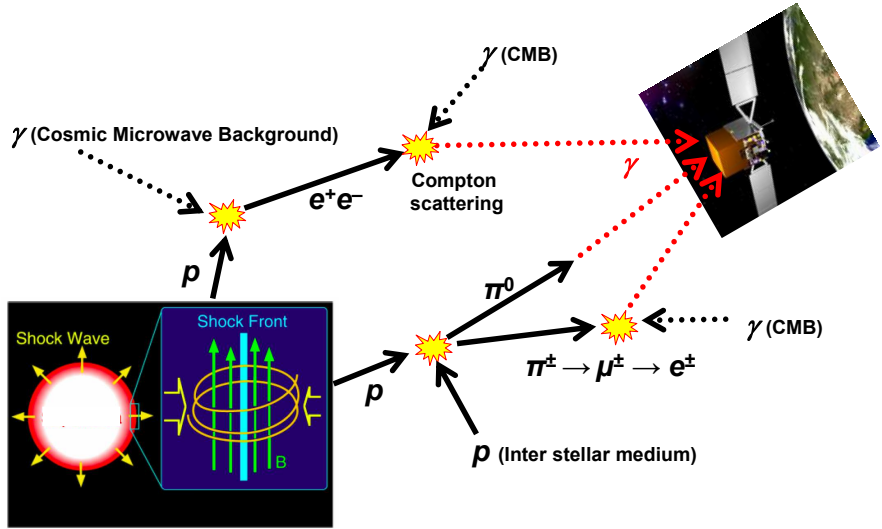}
\end{center}
\caption{A hadronic acceleration model for the production of very high energy gamma rays from extragalactic sources. High-energy Protons which have been accelerated in AGNs or other sources in the presence of magnetic cosmic fields, can interact with the protons of the interstellar medium to give rise to pions (neutral $\pi^0$ and charged $\pi^{\pm}$). Photons from neutral pion decays($\pi^0 \rightarrow 2 \gamma$) could be detected together with those coming from charged-pion-to-muon conversion processes $\pi^{\pm} \rightarrow \mu^{\pm} \rightarrow e^{\pm}$, which eventually yield electrons or positrons that scatter \'a la Inverse Compton with low-energy CMB photons resulting in detectable (on Earth or satellite experiments) high-energy photons (picture taken from H. Tajima SLAC-DOE (USA) Programme review talk (June 7, 2006),
(\texttt{www-conf.slac.stanford.edu/programreview/2006/Talks})
).}
\label{fig:hadronic}
\end{figure}
\emph{Hadronic Acceleration:} A prominent way of producing high energy Gamma Rays of extragalactic origin, is the scattering of very high energy protons (produced in the jet of the AGNs by means of gravitational energy conversion mentioned above (\emph{c.f.} fig.\ref{fig:crproduction}))
off protons in the \emph{interstellar} medium. Such collisions result in pion production, of which neutral pions decay ($\pi^0 \to 2\gamma$) and give rise to very high energy photons
that are detected directly. The charged pions on the other hand are converted to muons, whose decays produces electrons or positrons ($\pi^{\pm} \rightarrow \mu^{\pm} \rightarrow e^{\pm}$); the scattering of the latter with low-energy photons of the cosmic microwave background radiation (CMB) then results, through (inverse) Compton scattering, in photons being detected on Earth or satellites (\emph{c.f.} fig.~\ref{fig:hadronic}).
\begin{figure}[H]
\begin{center}
  \includegraphics[width=7cm, angle=-90]{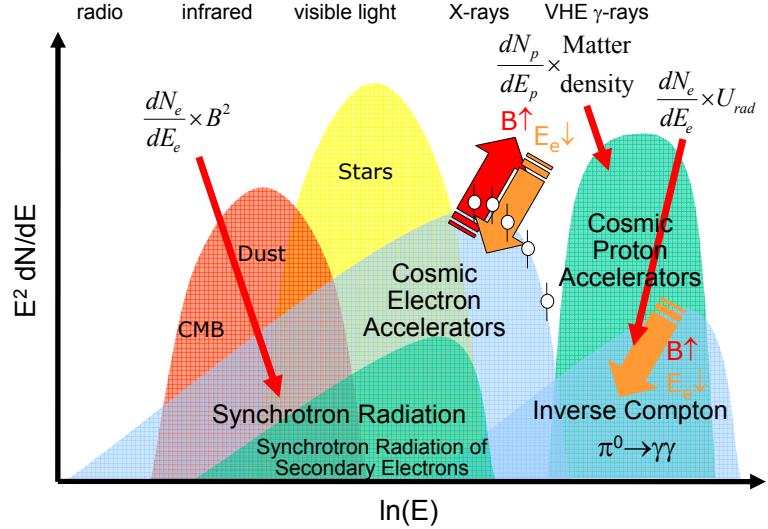}
\end{center}
\caption{Comparing the spectra of very high energy gamma rays in models of leptonic and hadronic acceleration of cosmic particles. In the figure, $E$ denotes the photon energy (in some generic units) and $N$ the observed cosmic photon number, while the suffix ``e'' (``p'') denotes quantities pertaining to
electrons (protons) and $B$ denotes the magnetic field. From the differences in the shape and position of these spectra
one can get information on the kind of acceleration that takes place (picture taken from talk of
M. de Naurois, Workshop of HSSHEP, April 2008, Olympia (Greece), (\texttt{http://www.inp.demokritos.gr/conferences/HEP2008-Olympia/})).}
\label{fig:hadron}
\end{figure}

\begin{figure}[H]
\begin{center}  \includegraphics[width=7cm]{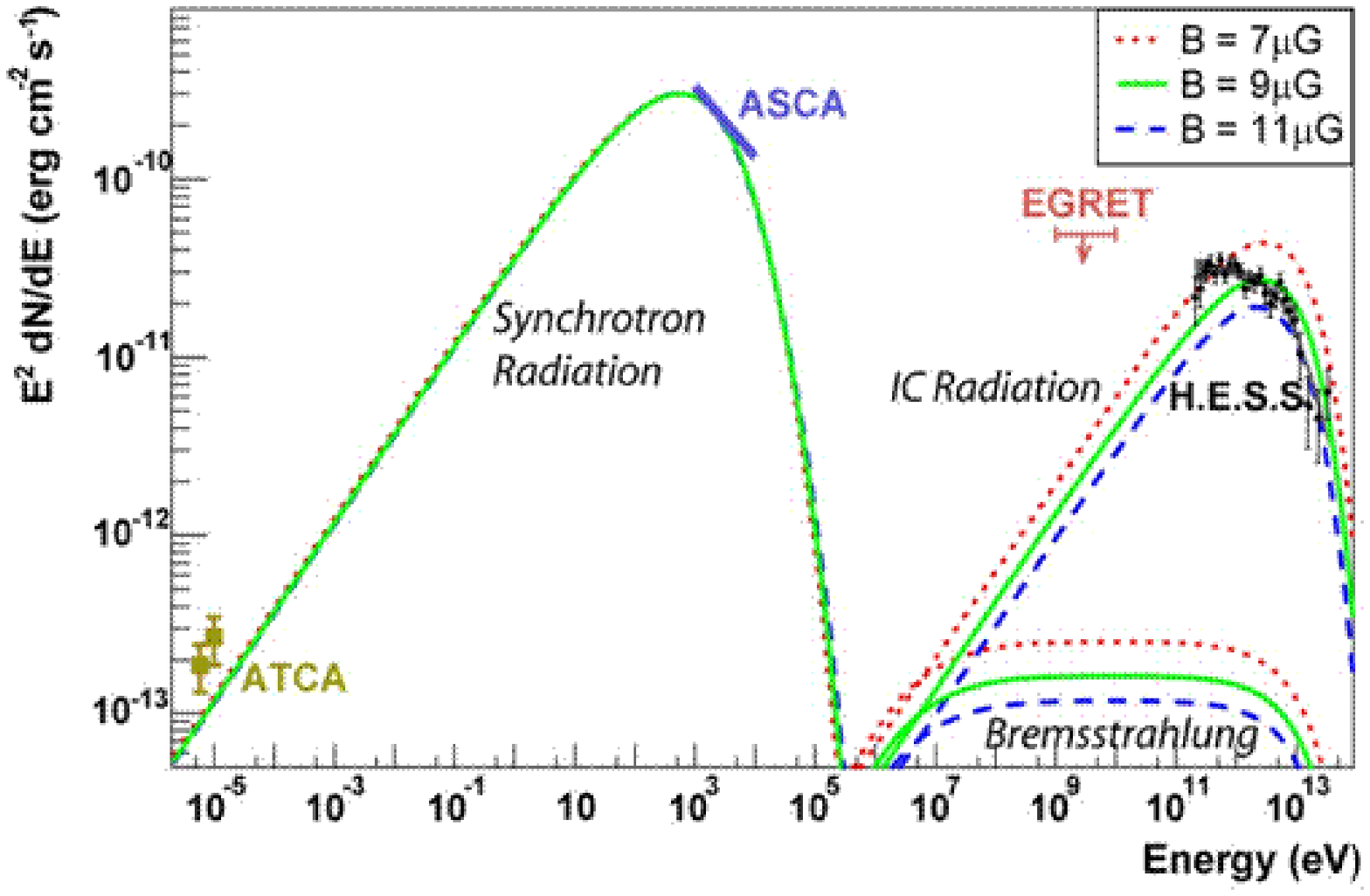} \hfill \includegraphics[width=7cm]{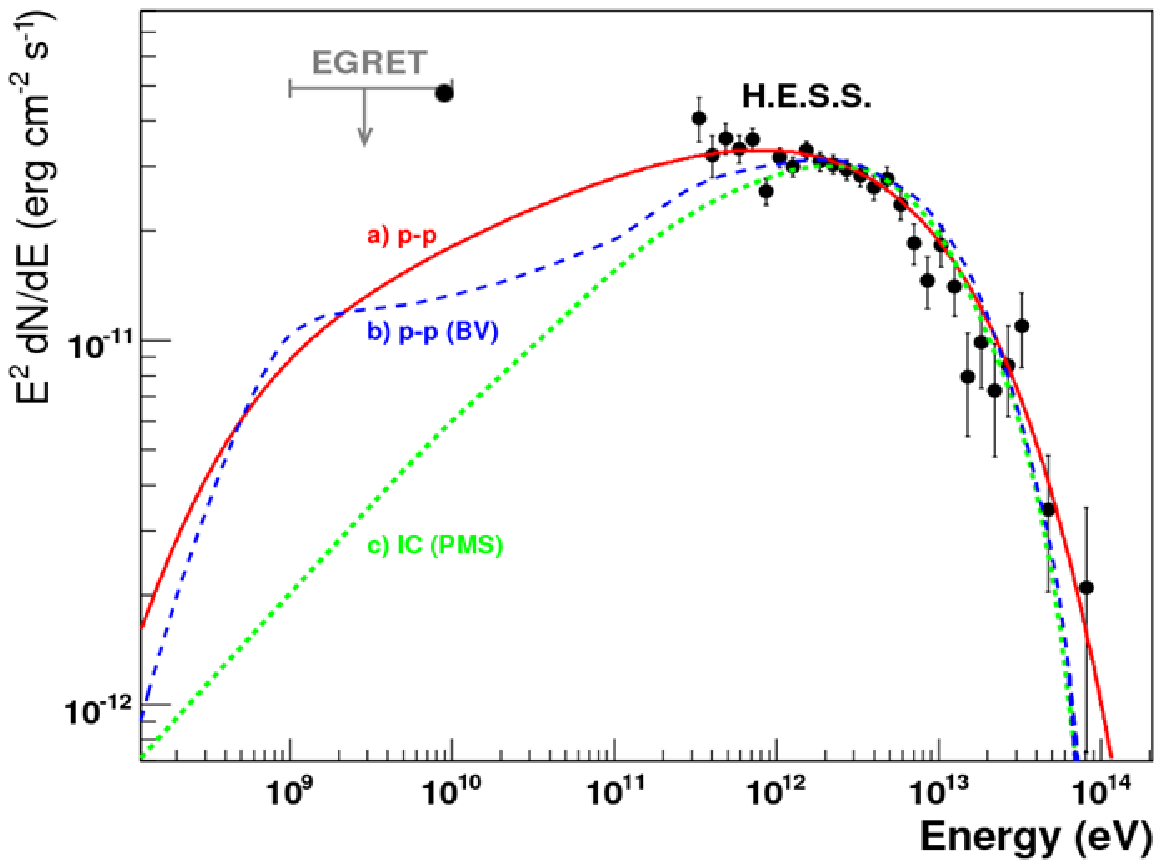}
\end{center}
\caption{ Recent data on Gamma ray spectra. The symbols in the axes have the same meaning as in  fig.~\ref{fig:hadron}. To be sure on the kind of cosmic acceleration taking place (\emph{i.e.} leptonic or hadronic) one needs data at the lower energy part of the spectrum. Such data can be provided, for instance, by the FERMI (GLAST) satellite, which was recently launched (picture taken from talk of
M. de Naurois, Workshop of HSSHEP, April 2008, Olympia (Greece), (\texttt{http://www.inp.demokritos.gr/conferences/HEP2008-Olympia/})).}
\label{fig:hess}
\end{figure}

The energy spectra of hadronic acceleration models are different in shape and location of their peaks on the energy axis from those of leptonic acceleration, as can be seen in fig.~\ref{fig:hadron}, where various photon spectra in the Universe are superimposed for comparison. It is worthy of mentioning that in order to interpret the current high energy gamma ray data using leptonic inverse Compton measurements, one has to assume that the galactic magnetic fields are of low intensity. From measurements of high energy Gamma ray spectra from AGNs or other extragalactic sources, such as Gamma Ray Bursts (GRB), we can then soon get sufficient information into the precise way of acceleration of cosmic particles. The current experimental knowledge on
cosmic high-energy gamma ray spectra can be summarised in fig.~\ref{fig:hess}, from which it is clear that we need more measurements in the lower-energy part of the spectrum before conclusions can be drawn on the kind of cosmic acceleration taking place at various celestial sources.

\subsection{The MAGIC delays: Adding to the uncertainties on the VHE Gamma-Ray production mechanisms \label{sec:3magic}}

The observed time delays of order $4 \pm 1$ minutes between the most energetic (TeV) Gamma rays
from AGN Mkn 501 (\emph{c.f.} fig.~\ref{fig:magic2}), observed by the MAGIC Telescope~\cite{MAGIC,MAGIC2},
lead to further uncertainties in the production mechanism of such photons.

As discussed in \cite{MAGIC}, the conventional model of SSC used to explain the origin of VHE Gamma Rays as being due to an electronic \emph{uniform} acceleration in the AGN jet region, which finds a good application in other AGNs, such as Crab Nebula, fails to account for the observed time delay by MAGIC. The use of the acceleration parameters in the Crab Nebula AGN leads to only millisecond delays of the more energetic photons if applied to the Mkn 501 case.

This prompted speculations that the conventional SSC mechanisms involving
uniform acceleration of the relativistic blob of particles in the jet of the AGN (\emph{c.f.} fig.~\ref{fig:crproduction}) need to be modified for the Mkn 501 case. In fact, several propositions along this line have been made so far:

\begin{itemize}

\item{(i)} Particles inside the emission region moving with constant
Doppler factor need some time to be accelerated to energies that
enable them to produce $\gamma$ rays with specific high energies in the TeV region
\cite{MAGIC}.

\item{(ii)} The $\gamma$-ray emission has been captured in the initial
phase of the acceleration of the relativistic material blob in the jet of the AGN (\emph{c.f.}~fig.~\ref{fig:crproduction}), which at
any point in time radiates up to highest $\gamma$-ray energies
possible~\cite{bwhdgs}.

\item{(iii)} A version of the SSC scenario (termed \emph{one-zone SSC model}), which invokes a brief episode of increased particle injection at low energies \cite{mas08}. Subsequently,
    the particles are accelerated to high energies, which thus accounts for the observed delays, but they also emit synchrotron and SSC, thereby loosing energy. As in scenario (ii) above, also according to this model the MAGIC observations have caught the relativistic electrons in the jet of the AGN at their acceleration phase.

\end{itemize}

To the above I would also like to add the possibilities that some hadronic mechanisms might also be in operation here, which seems not to have been discussed by the community so far.
It therefore becomes clear from the above brief discussion that the situation concerning the delayed production of VHE Gamma Rays from the AGN Mkn 501 is far from being resolved by means of conventional (astro)physics at the source.

This brings us to the main topic of this article, which is a possible (albeit speculative at this stage) link of the MAGIC observation with more fundamental physics associated with the very structure of space time on which the \emph{propagation} of the VHE Gamma Rays takes place.

\subsection{Quantum-Gravity Space-Time Foam and the MAGIC delays: wild speculation or realistic possibility? \label{sec:3magicqg}}

In \cite{MAGIC2} it has been observed that, as a result of the ability of the experiment to measure individual (within the accuracy of the observations of course) photons from Mkn 501, of various energies, it should be possible to reconstruct the peak of the flare of July 9th 2005 using
dispersion relations of these individual photons during their journey from emission till observation. In fact, we went one step further and assumed sub-luminal modified dispersion relations of the type expected~\cite{aemn,horizons,robust,mitsou} to be encountered in a model of quantum-gravity (QG) induced space-time foam~\cite{wheeler} coming from string theory~\cite{emnnewuncert}. As we have already discussed in the previous section, the sub-luminality of the QG-induced refractive index in such models is guaranteed by the very nature of string theory, which respects the cornerstone of special relativity that the speed of light in vacuo is the maximal material velocity. In this respect the space-time foam in such theories leads to the \emph{absence of birefringence}, in other words the refractive index is the same for
both photon polarizations. This is an important feature, which allows the MAGIC results to be compatible with other stringent limits of Lorentz invariance from other astrophysical sources, as we shall explain below.

Although, as we have already argued, the observed time delays within the string model pertain to stringy uncertainties~\cite{emnnewuncert}, and as such may be disentangled from modified dispersion relations due to a breakdown of the local effective lagrangian formalism, nevertheless it is instructive to follow a historical path, and discuss first constraints on the photon dispersion relations coming from best fits to the observed delays and then proceed to the realistic string theory case. This will help the reader understand the failure of the former approach and the advantage of the string model over many approaches to QG based on local effective field theories.

In \cite{MAGIC2} we examined two cases of QG-induced modified dispersion for photons, stemming from (\ref{mdr}) upon assuming:

\begin{itemize}

\item{Case I:} Photon Refractive index  \emph{suppressed  Linearly} by the QG energy scale $M_{{\rm QG1}}$, \emph{i.e.} only the coefficient $c_1 > 0$ in the series of Eq.~(\ref{mdr}) is non zero, its positivity being required by the sub-luminal nature assumed for the propagation, as ensured by the string theory underlying model~\cite{aemn,mitsou,emnnewuncert}.

\item{Case II:} Photon Refractive index  \emph{suppressed  Quadratically}  by the QG energy scale $M_{{\rm QG2}}$, \emph{i.e.} only the coefficient $c_2 > 0$ in the series of Eq.~(\ref{mdr}) is non zero, its positivity again being linked to the sub-luminal nature assumed for the propagation. String theory models with this kind of quadratic suppression also exist in the modern approach to string theory, including representation of our world as a brane (domain wall hyperplane)~\cite{pasipoularides}, but will not be discussed here.
\end{itemize}

The method of reconstructing the peak (``most active part'') of the flare by implementing
modified dispersion relations for individual photons is based on the following
well known fact of classical electrodynamics~\cite{jac}: a pulse of electromagnetic radiation
propagating through a linearly-dispersive medium, as postulated above, becomes
diluted so that its power (the energy per unit time) decreases. The
applicability of classical electrodynamics for estimating the low-energy
behavior induced by space-time foam and
the corresponding pulse-broadening effect have been discussed in ref.~\cite{mitsou}, where we refer the interested reader for further details and explicit examples.

The dilution effects
for the linear or quadratic cases may easily be obtained as described in
\cite{jac} by applying the dispersion laws
\begin{equation}
\omega(k)=k[1-k/(2M_\mathrm{QG1})~,  ~\quad {\rm or} \quad ~
\omega (k)=k[1-k^2/(3M_\mathrm{QG2}^2)]~,
\label{mqndisp}
\end{equation}
where $\omega $ denotes the frequency of the photon, with wave vector $k$.
Any transformation of a signal to reproduce the undispersed signal tends to
recover the original power of the pulse. If the parameter $M_{\rm QGn}~, n=1,2$
is \emph{chosen correctly}, the power of the recovered pulse is \emph{maximized}.

\begin{figure}[H]
\begin{center}\includegraphics[width=0.4\linewidth]{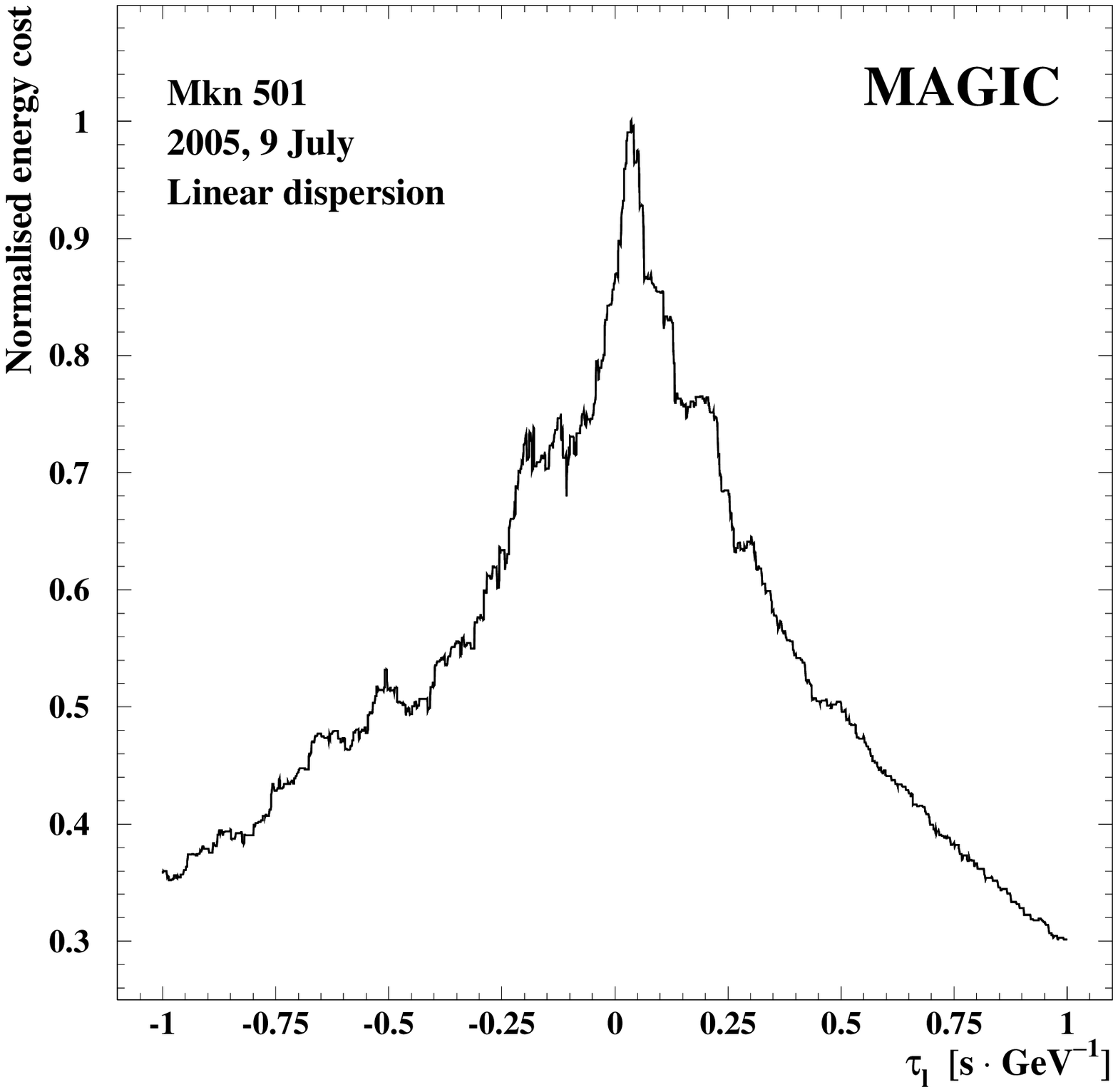}\hfill\includegraphics[width=0.4\linewidth]{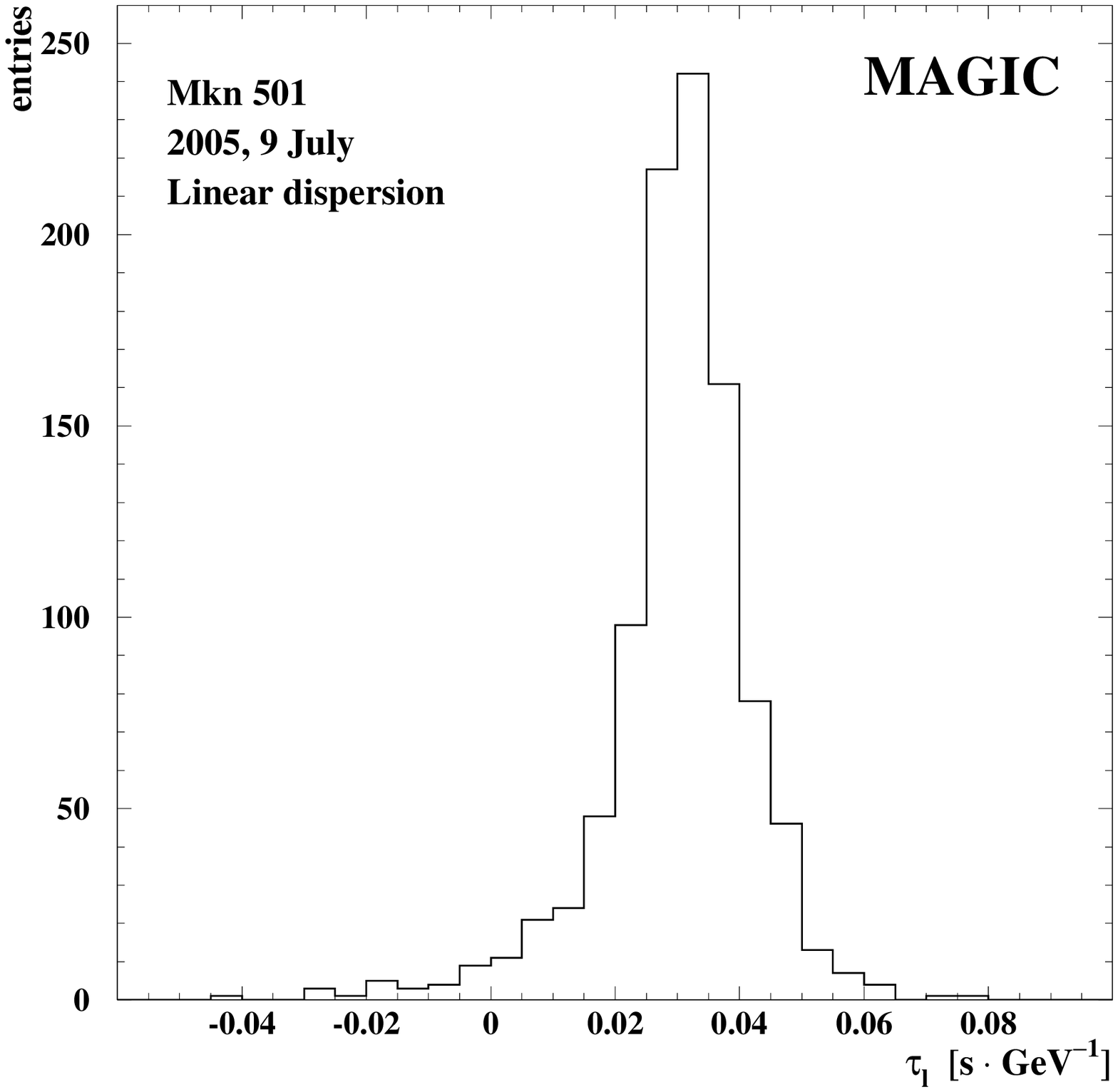}
\end{center}\caption{The \emph{Left figure} shows the Energy Cost Functional (ECF) from one realization of the MAGIC measurements~\protect\cite{MAGIC2} with photon energies smeared by Monte Carlo, for the case of a vacuum refractive index that is linear in the photon energy (Case I). The ECF
exhibits a clear maximum, whose position may be estimated by fitting it with a
Gaussian profile in the peak vicinity. The \emph{Right figure} shows the results of
such fits to the ECFs, specifically the $\tau_l$ distribution for the 1000 energy-smeared realizations of
the July~9 flare, where $\tau_l$ is defined through the representation of the QG scale as $M_{\rm QG1}=1.445\times10^{16}\,{\rm s}/\tau_l$. From this distribution we derive the value $\tau_l=(0.030\pm
0.012)$~s/GeV,  leading
to a lower limit $M_{\rm QG1} > 0.21 \times 10^{18}$~GeV at the 95\%
C.L.}
\label{fig:reconstr}
\end{figure}

This observation has been implemented in the analysis of \cite{MAGIC2}
by appropriately choosing (using statistical-analysis techniques)
a time interval $(t_1;t_2)$ containing the most active part of the flare.
For the record, we mention that this procedure has been applied in \cite{MAGIC} to 1000 Monte Carlo (MC) data samples generated by applying to the measured photon energies the (energy-dependent)
Gaussian measurement errors. The results of the reconstruction of the peak (``most active part'') of the flare using the linear Case I are demonstrated in fig.~\ref{fig:reconstr} for completeness. In a similar way one gets bounds on the quadratic Case II. These results have also been confirmed using different statistical analysis techniques, independent of the ECF. The interested reader is referred to \cite{MAGIC2} for details of the analysis.

Taking into account the \emph{uncertainties in the production mechanism} at the source,
which could also contribute as we have discussed before to the observed delays, we
can only place lower bounds on the \emph{quantum gravity energy scale} from such an analysis.
In fact for the linear and quadratic cases we obtain the lower bounds
(\emph{c.f.} fig~\ref{fig:reconstr}):
\begin{eqnarray}
M_{\rm QG1} &>& 0.21 \times 10^{18}~ {\rm GeV }~~{\rm at~95\%~Confidence~Level~(C.L.)},~\nonumber \\
M_{\rm QG2} &>& 0.26 \times 10^{11}~ {\rm GeV}~~ {\rm at~95\%~C.L}..
\label{mq12bound}
\end{eqnarray}
It is important to notice that, had the mechanism at the source been understood, and the emission of the different photons been
more or less simultaneous, \emph{i.e.} by at least two orders of magnitude smaller as compared to the observed delays, then the above lower bounds could be turned into a real measurement of the
quantum gravity scale. It is surprising, therefore, that in such a situation Case I can reproduce the observed delays, provided $M_{{\rm QG1}}$ is of order of the so-called \emph{reduced Planck mass}, which is an energy scale characterizing conventional string theory models.

In the analysis of \cite{MAGIC2} it was possible to exclude the possibility that the observed time delay may be due
to a conventional QED plasma refraction effect induced as photons propagate
through the source. From the discussion in sub-section \ref{sec:ntv}, in
particular eq.~(\ref{highenergy}), it becomes clear that
if the delay would be due to plasma effects at the source region, then
this would induce $$\Delta t = D (\alpha^2
T^2/6k^2) \ln^2(kT/m_e^2)~,$$ where $\alpha$ is the fine-structure constant, $k$
is the photon momentum, $T$ is the plasma temperature, $m_e$ is the mass of
electron, $D$ is the size of the plasma, and we use natural units: $c, \hbar
=1$. Plausible numbers such as $T \sim 10^{-2}$~MeV and $D \sim 10^9$~km
(for a review see \cite{hillas}) yield a negligible effect for $k \sim 1$~TeV, which
is of order of the photon momenta relevant to the MAGIC experiment.
Exclusion of other source effects, such as time evolution in the mean emitted
photon energy, might be possible with the observation of more flares, \emph{e.g.}, of
different AGNs at varying redshifts. Observations of a single flare cannot
distinguish the quantum-gravity scenarios considered here from modified
synchrotron-self-Compton mechanisms.

\subsection{H.E.S.S. and FERMI Observations and Quantum-Gravity Scale Bounds}

However, the above-described
pioneering study demonstrates clearly the potential scientific value of an
analysis of multiple flares from different sources. The most promising
candidate for applying the analyses proposed here is the flare
from the Active Galaxy PKS 2155-304 detected recently by H.E.S.S. (\emph{H}igh \emph{E}nergy \emph{S}tereoscopic \emph{S}ystem) Collaboration~\cite{hess2155}, another
experiment involving arrays of Cherenkov Telescopes in Namibia (Africa). This galaxy lies further than Mk501 at redshift $z = 0.116$ and there is a much higher statistics of photons at energy ranges of a few TeV.

In fact H.E.S.S. collaboration published their measurements~\cite{hessnew} on the arrival time of photons from PKS 2155-304.
However, unlike the MAGIC observations from Mk501 Galaxy, there was no time lag found between higher- and lower-energy photons in this case. These results can thus place only bounds on the quantum gravity scale,
if space-time foam is assumed to affect the photon propagation. The bounds are similar to the MAGIC case (\ref{mq12bound}) above.

The H.E.S.S. result can mean several things, and certainly points towards different source mechanisms for the acceleration of photons between the two galaxies Mkn 501 and PKS 2155-304. However, this second measurement by H.E.S.S. cannot still rule out the possibility that quantum gravity plays a r\^ole in photon propagation. For instance, one cannot exclude the (admittedly remote) possibility that, due to a still unknown source effect, the high-energy photons in the PKS 2155-304 Galaxy are emitted first, in contrast to the Mkn 501 case, in such a way that a sub-luminal vacuum refractive index quantum-gravity effect, of a strength appropriate to produce the delays observed by the MAGIC experiment, ``slowed these photons down'' as compared to the lower-energy ones, so that there are no observable delays in the arrival times between the higher and lower energy photons in the H.E.S.S. experiment. Of course, one cannot exclude the possibility that the conditions of this set of measurements, for some reason, prohibited the detection of an observable time lag between high and low energy photons.

\begin{figure}[H]
\centering
\includegraphics[width=7.5cm,angle=-90]{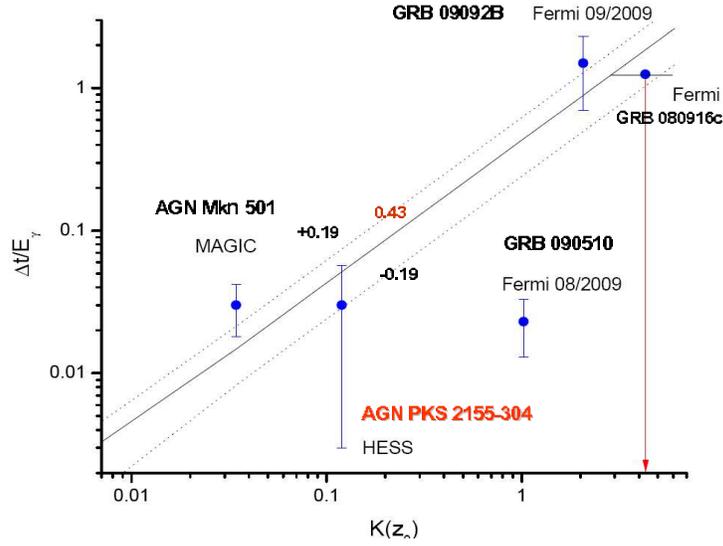}
\caption{\it Comparison of data on delays
$\Delta t$ in the the arrival times of energetic gamma rays from various astrophysical sources with
models in which the velocity of light is reduced by an amount linear in the photon
energy~\protect\cite{emnfit}. The graph plots on a logarithmic scale the quantity $\Delta t/E$ and a function of the
red-shift, $K(z)$, which is essentially the distance of the source from the observation point.
The data include two AGNs, Mkn 501~\protect\cite{MAGIC,MAGIC2} and PKS
2155-304~\protect\cite{hessnew}, and three GRBs observed by the Fermi satellite,
090510~\protect\cite{grb090510}, 09092B~\protect\cite{grb09092b} and 080916c~\protect\cite{grbglast}.}%
\label{fig:data}%
\end{figure}

Although the available sample of AGNs is still not large enough for a robust
analysis on bounds of Quantum-Gravity medium effects~\cite{robust,mitsou}, nevertheless, one can at least check for consistency between the
available MAGIC and HESS results, and gauge the magnitude of possible intrinsic
fluctuations in the AGN time-lags. Comparing the time-lag measured by MAGIC for
Mkn 501 at redshift $z = 0.034$: $\Delta t/E_\gamma = 0.030 \pm 0.012$~s/GeV, with that
measured for PKS 2155-304 at $z = 0.116$: $\Delta t/E_\gamma = 0.030 \pm 0.027$~s/GeV,
we see that they are compatible with a common, energy-dependent {\it intrinsic} time-lag
at the source.
On the other hand, they are also compatible with a universal redshift- and energy-dependent
{\it propagation} effect:
\begin{equation}
\Delta t/E_\gamma = (0.43 \pm 0.19) \times K(z) {\rm s/GeV} ,~\quad
K(z) \equiv   \int_0^z \frac{(1 + z)dz}{\sqrt{\Omega_\Lambda + \Omega_m (1 + z)^3}} ,
\label{bestfit}
\end{equation}
assuming an expanding Universe within the framework of the standard Cosmological-constant-Cold-Dark-Matter ($\Lambda$CDM) model.
The best fit~\cite{emnfit} of the MAGIC and HESS data based on (\ref{bestfit}) leads to the following result for the
Quantum Gravity scale, assuming that it is the dominant cause of the delay:
$M_{QG1} = (0.98^{+0.77}_{-0.30}) \times 10^{18}$~GeV. The situation is depicted in fig.~\ref{fig:data}.

With measurements from only a few available flares from AGN it is,
therefore, not possible to disentangle with any certainty source from propagation effects.
For this we need statistically significant
populations of available data~\cite{mitsou}.
Unfortunately the occurrence of fast flares in AGNs is currently unpredictable,
and since no correlation has yet been established with observations in other
energy bands that could be used as a trigger signal, only serendipitous
detections are currently possible.
It seems unlikely that the relatively rare
and unpredictable sharp energetic flares produced
only occasionally by AGNs, which have a relatively restricted redshift range and hence a
small lever arm, will soon be able to provide the desired discrimination.

On the other hand, Gamma Ray Bursts (GRBs) are observed at a relatively high rate, about one a day, and
generally have considerably larger redshifts.
The advent of the FERMI (n{\' e}e GLAST) Telescope with its large acceptance offers the
possibility of achieving the required sensitivity. Indeed, the FERMI Collaboration has
already made a report of GeV-range $\gamma$ rays from the GRB 080916c~\cite{grbglast}.
In this GRB, there is a 4.5-second time-lag between the onsets
of high- ($> 100$~MeV) and low-energy ($< 100$~KeV)
emissions. Moreover, the highest-energy photon GRB 080916c measured by the FERMI $\gamma$-ray telescope had an energy $E = 13.2^{+0.70}_{-1.54}$~GeV, and was
detected $\Delta t = 16.5$~s after the start of the burst.
Spectroscopic information has been used by the GROND Collaboration~\cite{GROND} to estimate
the redshift of GRB 080916c as $z = 4.2 \pm 0.3$~\cite{grbglast}. Assuming
this value of the redshift, the best fit (\ref{bestfit}) would correspond to
a time-lag
\begin{equation}
\Delta t = 25 \pm 11~{\rm s}
\end{equation}
for a 13~GeV photon from GRB 080916c. As discussed in \cite{emnfit,grbglast}, such time delays can fit excellently within the above-mentioned QG scenario for a subluminal refractive index for photons, with the following lower bound for the QG scale
$M_{QG1} > 1.50 \pm 0.20 \times 10^{18}~{\rm GeV}$, where the inequality is due to the ignorance of the source mechanism (\emph{c.f}. fig.~\ref{fig:data}).
This bound is consistent with the MAGIC
and HESS results stated previously. The reader should also bear in mind that
the 4.5-second time-lag
observed for $\sim 100$~MeV photons could not be explained by a propagation
effect that depends linearly on the energy~\cite{grbglast}.  It is clear therefore that the FERMI Telescope has already demonstrated the sensitivity to probe a possible linearly
energy-dependent propagation effect at the level reached by the available AGN data,
and it is appropriate and possibly helpful to consider how such an effect could be probed
in the future.

The analysis of \cite{emnfit} has also demonstrated that although the three above-mentioned sets of data, from MAGIC, FERMI and HESS Collaborations, can be explained simultaneously by a linear in energy vacuum refractive index, suppressed by a single power of the quantum gravity scale, this is not the case for a refractive index scaling quadratically with the photon energy, for instance of the type encountered in some brane models with asymmetric warp factors, as in ref.~\cite{pasipoularides}. Indeed, on assuming that the quadratic refractive index is the sole cause of the observed delays in the arrival of high- \emph{vs}. low- energy photons
in \emph{both} the MAGIC and FERMI cases, this would imply a time delay of order $0.24 \pm 0.16$ s for the most energetic photon (13.22 GeV) of the GRB 080916c, for a quantum gravity scale that saturates $M_{\rm QG2}$ in (\ref{mq12bound}).
This is two orders of magnitude smaller than the measured time-lag (16.5 s) by the FERMI Collaboration~\cite{grbglast}.

\subsection{Trouble with GRB 090510 - non-uniform D-particle foam? \label{sec:trouble}}

In figure \ref{fig:data} we also depict the data from the short burst GRB 090510~\cite{grb090510}, observed by the FERMI Telescope, which so far we did not include in our linear in energy fit. It is clear that a fit with a linearly modified dispersion relation  does not work in this case for the values of the Quantum Gravity scale $M^{\rm MAGIC}_{\rm QG\,1} \sim 10^{18}$~GeV that fit the MAGIC, H.E.S.S. and the other FERMI data. Indeed, the observed short delays of this burst can be explained on the basis of linearly modified dispersion relations only if the quantum gravity scale is \emph{larger} than $M_{\rm QG\,1} \simeq 1.2 M_P$ where $M_P = 1.22 \times 10^{19}$~GeV is the Planck mass, which thus is incompatible with the value of the quantum gravity scale for the MAGIC data fit.
This has been used by the Fermi collaboration as a strong indication that Lorentz violating models entailing modified dispersion relations with linear suppression are ruled out on naturalness grounds, since the so-obtained quantum gravity scale for the GRB 090510 is about 1.2 times larger than the four-dimensional Planck mass $M_P$.

We would not agree with this statement. Leaving aside the fact that from a single measurement, with uncertainties on the precursor of the GRB, one cannot draw safe conclusions, we mention that what one calls a \emph{natural
scale of QG  } is highly model dependent. For instance, as we have seen in the previous subsection,
in the string foam model,  the relevant ``QG scale'', that dictates the order of the foam-induced time delays of photons, is a complex function of the model's parameters and is not simply given by the string scale $M_s$~\cite{emndvoid}. Indeed, as becomes clear from (\ref{totaldelay}) (or (\ref{redshift})) the relevant scale is not simply the D-particle mass, $M_s/g_s$, but a combination
\begin{equation}
M_{\rm QG-D-foam} \sim \frac{M_s}{g_s\, n^*(z)}
\end{equation}
 involving the linear number density $n^*(z)$ of the foam defect, encountered by the photon during its propagation from the source till observation. This quantity depends on the bulk density of the D-particles, which in the model of \cite{emnw} is a free parameter.

Inhomogeneous bulks are perfectly consistent background configurations for our brane world scenario~\cite{emndvoid}.
In order to match the photon delay data of Fig.~\ref{fig:data} with the D-foam model,
we need a reduction of the \emph{linear} density of defects encountered by the photon by
about two orders of magnitude in the region $0.2 < z < 1$,
whereas for $z < 0.2 $ there must be, on average, one D-particle defect per unit string
length $\ell_s$. We therefore assume that our D-brane encountered a D-void when $0.2 < z < 1$,
in which there was a significant reduction in the bulk nine-dimensional density of defects.
Such assumptions can be consistent with cosmological considerations on the dark sector of the model, as explained in \cite{emndvoid}. In particular, an important parameter for the cosmology of the model is the propagation velocity $v$ of the brane world in the bulk (\emph{c.f}. fig.~\ref{dfoam:fig}). As we have discussed previously (\ref{nofz}) this is related~\cite{emndvoid} to the density of the defects near the brane world, $n^{\rm short}(z)$, and hence to the linear density of defects encountered by the photon on the brane world.

To have a reduced density by two order of magnitude at redshifts $z=0.9$, while having a density of defects of $O(1)$ per string length at redshifts $z < 0.1$, one  must consider
the magnitude of $v$ as well as the string scale $\ell_s$.
For instance, it follows from (\ref{nofz}) that for string energy scales of the order of TeV, \emph{i.e.} string time scales $\ell_s/c = 10^{-27}~s$, one must consider a brane velocity $v \le \sqrt{10} \times 10^{-11} \, c$,
which is not implausible for a slowly moving D-brane at a late era of the Universe~\footnote{Much
smaller velocities are required for small string scales that are
comparable to the four-dimensional Planck length.}.
This is compatible with the constraint on $v$ obtained from inflation in~\cite{brany},
namely $v^2 \le 1.48 \times 10^{-5}\,g_s^{-1}$, where $g_s < 1$ for the
weak string coupling we assume here. In  our model, due to the friction induced on the
D-brane by the bulk D-particles, one would expect that the late-epoch brane velocity
should be much smaller than that during the inflationary era immediately following
a D-brane collision~\cite{brany}.

The above considerations
can be extended~\cite{emndvoid} to type-IIB constructions of D-foam~\cite{li}.
If there was a depletion of D-particle defects
in a certain range of redshifts in the past, \emph{e.g}., when $z \sim 0.9$,
the volume $V_{A3}$ in (\ref{coupl-N}) would have
been larger than the corresponding volume at $z \ll 1$,
and the corresponding coupling reduced, and hence also the corresponding cross section
$\sigma$ describing the probability of interaction of a photon with the D-particle in the foam.

The \emph{effective} linear density of defects $n^*(z)$ appearing in (\ref{totaldelay})
can then be related to the number density $\tilde{n}^{(3)}$ per unit three-volume on the
compactified D7-brane via $ n^*(z) \sim \sigma \tilde{n}^{(3)}$. The quantity $\tilde{n}^{(3)}$ is,
in turn, directly related to the bulk density of D-particles in this model, where there is no
capture of defects by the moving D-brane, as a result of the repulsive forces between them.
The cross section $\sigma$ is proportional to the square of the string amplitude (\ref{4ampl})
describing the capture of an open photon string state by the D-particle in the model,
and hence is proportional to $g_{37}^4$, where $g_{37}$ is given in (\ref{coupl-N}).
If recoil of the D-particle is included,
the amplitude  is proportional to the effective string coupling: $g_s^{\rm eff} \propto g_{37}^2$,
where $g^{\rm eff} = g_s (1 - |\vec u|^2)^{1/2}$, with $\vec u \ll 1$  the recoil velocity
of the D-particle~\cite{emnnewuncert,li}.
One may normalize the string couplings in such a way that $n^*(z=0)={\cal O}(1)$,
in which case $n^*(z\simeq 0.9) \ll 1$ by about two orders of magnitude.
However, in general the induced gauge string couplings depend crucially on the
details of the compactification as well as the Standard Model phenomenology on the
D-brane world. Hence the precise magnitude of the time delays is model-dependent,
and can differ between models.

The above discussion hopefully demonstrates that the question: ``what is a \emph{natural scale} for QG?''  is highly dependent on the details of the microscopic model and does not admit a straightforward answer. Many more measurements, at various cosmological distances (red shifts) are needed before we falsify a model, such as the string foam model, for instance, discussed here. Nevertheless, with the progress in extraterrestrial instrumentation, and with the launch of facilities like FERMI or future Cherenkov Telescope Arrays (CTA)~\cite{cta}, this will hopefully become possible in the foreseeable future.

\subsection{Other (astrophysical) constraints on quantum-gravity foam \label{sec:biref}}

The sensitivity of the MAGIC (and FERMI) observations to Planck scale physics (\ref{mq12bound}), at least
for linearly suppressed modified dispersion relations, calls for an immediate comparison with other sensitive probes of non-trivial optical properties of QG medium.

Indeed, from the analysis of \cite{MAGIC2}, there was no microscopic model dependence of
the induced modifications of the photon dispersion relations, other than the sub-luminal nature of the induced refractive index and the associated absence of birefringence, that is the independence
of the refractive index on the photon polarization. The latter feature avoids the otherwise very stringent constraints on the photon dispersion relation imposed by astrophysical observations, as we now come to discuss.

We shall be very brief in our description of the complementary astrophysical tests on Lorentz invariance and quantum-gravity modified dispersion relations, to avoid large diversion from our main point of this review article which is string theory.

There are three major classes of complementary astrophysical constraints, to be considered in any attempt to interpret the MAGIC, FERMI or more general $\gamma$-ray Astrophysics results in terms of quantum-gravity induced anomalies in photon dispersion.

\begin{itemize}

\item{\emph{Birefringence and strong constraints on QG-induced photon dispersion:}}

In certain models of quantum gravity, with modified dispersion relations, for instance the so-called loop-quantum gravity~\cite{gambini}, the ground state breaks reflexion symmetry (parity) and this is one of the pre-requisites for a dependence of the induced refractive index on the photon polarization, \emph{i.e}. birefringence. We remind the reader that in birefringent materials this is caused precisely by the existence of some kind of anisotropies in the material. The velocities of the two photon polarizations (denoted by $\pm$) in such QG models may be parametrised by:\begin{equation}
v_{\pm} = c\left(1 \pm \xi (\hbar \omega /M_P )^n \right)
\label{birefrrel}
\end{equation}where $M_P = 1.22 \times 10^{19}$ GeV is the Planck energy scale, and $\xi$ is a parameter
following from the underlying theoretical model, which is related with the modifications
of the pertinent dispersion relations for photons. The order of suppression
of these effects is described by $n$ which in the models of \cite{gambini} assumed the value $n=1$, but in general one could have higher order suppression, as we have discussed in (\ref{mdr}).

 Vacuum QG birefringence should have showed up in optical measurements from remote astrophysical sources, in particular GRBs.
Ultraviolet (UV) radiation measurements from distant galaxies~\cite{uv} and UV/optical polarization measurements of light from  Gamma Ray Bursters~\cite{grb} rule out
birefringence unless it is induced at a scale (way) beyond the Planck mass (for linear models,
the lower bound on the QG scale in such models can exceed the Planck scale ($\sim 10^{19}$ GeV) by as much as \emph{seven orders of magnitude}). Indeed,  in terms of the parameter $\eta$ introduced above (\emph{c.f.} (\ref{birefrrel})), for the case $n=1$ of \cite{gambini}, one finds from
 optical polarization observations that the absence of detectable birefringence effects imply the upper bound  $|\xi| < 2 \times 10^{-7}$, which is incompatible with the MAGIC observed delays,
 saturating from below the bounds (\ref{mq12bound}).

 At this point, we wish to mention that, using recent polarimetric observations of the Crab Nebula in the hard X-ray band by INTEGRAL~\cite{integral}, the authors of \cite{macio} have demonstrated  that the absence of vacuum birefringence effects constrains linearly suppressed Lorentz violation in quantum electrodynamics to the level $|\xi | < 6 \times 10^{-10}$ at 95\% C.L., thereby tightening by about three orders of magnitude the above-mentioned constraint.

\item{\emph{Synchrotron Radiation and further stringent constraints for electronic QG-induced anomalous dispersion in vacuo.}}

\begin{figure}[ht]
\begin{center}
\includegraphics[width=0.2\linewidth]{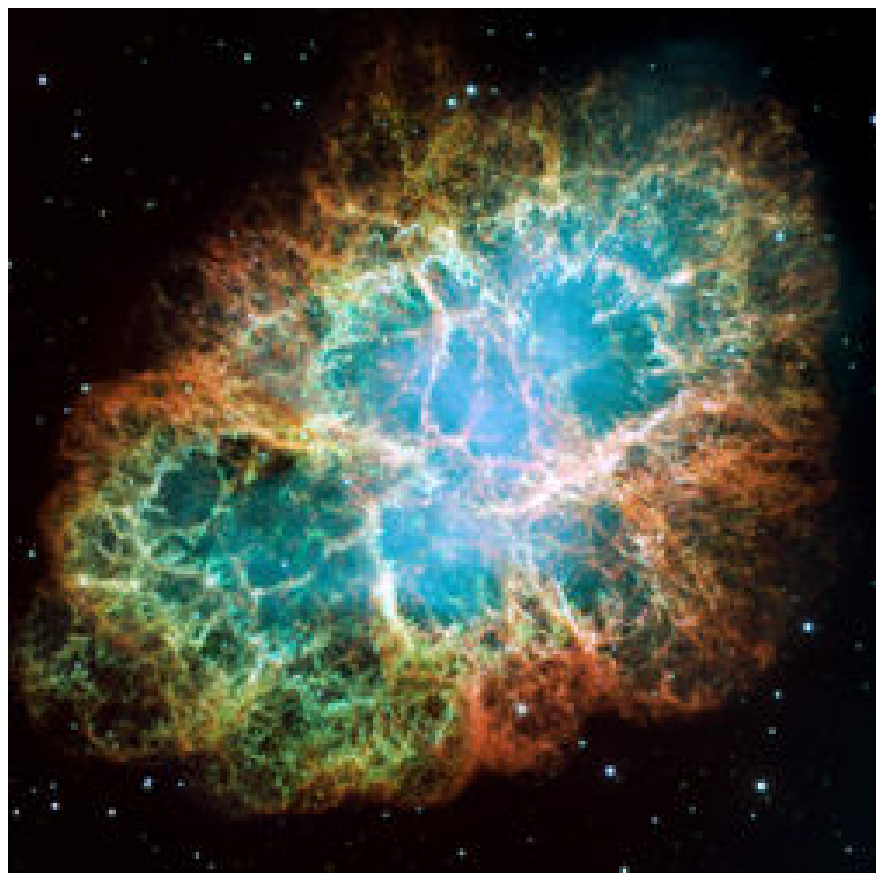}\hfill \includegraphics[width=0.2\linewidth]{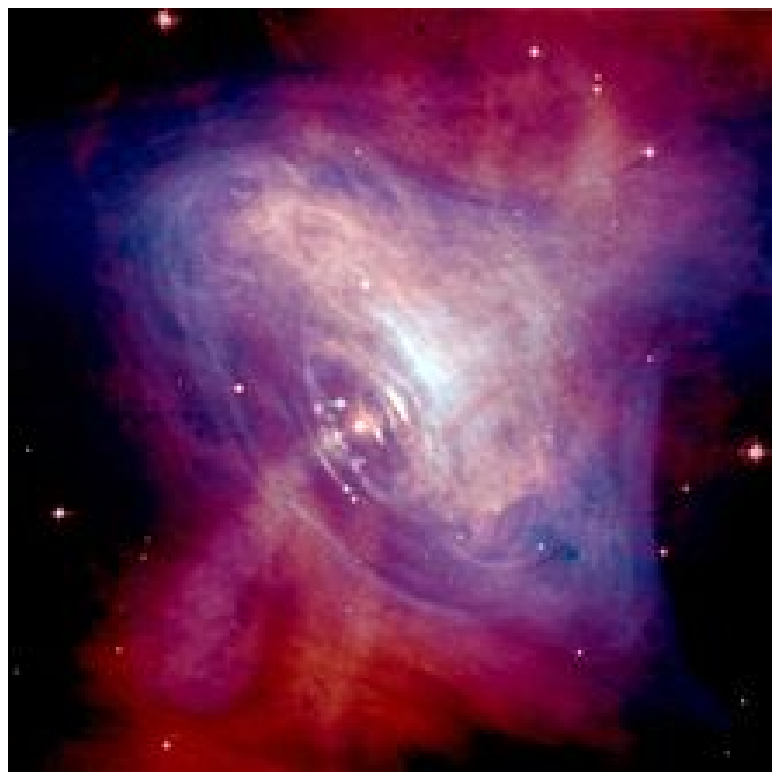}\hfill
\includegraphics[width=0.4\linewidth]{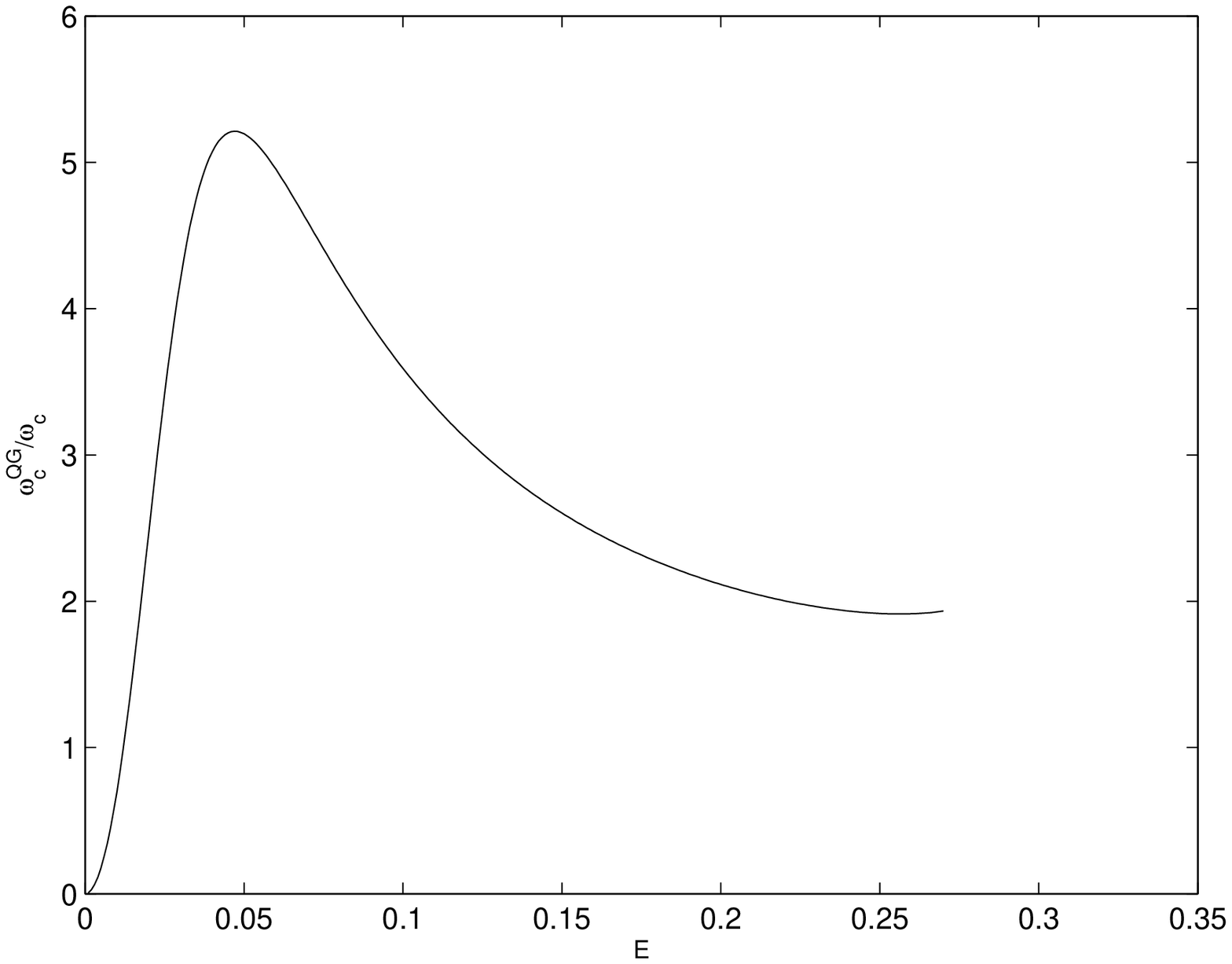}
\end{center}
\caption{The Crab Nebula (left image) is a supernova remnant, with a rotating neutron star (the Crab Pulsar) at its centre (middle image). Observations of synchrotron radiation (right image, radiation spectrum (arbitrary units)) from such celestial objects place very stringent constraints on quantum gravity models with anomalous dispersion relations for electrons (images taken from NASA, \texttt{http://www.nasa.gov/} (Crab Nebula, left), wikipedia, \texttt{http://en.wikipedia.org/wiki/CrabNebula} (Crab Pulsar, middle) and the first of Ref.~\protect\cite{ems} (right)).}
\label{fig:crab}
\end{figure}
Another important experimental constraint on models with QG-induced anomalous dispersion relations
comes from observations of synchrotron radiation from distant galaxies~\cite{crab,ems,crab2}, such as Crab Nebula (\emph{c.f.} fig.~\ref{fig:crab}). As mentioned previously, the magnetic fields at the core regions of galaxies curve the paths of (and thus accelerate) charged particles, in particularly electrons (which are stable and therefore appropriate for astrophysical observations), and thus, on account of energy conservation this results in synchrotron radiation.

In standard electrodynamics~\cite{jac}, electrons in an
external magnetic field ${H}$, follow helical orbits
transverse to the direction of ${H}$. The so-accelerated electrons in a magnetic field
emit synchrotron radiation with a spectrum that cuts off sharply at a
frequency $\omega_c$ (\emph{c.f.} fig.~\ref{fig:crab}, right panel):
\begin{equation}
\omega^{LI}_c = \frac{3}{2}\frac{eH}{m_0}\frac{1}{1 - \beta^2},
\label{sync}
\end{equation}
where $e$ is the electron charge, $m_0$ its mass, and $\beta_\perp \equiv v_\perp $ is the component of the velocity of the electron perpendicular to the direction of the
magnetic field.
The superfix $LI$ in (\ref{sync})
stresses that this formula is based on a LI approach, in which one
calculates the electron trajectory in a given magnetic field $H$ and the
radiation produced by a given current, using the relativistic relation
between energy and velocity.

All these assumptions are
affected by violations of Lorentz symmetry, such as those encountered in quantum-gravity space-time foam models, leading to modified dispersion relations
of the form:
\begin{eqnarray}
 \omega^2(k)&=& k^2 + \xi_\gamma \frac{k^{2 + \alpha}}{M_P^\alpha},
 \label{eq:pdr}\\
 E^2(p)&=& m_0^2+ p^2 + \xi_e \frac{p^{2 + \alpha}}{M_P^\alpha},
 \label{eq:mdr}
\end{eqnarray}
for photons (\ref{eq:pdr}) and electrons (\ref{eq:mdr}), where $\omega$ and $k$ are the
photon frequency and wave number, and $E$ and $p$ are the electron energy
and momentum, with $m_0$ the electron (rest) mass. In the spirit of the MAGIC observation analysis above, we assume here linear ($\alpha =1$) or quadratic QG ($\alpha = 2$)
effects, characterized by parameters $\xi_\gamma$ and $\xi_e$, extracting
the Planck mass scale $M_{P}=1.22\times 10^{19}$ GeV. In fact one can do the analysis~\cite{ems} for a general $\alpha$ (single power) and attempt to extract limits on this parameter by matching with observations.

A detailed analysis~\cite{crab,ems}, including the modifications in the electron's trajectories due to space-time foam~\cite{ems}, yields:
\begin{eqnarray}
&& \omega_c^{QG} \propto \omega_c^{LI} \frac{1}{
(1 + \sqrt{2 -1/{\cal \eta}^2})^{1/2}
\left(\frac{m_0^2}{E^2} + (\alpha + 1)\left(\frac{E}{{\cal M}}\right)^\alpha
\right)}~, \nonumber \\
&& {\cal M} \equiv M_P/|\xi_e|~, \quad \eta \equiv
1 - (E/{\cal M})^\alpha~,
\label{modqg}
\end{eqnarray}
where $\omega_c^{LI}$ is given in (\ref{sync}) and the superscript ``QG'' indicates that the QG-modified dispersion relations (\ref{eq:mdr})
are used. This function is plotted schematically (for $\alpha =1$)
in fig.~\ref{fig:crab} (right panel).

In \cite{crab,ems},
the above QG-modified dispersion relations have been tested using observations
from Crab Nebula.
It should be emphasized that the
estimate of the end-point energy of the Crab synchrotron spectrum
and of the magnetic field used above are indirect values based on
the predictions of the Synchrotron Self-Compton (SSC) model
of very-high-energy emission from Crab Nebula~\cite{reports}. In \cite{ems}
the choice of parameters used was the one that gives good agreement
between the experimental data on high-energy emission and the
predictions of the SSC model \cite{reports,hillas}.
Estimating the magnetic
field of Crab Nebula in the region
$160 \times 10^{-6}~{\rm Gauss} ~ < ~ H ~ < ~ 260 \times 10^{-6} ~{\rm Gauss}$,
and requiring $|\xi_e| \le 1$ (which thus sets the quantum gravity scale as at least $M_P$)
one obtains
the following bounds on the exponent $\alpha $ of the dispersion relations (\ref{eq:mdr})~\cite{ems}:
\begin{equation}
\alpha \ge \alpha_c~: \qquad 1.72 < \alpha_c < 1.74
\label{alphabounds}
\end{equation}
These results
imply already a sensitivity
to quadratic QG corrections with Planck mass suppression $M_P$.

However, for photons there are no strong constraints on $\xi_\gamma$ coming from synchrotron radiation studies, unless in cases where QG models entail  birefringence~\cite{crab2}, where, as we discussed above, strong constraints on the photon dispersion are expected at any rate from optical measurements on GRBs.
In this sense, the result (\ref{alphabounds}) \emph{excludes the possibility} that the MAGIC observations
leading to a four-minute delay of the most energetic photons are due to a quantum foam that acts \emph{universally} among photons and electrons. However, the synchrotron radiation measurements cannot exclude anomalous photon dispersion with linear Planck-mass suppression, leading to a saturation of the lower bound (\ref{mq12bound}), in models where the foam is transparent to electrons, as in the string foam case.

\item{\emph{Strong constraints from Ultra-high-energy Cosmic photon annihilation}}

Further strong constraints on generic modified dispersion relations for photons, like the ones used in the aforementioned QG-interpretation of the MAGIC results~\cite{MAGIC2}, comes from processes of scattering of ultra-high-energy photons, with energies above $10^{19}$ eV, off
very-low energy cosmic photons, such as the ones of the cosmic microwave background (CMB) radiation that populates the Universe today, as a remnant from the Big-Bang epoch.
In \cite{sigl} it has been argued that the non-observation of such ultra-high energy (UHE) photons
places very strong constraints on the parameters governing the modification of the photon dispersion relations, that are several orders of magnitude smaller than the values required to reproduce the MAGIC time delays, should the effect be attributed predominantly to photon propagation in a QG dispersive medium.

The main argument relies on the fact that
an ultra-high-energy photon would interact with a low-energy (``infrared'') photon of the CMB background (with energies in the eV range) to produce electron prositron pairs, according to the reaction:
\begin{equation}
  \gamma_{\rm UHE} \quad + \quad \gamma_{\rm CMB} \quad \Rightarrow \quad e^+~e^- ~.
\label{heirreact}
\end{equation}
The basic assumption in the analysis is the strict energy and momentum conservation in the
above reaction, despite the modified dispersion relations for the photons.
Such an assumption stems from the validity of a \emph{local-effective-lagrangian description}
of QG foam effects on particles with energies much lower than the QG energy scale (assumed close to Planck scale $M_{P} = 10^{19}$ GeV). In this formalism, one can represent effectively the foam \emph{dispersive effects}  by higher-derivative \emph{local operators} in a \emph{flat-space-time} Lagrangian. The upshot of this is the modification of the pertinent equations of motion for the photon field (which in a Lorentz-invariant theory would be the ordinary Maxwell equations) by higher-derivative terms, suppressed by some power of the QG mass scale.

One considers the modified dispersion relations (\ref{eq:pdr}), (\ref{eq:mdr}), which in the notation of \cite{sigl}, taking explicit account of the various polarizations and helicities, can be written as:
\begin{eqnarray}\label{siglmdr}
  && \omega^2_{\pm} = k^2 + \xi_n^{\pm} k^2 \left(\frac{k}{M_{P}}\right)^n~, \qquad \omega_b^2 = k_b^2~, \nonumber \\
 &&  E_{e,\pm}^2 = p_e^2 + m_e^2 + \eta^{e,\pm}_n p_e^2 \left(\frac{p_e}{M_{P}}\right)^n~
\end{eqnarray}
with $(\omega, \vec k)$ the four-momenta for photons, and $(E, \vec{p}_e $ the corresponding four-momentum vectors for electrons; the suffix $b$ indicates a low-energy CMB photon, whose dispersion relations are assumed approximately the normal ones, as any QG correction is negligible due to the low values of energy and momenta. The +(-) signs indicate left(right) polarizations (photons) or helicities (electrons). Positive (negative) $\xi$ indicate subluminal (superluminal) refractive indices.
Upon the assumption of energy-momentum conservation in the process,  one arrives at kinematic equations for the threshold of the reaction (\ref{heirreact}), that is the minimum energy of the high-energy photon required
to produce the electron-positron pairs.

For the linear- or quadratic- suppression case (Cases I and II in the MAGIC analysis above, for which $n=1,2$ respectively (\ref{siglmdr})), one finds that, for the relevant subluminal photon refractive indices corresponding to the saturation of the $M_{QG1}$ lower bound in (\ref{mq12bound}), the threshold for pair production disappears for ultra-high-energy photons, and hence such photons should have been observed.
The non-observation of such photons implies constraints for the relevant parameters $\xi, \eta$
which are stronger by \emph{several orders of magnitude} than the bounds (\ref{mq12bound}) inferred from the MAGIC observations.

From the analysis of \cite{sigl} one concludes that in the case of linear Planck-mass suppression of the sub-luminal QG-induced modified dispersion relations for photons, of interest for   the QG-foam interpretation of the MAGIC results~\cite{MAGIC2}, parameters with size $\xi_1 > 10^{14}$ are ruled out. This exceeds the sensitivity of the MAGIC experiment (\emph{c.f.} (\ref{mq12bound})) to such Lorentz-symmetry violating effects by fifteen orders of magnitude !
Similar strong constraints are also obtained from the non observations of \emph{photon decay} ($\gamma \to e^+ e^-$), a process which, if there are modified dispersion relations, is in general allowed~\cite{sigl}.

\end{itemize}

\subsection{The String Foam Models Evade the above Constraints}

From the above discussion it becomes clear that any model
of refraction in space-time foam that exhibits effects at the level of the MAGIC sensitivity~\cite{MAGIC2}  (\ref{mq12bound}) should be characterised by the following specific properties:
\begin{itemize}

\item{(i)} photons are \emph{stable} (\emph{i.e.} do \emph{not} decay) but should exhibit a modified \emph{subluminal} refractive index with Lorentz-violating corrections that grow linearly with $E/(M_{\rm QG\gamma}c^2)$, where $M_{\rm QG\gamma}$ is close to the Planck scale,

\item{(ii)} the medium should not refract electrons, so as to avoid the synchrotron-radiation
constraints~\cite{crab,ems}, and

\item{(iii)} the coupling of the photons to the medium must be independent of photon polarization, so as not to have birefringence, thus avoiding the pertinent stringent constraints~\cite{uv,grb,crab2,macio}.

\item{(iv)} The formalism of local effective lagrangians should break down, in the sense that
there are quantum fluctuations in the total energy in particle interactions,
due to the presence of a quantum gravitational `environment', such that stringent constraints, which otherwise would have been imposed from the non-observation of ultra-high energy photons ($\hbar \omega > 10^{19}$ eV), are evaded.
\end{itemize}

 The string-foam models (both type IIA and type IIB) are characterized by all these properties~\cite{emnw,ems,emnnewuncert}, and thus avoid the stringent constraints. The absence of birefringence and the transparency  of foam to electrons or charged probes, for reasons of charge conservation, make the models surviving the stringent constraints from synchrotron radiation
 from distant Nebulae~\cite{crab,crab2}.

Moreover, the analysis in \cite{sigl} is based on exact energy momentum conservation in the process (\ref{heirreact}), stemming from the assumption of the local-effective lagrangian formalism for QG-foam. As we discussed in \cite{emngzk,emncomment}, and mentioned in section \ref{sec:nonlocaleft}, however, such a formalism is not applicable in the case of the recoiling D-particle space-time foam model, where the fluctuations of space-time or other defects of gravitational nature paly the r\^ole of an external environment, resulting in \emph{energy fluctuations} in the reaction (\ref{heirreact}). The presence of such fluctuations does affect the relevant energy-threshold equations, for the reaction to occur, which stem from kinematics, in such a way that the above stringent limits are no longer valid. In particular, we have seen that the pertinent anomalous dispersion terms are induced by the Finsler-like metric (\ref{opsmetric2}), which in turn arises by the distortion of space time due to the recoil of the D-particle space-time defect during its interaction with the photon. This metric is quadratically suppressed by the string scale, and hence the so-induced modified dispersion relations (\ref{drps}) contain anomalous terms quadratically suppressed by the QG scale, a case for which the constraints coming from ultra high energy cosmic rays are weak. It should be stressed again that in the string model the modified dispersion relations are disentangled from the time delays (\ref{timerecoil}) of the more energetic photons, which are linearly suppressed by the string scale. These delays are associated with the string uncertainty principle~\cite{yoneya} and thus are not represented within the local effective field theory framework.
We have also seen in section \ref{sec:trouble} that this disentanglement allows for a consistent interpretation of the MAGIC delays with those of the short GRB 090510 observed by FERMI within the string D-foam framework, provided of course
the foam is inhomogeneous.

In addition, as discussed in section \ref{sec:ncrecoil}, the form of the effective string coupling (\ref{effstringcoupl2}), and the induced metric (\ref{opsmetric2}), imply
an upper bound in the photon momentum transfer (\ref{novelgzk}), which characterizes the D-foam model as a result of the sub-luminal nature of the D-particle recoil velocity. These considerations imply that very high energy cosmic rays, when interacting with the D-foam, will have an extremely suppressed interaction rate (the string amplitudes are vanishing when
the recoil velocity approaches the speed of light), thus providing additional reasons~\cite{emncomment} for evading the strong constraints of \cite{sigl}. There are also phenomenologically realistic brane models, with large extra dimensions~\cite{pioline}, for which the ratio $M_s/g_s$ (\emph{i.e.} in our context the D-particle mass) may be of the order of the conventional GZK cutoff~\cite{GZK}, $10^{20}$ eV. In these models, in view of (\ref{novelgzk}), there would be no photons with energies higher than this that could not be completely absorbed when interact with the D-particle foam, thereby providing an explanation for their absence, in accordance with observations.

\section{Conclusions and Outlook \label{sec:5}}

In this review we have discussed a stringy version of a (Lorentz-Invariance-Violating) space-time foam model
as a candidate theory that provides an explanation for
the delays of the more energetic photons from celestial sources, as observed by MAGIC and FERMI Telescopes, in agreement with all other current astrophysical tests of Lorentz Violation.
 This of course does not mean that there are no conventional astrophysics explanations for these delayed photon arrivals, but it demonstrates clearly that  string theory (or better its modern version involving D-brane defects) is capable of explaining the observed photon delays, in agreement with all the other astrophysical data currently available.
 This is at least amusing, since it provides a framework for experimentally testing some models of string theory (entailing Lorentz Violation) at present or in the foreseeable future.

 The key point in the approach is the existence of space-time defects in the ground state of the model, whose topologically non-trivial interactions with the
 string states, via string-stretching during the capture process (\emph{c.f.} fig.~\ref{fig:restoring}),
 are mainly responsible for the observed delays. The latter are found proportional to the incident photon energy.
The peculiarity of the D-particle foam in being \emph{transparent} to charged particles (as a result of electric charge conservation requirements), evades the stringent constraints on linear Planck scale suppression refractive indices that would be induced by electron synchrotron radiation studies from Crab Nebula~\cite{crab,crab2}. Moreover, the absence of birefringence avoids the similarly stringent constraints on such models that otherwise would be imposed by galactic measurements~\cite{crab2}.
Finally, it worths mentioning that the string-stretched linear in energy time delays (\ref{delayonecapture}),(\ref{redshift}), when applied to neutrinos, can be flavour (\emph{i.e.} neutrino-species) independent, thus avoiding~\cite{ellis08} the stringent constraints that would be obtained from models of quantum gravity with flavour-dependent modifications of neutrino propagation and thus modifications in their oscillations~\cite{brustein}.

As already mentioned, the MAGIC result needs confirmation by other experiments like H.E.S.S.~\cite{hess2155,hessnew}, or other photon dedicated experiments, like FERMI (formerly GLAST)~\cite{glast}, where photons from Gamma Ray bursts will be observed. FERMI observed an extremely intense and short burst, GRB 090510, which imposes a stringent constraint on the allowed effective quantum gravity scale for linear anomalous refraction effects, $M_{\rm QG} > 1.2 M_P$.
Assuming that the measurement is correct, we could incorporate such a stringent constraint into our D-foam model, in a way consistent with the MAGIC and the other FERMI measurements on delayed photon arrivals, by invoking inhomogeneous D-particle foam models, exhibiting voids around the red-shift region of the GRB 090510. Such models can provide consistent cosmologies within our framework.

We must stress once more, however, that despite the apparent ``success'' of the D-foam model in dealing with
 stringent astrophysical constraints so far, one cannot as yet draw any safe conclusions on the validity of the model in Nature.
 Many more high-energy astrophysical photon measurements are needed in order to disentangle source from possible propagation effects due to fundamental physics.
 If a statistically significant population of data on photons from cosmic sources is collected, exhibiting refractive indices varying linearly with the distance of the source~\cite{mitsou}, as well as the photon energy, then this would be a very strong confirmation of the D-particle foam model, for reasons explained above. However, it must be noted that GRB's, which are expected to lead to statistically significant data in the next few years, will produce photons much lower in energies than the flares observed in AGN, and this could be a drawback. At any rate there are attempts to claim that observations from FERMI will have sensitivity close to the Planck scale~\cite{lamon} for such linear-suppression models soon. The case of GRB090510 is a perfect example of how a single measurement of a short, intense high energy burst, can place stringent limits on Lorentz Violation at the Planck scale. However, caution should be exercised here as to what one means by ``sensitivity at the Planck scale''. As demonstrated above (\emph{c.f}. (\ref{totaldelay})), the effective quantum gravity scale is actually a complicated function of many microscopic parameters in the model.
 Nevertheless, if many observations at various redshift regimes on delayed arrivals of cosmic photons become available in the future, then we shall be able to make some definite conclusions regarding the order of magnitude of possible quantum gravity effects and thus falsify models, such as the D-particle space-time foam.

From the above discussion it becomes hopefully clear to the reader that experimental searches for quantum gravity, if the latter is viewed as a medium, are highly model dependent as far as the sensitivity
to experimental falsification of the predictions of the underlying model is concerned. However, we are entering an era where low-energy (compared to Planck scale) physics experiments can already provide valuable information on the structure of space-time at the scales where Quantum Gravity is expected to set in. Very- and Ultra- high energy Astrophysics is at the forefront of such fundamental research. We therefore expect that, for the years to come, this branch of physics will proceed in parallel with terrestrial high energy experiments, such as the Large Hadron Collider launched at CERN recently, and be able to provide us soon with complementary
important information on the underlying fundamental structure of our Cosmos.
Time will then show whether quantum gravity can be finally put to experimental confirmation.

\section*{Acknowledgements}

I would like to thank the editors (K.K. Phua) of International Journal of Modern Physics A for the invitation to write this review.
It is a real pleasure to thank J. Ellis, K. Farakos, V.A. Mitsou, D.V. Nanopoulos, M. Sakellariadou, A. Sakharov, Sarben Sarkar, R. Szabo and R. Wagner for the collaboration and numerous discussions, especially on astro-particle physics issues covered in this review.
I would also like to thank S. Hossenfelder for the invitation to participate to the Workshop on
\emph{Experimental Search for Quantum Gravity} (Nordita, Stockholm (Sweden), July 12-16 2010), where some issues discussed in this review have been presented.
This work is partially supported by the European Union
through the Marie Curie Research and Training Network \emph{UniverseNet}
(MRTN-2006-035863).

\end{document}